\documentclass[10pt,a4paper]{article}
\usepackage{amssymb,amsmath} 
 \usepackage{amsthm}
\usepackage{mathrsfs}
\usepackage{graphicx}
\usepackage{fullpage} 
\usepackage{color}
\usepackage{bm}
\usepackage[all]{xy} 
\usepackage{makeidx}

\makeindex
\begin{document} 
\newtheorem{Def}{Definition}[section]
\newtheorem{Thm}{Theorem}[section]
\newtheorem{Proposition}{Proposition}[section]
\newtheorem{Lemma}{Lemma}[section]
\theoremstyle{definition}
\newtheorem*{Proof}{Proof}
\newtheorem{Postulate}{Postulate}[section]
\newtheorem{Corollary}{Corollary}[section]
\theoremstyle{remark}
\newtheorem{Example}{Example}[section]
\newtheorem{Remark}{Remark}[section]
\newcommand{\beq}{\begin{equation}}
\newcommand{\beqa}{\begin{eqnarray}}
\newcommand{\eeq}{\end{equation}}
\newcommand{\eeqa}{\end{eqnarray}}
\newcommand{\non}{\nonumber}
\newcommand{\fr}[1]{(\ref{#1})}
\newcommand{\abs}{\mathrm{abs}}
\newcommand{\Exp}{\mathrm{Exp}}
\newcommand{\tot}{\mathrm{tot}}
\newcommand{\Inv}{\mathrm{\,Inv}}
\newcommand{\B}{\mathrm{B}}
\newcommand{\C}{\mathrm{C}}
\newcommand{\G}{\mathrm{G}}
\renewcommand{\L}{\mathrm{L}}
\newcommand{\can}{\mathrm{can}}
\newcommand{\Diff}{\mbox{Diff}}
\newcommand{\Id}{\mathrm{Id}}
\newcommand{\bC}{\mbox{\boldmath {$C$}}}
\newcommand{\bK}{\mbox{\boldmath {$K$}}}
\newcommand{\bp}{\mbox{\boldmath {$p$}}}
\newcommand{\bx}{\mbox{\boldmath {$x$}}}
\newcommand{\by}{\mbox{\boldmath {$y$}}}
\newcommand{\bz}{\mbox{\boldmath {$z$}}}
\newcommand{\bF}{\mbox{\boldmath {$F$}}}
\newcommand{\bJ}{\mbox{\boldmath {$J$}}}
\newcommand{\bT}{\mbox{\boldmath {$T$}}}
\newcommand{\bR}{\mbox{\boldmath {$R$}}}
\newcommand{\bDelta}{\mathbf{\Delta}}
\newcommand{\ve}{{\varepsilon}}
\newcommand{\e}{\mathrm{e}}
\newcommand{\dr}{\mathrm{d}}
\newcommand{\Dr}{\mathrm{D}}
\newcommand{\F}{\mathrm{F}}
\newcommand{\mbbD}{\mathbb{D}}
\newcommand{\mbbE}{\mathbb{E}}
\newcommand{\mbbP}{\mathbb{P}}
\newcommand{\mbbR}{\mathbb{R}}
\newcommand{\mbbT}{\mathbb{T}}
\newcommand{\tA}{\widetilde A}
\newcommand{\tB}{\widetilde B}
\newcommand{\tC}{\widetilde C}
\newcommand{\chX}{\check{X}}
\newcommand{\cA}{{\cal A}}
\newcommand{\cB}{{\cal B}}
\newcommand{\cC}{{\cal C}}
\newcommand{\cD}{{\cal D}}
\newcommand{\cE}{{\cal E}}
\newcommand{\cF}{{\cal F}}
\newcommand{\cG}{{\cal G}}
\newcommand{\cH}{{\cal H}}
\newcommand{\cI}{{\cal I}}
\newcommand{\cK}{{\cal K}}
\newcommand{\cL}{{\cal L}}
\newcommand{\cM}{{\cal M}}
\newcommand{\cN}{{\cal N}}
\newcommand{\cO}{{\cal O}}
\newcommand{\cP}{{\cal P}}
\newcommand{\cR}{{\cal R}}
\newcommand{\Reeb}{{\cal R}}
\newcommand{\cS}{{\cal S}}
\newcommand{\cT}{{\cal T}}
\newcommand{\cU}{{\cal U}}
\newcommand{\cV}{{\cal V}}
\newcommand{\cY}{{\cal Y}}
\newcommand{\cZ}{{\cal Z}}
\newcommand{\tcA}{\widetilde{\cal A}}
\newcommand{\cAone}{   \underset{(1)}{ {\cal A} }     }
\newcommand\wh[1]{\widehat{#1}}
\newcommand{\DD}{{\cal D}}
\newcommand{\sympro}{\overset{s}{\otimes}}
\newcommand{\hash}{\#}
\newcommand{\GamCLamM}[1]{{\Gamma\mathbb{C}\Lambda^{{#1}}\cal{M}}}
\newcommand{\GamLamA}[1]{{\Gamma\Lambda^{{#1}}\cal{A}}}
\newcommand{\GamLamB}[1]{{\Gamma\Lambda^{{#1}}\cal{B}}}
\newcommand{\GamLamC}[1]{{\Gamma\Lambda^{{#1}}\cal{C}}}
\newcommand{\GamLamS}[1]{{\Gamma\Lambda^{{#1}}\cal{S}}}
\newcommand{\GamLamM}[1]{{\Gamma\Lambda^{{#1}}\cal{M}}}
\newcommand{\GamLamN}[1]{{\Gamma\Lambda^{{#1}}\cal{N}}}
\newcommand{\GTM}{{\Gamma T\cal{M}}}
\newcommand{\GTB}{{\Gamma T\cal{B}}}
\newcommand{\GTC}{{\Gamma T\cal{C}}}
\newcommand{\GT}[1]{{\Gamma T {#1}}}
\newcommand{\normM}[2]{\left(  {#1}\, , \, {#2} \right)}
\newcommand{\normU}[2]{\left\{ {#1}\, , \, {#2} \right\}}
\newcommand{\diag}[1]{\mbox{diag}\{\, {#1}\,\}}
\newcommand{\GtC}[2]{\Gamma T^{#1}_{#2}{\cal C}}
\newcommand{\GtM}[2]{\Gamma T^{#1}_{#2}{\cal M}}
\newcommand{\GtN}[2]{\Gamma T^{#1}_{#2}{\cal N}}
\newcommand{\Gt}[3]{\Gamma T^{#1}_{#2}{#3}}
\newcommand{\equp}[1]{\overset{\mathrm{#1}}{=}}
\newcommand{\lequp}[1]{\overset{\mathrm{#1}}{\leq}}
\newcommand{\gequp}[1]{\overset{\mathrm{#1}}{\geq}}
\newcommand{\lup}[1]{\overset{\mathrm{#1}}{<}}
\newcommand{\gup}[1]{\overset{\mathrm{#1}}{>}}
\newcommand{\rightup}[1]{\overset{\mathrm{#1}}{\longrightarrow}}
\newcommand{\leftup}[1]{\overset{\mathrm{#1}}{\Longleftarrow}}
\newcommand{\Leftrightup}[1]{\overset{\mathrm{#1}}{\Longleftrightarrow}}
\newcommand{\leftrightup}[1]{\overset{\mathrm{#1}}{\longleftrightarrow}}
\newcommand{\nLongleftup}{\Longleftarrow\hspace{-4.5mm}\diagup}
\newcommand{\fun}[1]{{#1}^{\hash}}
\newcommand{\wt}[1]{\widetilde{#1}}
\newcommand{\ol}[1]{\overline{#1}}
\newcommand{\sff}[1]{ {\sf{#1}}}
\newcommand{\sfL}{ {\sf L}}
\newcommand{\ddiv}{\mbox{div}}
\newcommand{\rank}{\mbox{rank}\,}
\newcommand{\bkt}[2]{\left\{\,{#1}\, ,\,{#2}\,\right\}}
\newcommand{\inp}[2]{\left\langle\,{#1}\, ,\,{#2}\,\right\rangle}
\newcommand{\inpr}[2]{\left(\,{#1}\, ,\,{#2}\,\right)}
\newcommand{\ave}[1]{\left\langle\,{#1}\, \right\rangle}
\newcommand{\avgg}[2]{\left\langle\,{#1}\, \right\rangle_{#2}}
\newcommand{\Leg}{{\mathfrak{L}}}
\newcommand{\ic}{\sqrt{-1}\,}
\newcommand{\ii}{\imath}
\newcommand{\bracket}[3]{\left[\,{#1},{#2}\,\right]_{#3}}
\newcommand{\LM}[2]{{\Lambda^{{#1}}_{{#2}}\,\cal{M}}}
\newcommand{\TM}[2]{{T^{{#1}}_{{#2}}\,\cal{M}}}
\newcommand{\nequiv}{\equiv\hspace{-.85em}/\,\,}

\renewcommand{\labelenumi}{{\bf\arabic{enumi}. }}
%

 

\title{
Contact geometric descriptions of vector fields on\\
dually flat spaces and their applications in electric circuit models
and nonequilibrium statistical mechanics
%
%
}

\author{Shin-itiro Goto\footnote{
Present affiliation : Department of Applied Mathematics and Physics, 
Graduate School of Informatics, Kyoto University,
Yoshida-Honmachi, Sakyo-ku, Kyoto 606-8501, Japan
}\\
Institute for  Molecular Science, \\
38 Nishigo-Naka, Myodaiji, Okazaki 444-8585, Japan
}

\maketitle
\begin{abstract}%
Contact geometry has been applied to various mathematical sciences,  
and it has been proposed that a contact manifold and a strictly 
convex function induce a dually flat space 
that is used in information geometry.  
Here, such a dually flat space is related to a Legendre submanifold in a 
contact manifold. 
In this paper contact geometric descriptions of vector fields on 
dually flat spaces are proposed on the basis of the theory of 
contact Hamiltonian vector fields. Based on these descriptions, two ways of  
lifting vector fields on Legendre submanifolds to contact manifolds  
are given. For some classes of these lifted vector fields, 
invariant measures in contact manifolds and  
stability analysis around Legendre submanifolds 
are explicitly given. 
Throughout this paper, Legendre duality is explicitly stated. In addition,   
to show how to apply these general methodologies 
to applied mathematical disciplines, 
electric circuit models and 
some examples taken from nonequilibrium statistical mechanics are analyzed.
\end{abstract}
\section{Introduction}
Contact geometry is often referred to as an odd-dimensional counterpart of 
symplectic geometry and then it has been studied from purely mathematical 
viewpoints\cite{Arnold1976}. 
Aside from its purely mathematical interest, 
there are several applications in 
science and foundation of engineering. These applications include 
equilibrium thermodynamics\cite{Hermann1973}\cite{MrugalaX}\cite{Schaft2007}, 
nonequilibrium thermodynamics\cite{Bravetti-Sep-2014}\cite{Goto2015},   
statistical mechanics\cite{Mrugala1990}\cite{Jurkowski2000}\cite{Bravetti-Aug-2014}, 
fluid mechanics\cite{Ghrist2007}, 
electromagnetism\cite{Dahl2004}, 
control theory\cite{Ohsawa2015}, 
statistical theory for non-conservative system\cite{Bravetti-JPhysA2014} 
and so on.
In general, if geometric theories of mathematical disciplines are ascribed to
a same geometry, then it can be expected that there are links among these 
disciplines. These links may give a unified picture of such disciplines.  
Such an example is found in contact geometry, where 
information geometry is linked to 
contact geometric thermodynamics\cite{Goto2015}\cite{Mrugala1990}.

There are at least three key objects in contact geometry 
for the applications mentioned above. 
The first one is a class of vector fields, the second one  
a class of submanifolds, and the third one the total Legendre transform.  
The first one 
is canonical vector field called 
contact Hamiltonian vector field. Such vector fields 
are in some sense analogous to Hamiltonian 
vector fields in symplectic geometry. 
It has been known 
that such a vector field is specified by a function defined on a contact 
manifold. This function is called a contact Hamiltonian.     
With this class of vector fields, one is able to describe time dependent  
phenomena by identifying a parameter of an integral curve with time. 
The second one is a class of Legendre submanifolds. Here, 
Legendre submanifold is defined as a  
maximal dimensional integrable submanifold of a contact manifold. 
Such a submanifold can locally be specified with a function on 
a contact manifold, called a generating function of a Legendre submanifold. 
Thus, integrable submanifolds in contact manifolds can locally be 
described by generating functions.
%
The third one is the total Legendre transform, and is well-known 
in mathematical sciences and foundation of engineering.  
This transform acting on functions enables one to
change coordinates systematically, and gives different views  
such as Hamiltonian and Lagrangian formulations in classical mechanics. 
When there is a pair of mathematical statements ascribed to the total  
Legendre transform, such a pair is referred to as Legendre duality. 
In geometric thermodynamics formulated with 
contact geometry, these three keys are essential. Contact Hamiltonian 
vector fields are used for describing thermodynamic processes,  
Legendre submanifolds are used for specifying a subspace where the first law of 
thermodynamics is satisfied, and the total Legendre transform 
is used as well as 
non-geometric thermodynamics. 

An approach to develop such contact geometric theories 
of mathematical sciences 
is therefore to give various relations on these three keys. 
One subject expected to be developed with the three keys in 
contact geometry is   
information geometry, where 
information geometry is a geometrization of mathematical statistics\cite{AN}.   
As briefly mentioned earlier in this section, 
it has been pointed out that 
information geometry is connected to contact geometry. Although 
ideas in information geometry and those in contact geometry 
are well-developed,   
they have not been communicated.  
We then feel that a way to connect between them  should be explored. 
Another 
approach to develop contact geometric theories  
is to give 
lifted vector fields on contact manifolds from those on Legendre submanifolds. 
This is because in contact geometric thermodynamics and in 
information geometry, vector fields are often studied only on 
manifolds that are effectively Legendre submanifolds of contact 
manifolds, and some extension of vector fields, 
not only on Legendre submanifolds, 
is expected to new applications.    
This class of lifts has been discussed in the literature. 
In Ref.\,\cite{Favache2009}, an equivalence of vector fields lifted  to  
a contact manifold and stability of fixed points of contact vector fields 
being restricted to the Legendre submanifold has been discussed. 
In Ref.\,\cite{Estay2011},  it has been proposed 
how to solve the problem of matching  
two vector fields in thermodynamic systems, and 
stability of a class of dynamical systems lifted  from a 
Legendre submanifold has been analyzed without any Lyapunov function on 
a contact manifold.  

In this paper, some basic notations are fixed in \S\ref{sec-review}.  
In \S\ref{sec-contact-geometric-description-dually-flat-space},
it is shown that vector fields on dually flat spaces can be described as 
restricted contact Hamiltonian vector fields on Legendre submanifolds, 
where such restricted vector fields have been developed in the 
context of contact geometric thermodynamics. 
In this way, some theorems in information geometry are rewritten in terms of 
a contact geometric language. 
By showing that electric circuit models  
can be seen as dynamical systems  
on dually flat spaces, we argue how the present contact geometric 
and information geometric methodologies 
apply to engineering problems. 
In \S\ref{sec-lift},  
it is shown that a class of flows   
defined on Legendre submanifolds is lifted to a contact manifold, 
and shown that lifted flows of some classes asymptotically 
approach to the Legendre submanifolds. 
Statistical properties such as phase compressibilities and 
invariant measures for lifted vector fields 
are argued as well. 
To show how these general theories can be used in physics, 
examples are constructed in nonequilibrium statistical mechanics. 
Since a relaxation dynamics of 
a spin model in contact with time-independent 
heat bath and time-independent external magnetic field 
has been well-described with contact geometry in Ref.\,\cite{Goto2015}, 
an extension of this dynamics may be of interest. 
A modification of such system is considered in this article, and  
it is shown that a relaxation process can be described in a system
that is in contact with a time-dependent 
external magnetic field and time-independent heat bath. As another example, 
a class of phenomenological equations with the Onsager coefficients is 
also discussed.   
In \S\ref{sec-generating-function-preserving-lifts},
another scheme of lifting vector fields is shown where
the values of  generating functions 
for Legendre submanifolds 
are conserved
along vector fields. 
In this scheme, 
higher dimensional contact manifolds are to be employed.  
Statistical properties of such lifted vector fields are discussed as well. 
To show 
how this scheme is applied to engineering problems,
electric circuit models in a thermal 
environment are analyzed. 
Throughout this paper, Legendre duality for mathematical statements  
is explicitly stated. 


\section{Preliminaries}
\label{sec-review}
In this section we give a brief summary of contact geometry 
and information geometry  
in order to describe statements that will be shown in 
Secs.\ref{sec-contact-geometric-description-dually-flat-space}--\ref{sec-generating-function-preserving-lifts}.
\subsection{Mathematical symbols, definitions, and known facts}
\label{sec-Mathematical-preliminaries}
Throughout this paper, geometric objects are assumed smooth.  
A point on an $n$-dimensional manifold $\xi\in\cM$ 
is often identified with a set of values of 
local coordinates $x(\xi)=\{\,x^{\,1}(\xi),\ldots,x^{\,n}(\xi)\,\}$.  
A set of vector fields on a manifold $\cM$ is denoted $\GTM$,
the tangent space at $\xi\in\cM$ as $T_{\xi}\cM$, 
the cotangent space at $\xi\in\cM$ as $T_{\xi}^{\,*}\cM$, 
a set of $q$-form fields $\GamLamM{q}$ with $q\in\{\,0,\ldots,\dim\cM\,\}$,  
and a set of tensor fields $\GtM{q'}{q}$ with 
$q,q'\in\{\,0,1,\ldots\,\}$. 
If a tensor field $g\in\GtM{0}{2}$ 
on an $n$-dimensional manifold $\cM$ satisfies 
(i) $g|_{\xi}(X,Y)=g|_{\xi}(Y,X)$, (ii) $g|_{\xi}(X,X)\geq 0$, 
(iii) $g|_{\xi}(X,X)=0$ iff $X=0$, 
for $X,Y\in T_{\,\xi}\cM$, then $g$ is referred to as 
a Riemannian metric tensor field. An $n$-dimensional manifold together with a 
Riemannian metric tensor field $g$ is denoted $(\cM,g)$ 
and this is referred to as an $n$-dimensional Riemannian manifold.
If a tensor field $g\in\GtM{0}{2}$ on an $n$-dimensional manifold 
$\cM$ satisfies 
(i) $g|_{\xi}(X,Y)=g|_{\xi}(Y,X)$, 
(ii) $g|_{\xi}(Z,Y)=0,(\forall Z\in T_{\xi}\cM) \Longrightarrow Y=0$, 
for $X,Y\in T_{\,\xi}\cM$, then $g$ is referred to as 
a pseudo-Riemannian metric tensor field. 
An $n$-dimensional manifold together with a 
pseudo-Riemannian metric tensor field $g$ is denoted $(\cM,g)$ 
and this is referred to as an $n$-dimensional pseudo-Riemannian manifold.
To express tensor fields the direct product is denoted $\otimes$.  
Einstein notation, when an index variables appear twice in a single 
term it implies summation of all the values of the index, 
is adopted. The Kronecker delta is denoted 
$\delta_{\,a}^{\,b},\delta_{\,ab},\ldots,$ 
and its value is unity when $a=b$, and zero when $a\neq b$.   
The exterior derivative acting on $\GamLamM{q}$ is denoted 
$\dr:\GamLamM{q}\to\GamLamM{q+1}$, and 
the interior product operator with $X\in\GTM$ as $\ii_X:\GamLamM{q}\to\GamLamM{q-1}$.  
Given a map $\Phi$ between two manifolds, the pull-back is denoted $\Phi^*$, 
and the push-forward $\Phi_*$.  
Then one can define the Lie derivative acting on tensor fields 
with respect to 
$X\in\GTM$ denoted $\cL_{X}:\GtM{q'}{q}\to\GtM{q'}{q}$. 
It follows that $\cL_{X}\beta=(\ii_X\dr+\dr\ii_X)\beta,$ 
for any $\beta\in\GamLamM{q}$, which is referred to as the Cartan formula. 
One can also define a derivative along a given vector field $X$, called the 
covariant derivative, denoted $\nabla_X:\GtM{0}{q}\to\GtM{0}{q}$. The action is 
explicitly given by specifying the connection coefficients 
$\Gamma_{ab}^{\ \ c}, (a,b,c\in\{\,1,\ldots,\dim\cM\,\})$ such that 
$\nabla_{X_{\,a}}X_{\,b}=\Gamma_{ab}^{\ \ c}X_{\,c}$ where 
$\{\,X_{\,1},\ldots,X_{\,n}\,\}\in\GTM$ 
is a basis. For a given $S\in\GtM{0}{q}$, an object $\nabla S\in\GtM{0}{q+1}$
is defined such that  
$(\nabla S)(X,Y_{\,1},\ldots,Y_{\,q})=(\nabla_{\,X}S)(Y_{\,1},\ldots,Y_{\,q})$, where  
$X,Y_{\,1},\ldots,Y_{\,q}\in \GTM$.  
For example it can be shown that $\nabla f=\dr f$ for $f\in\GtM{0}{0}$, 
and that $(\nabla\dr f)(Y,\partial/\partial x^{\,a})=Y^{\,b}(\partial^2\,f/\partial x^{\,a}\partial x^{\,b}-\Gamma_{ba}^{\ \ c}\,\partial f/\partial x^{\,c})$, where
$Y=Y^{\,b}\partial/\partial x^{\,b}\in\GTM$.

\begin{Def}
( Dynamical system and its flow ) : 
Let $\cM$ be an $n$-dimensional manifold, $x=\{\,x^{\,1},\ldots,x^{\,n}\,\}$ 
a local coordinate system,  
$\mbbT$ a subset of the real numbers $\mbbR$, and 
$F=\{\,F^{\,1},\ldots,F^{\,n}\,\}$ a set of functions 
on $\cM$.  
Consider a set of ordinary differential equations 
$$
\frac{\dr x^{\,a}}{\dr t}
=F^{\,a}(x).\qquad t\in \mbbT\subset\mbbR
$$
If there exists the unique solution $\phi(t,x_{\,0})$ with 
$x_{\,0}$ being the initial point at $t=0$ for all $x_{\,0}\in\cM$ and 
all $t\in \mbbT$,  
then this set of ordinary differential equations is referred to as a 
dynamical system on $\cM$.  Define $\phi_{\,t}:\cM\to\cM$ with a fixed 
$t\in \mbbT$ 
such that $\phi_{\,t}=\phi(t,-)$. Then, $\{\phi_{\,t}\}_{\,t\in\, \mbbT}$ 
is referred to as a flow. In addition, 
one can have a vector field $X=F^{\,a}\partial/\partial x^{\,a}$ on $\cM$. 
This $X$ is referred to as a 
vector field associated with a dynamical system.
\end{Def}

\begin{Def}
\label{def-volume-form-compressibility}
( Volume-form divergence and phase compressibility, \cite{Ezra02},\cite{Ezra04} ) : 
Let $\cM$ be an $n$-dimensional manifold, $X$ a vector field on $\cM$, and 
$\Omega$ a non-vanishing $n$-form. 
Define $\kappa_{\Omega}:T_{\xi}\cM\to\mbbR$, $(\xi\in\cM)$ such that 
\beq
\cL_{\,X}\Omega
=\kappa_{\Omega}(X)\,\Omega.
\label{def-phase-compressibility}
\eeq
Then, $\kappa_{\Omega}(X)$ 
is referred to as a volume-form divergence. 
If $X$ is a vector filed associated with a dynamical system on $\cM$, then $\kappa_{\Omega}(X)$
is referred to as a  phase compressibility.
\end{Def}
\begin{Remark}
In the context of dynamical systems theory, 
volume-form divergences are simply referred to as divergences, where 
a volume-form is defined in differential geometry 
as an $n$-form that does not vanish on any point on $\cM$. 
In this paper, another divergence in a different context 
will be introduced and that will be distinguished from the volume-form divergence.  
\end{Remark}
\begin{Def}
\label{def-invariant-measure-general}
( Continuity equation for volume form and invariant measure, 
\cite{Ezra02},\cite{Ezra04} ) : 
Let $\cM$ be an $n$-dimensional manifold, $\mbbT$ a subset of $\mbbR$, 
$\Omega$ a non-vanishing $n$-form that can depend on $t\in \mbbT$,  and 
$X$ a vector field associated with a dynamical system 
$\dr x^{\,a}/\dr t=F^{\,a}(x)$, $(a\in\{1,\ldots,n\})$ with 
$\{\,F^{\,1},\ldots,F^{\,n}\,\}$ being functions.  
The equation for $\Omega$ 
$$
\frac{\partial}{\partial t}\Omega+\cL_{X}\Omega
=0,
$$
is referred to as the continuity equation for $\Omega$. 
In addition, 
a solution that satisfies 
$$
\frac{\partial}{\partial t} \Omega=0,
\qquad\mbox{and}\qquad 
\cL_{\,X}\Omega 
=0,
$$
is referred to as an invariant measure for $X$.  
\end{Def}
\begin{Remark}
The equation for a function $f$ 
$$
\frac{\partial f}{\partial t}+(Xf)+\kappa_{\Omega}(X)f
=0,
$$
is referred to as the generalized Liouville equation. This is often studied in 
the literature, and is derived from the continuity equation for $\Omega$ 
as follows. 
Substituting  
$\Omega=f\,\Omega_{\,0}$ with $\Omega_{\,0}$ being a non-vanishing $n$-form 
that does not depend on $t$ 
into the continuity equation for $\Omega$, one has that  
$$
\left[\,\frac{\partial}{\partial t}f
+(Xf)\right]\,\Omega_{\,0}+f\,\cL_{X}\Omega_{\,0}
=0.
$$
With this, $\cL_{X}\Omega_{\,0}=\kappa_{\,\Omega_{\,0}}(X)\Omega_{\,0}$ 
as in \fr{def-phase-compressibility}, and $\Omega_{\,0}$ is then rewritten as 
$\Omega$,  
one has the generalized Liouville equation.  
\end{Remark}

\begin{Def}
\label{def-contact-manifold}
( Contact manifold ) : 
Let $\cC$ be a $(2n+1)$-dimensional manifold, and $\lambda$ a 
one-form on $\cC$ such that
$$
\lambda\wedge\underbrace{\dr \lambda\wedge\dr\lambda\cdots\wedge\dr\lambda}_{n}
\neq 0,
$$
at any point on $\cC$.
If $\cC$ carries $\lambda$, then $(\,\cC,\lambda\,)$ 
is referred to as a contact manifold and $\lambda$ a contact form. 
\end{Def}
It should be noted that there are other definitions of contact manifold. 
However the definition above is used in this paper.  
The following definition is not commonly used in the literature. 
\begin{Def}
( Standard volume-form ) : 
The $(2n+1)$-form in Definition\,\ref{def-contact-manifold} 
is referred to as the standard volume-form :  
\beq
\Omega_{\,\lambda}
=\lambda\wedge\dr\lambda\wedge\cdots\wedge\dr\lambda.
\label{def-standard-volume-form-contact-manifold}
\eeq
\end{Def}


There is a standard local coordinate system.
\begin{Thm}
\label{theorem-canonical-coordinates}
( Existence of particular coordinates ) : 
There exist local $(2n+1)$ coordinates 
$(x,p,z)$ with $x=\{\,x^{\,1},\ldots,x^{\,n}\,\}$ and 
$p=\{\,p_{\,1},\ldots,p_{\,n}\,\}$,  
in which $\lambda$ has the form 
\beq
\lambda
=\dr z-p_{\,a}\,\dr x^{\,a}.
\label{contact-form-standard-Darboux}
\eeq
\end{Thm} 
\begin{Def}
( Canonical coordinates or Darboux coordinates  ) : 
The $(2n+1)$ coordinates $(x,p,z)$ introduced in 
Theorem \ref{theorem-canonical-coordinates}
are referred to as the canonical coordinates, 
or the Darboux coordinates. 
\end{Def}
In addition to the above coordinates,
ones in which $\lambda$ has the 
form $\lambda=\dr z+p_{\,a}\,\dr x^{\,a}$ are also used in the literature.
In this paper \fr{contact-form-standard-Darboux} is used. 

Given a contact manifold, 
there exists a unique vector field that is defined as follows.
\begin{Def}
( Reeb vector field or characteristic vector field ) : 
Let $(\,\cC,\lambda\,)$ be a contact manifold, and $\Reeb$ a vector field 
on $\cC$. 
If $\Reeb$ satisfies 
\beq
\ii_{\,\Reeb}\, \dr\lambda
=0,\qquad\mbox{and}\qquad 
\ii_{\Reeb}\,\lambda
=1,
\label{def-Reeb-vector}
\eeq
then $\Reeb\in\GTC$ is referred to as the Reeb vector field, or  
the characteristic vector field.  
\end{Def}


When $\lambda$ is given, 
a coordinate expression for $\Reeb$ is given as follows.
\begin{Proposition}
( Coordinate expression of the Reeb vector field ) : 
Let $(\,\cC,\lambda\,)$ be a contact manifold, and $\Reeb$ 
the Reeb vector field.
If the canonical coordinates $(x,p,z)$ are such that 
$\lambda=\dr z\pm p_{\,a}\,\dr x^{\,a}$  
with $x=\{\,x^{\,1},\ldots,x^{\,n}\,\}$ and $p=\{\,p_{\,1},\ldots,p_{\,n}\,\}$, 
then 
\beq
\Reeb
=\frac{\partial }{\partial z}.
\label{coordinate-expression-Reeb}
\eeq
\end{Proposition}

The following submanifold plays various roles in applications of  
contact geometry.
\begin{Def}
\label{def-Legendre-submanifold}
( Legendre submanifold ) : 
Let $(\,\cC,\lambda\,)$ be a contact manifold, and 
$\cA$  a submanifold of $\cC$. 
If $\cA$ is 
a maximal dimensional integral submanifold of $\lambda$,  
then $\cA$ is referred to as a Legendre submanifold.
\end{Def}

The following theorem states the dimension of a Legendre submanifold for 
a given contact manifold.
\begin{Thm}
\label{thm:max-integral-submanifold-is-n}
( Maximal dimensional integral submanifold ) : 
On a $(2n+1)$-dimensional contact manifold $(\,\cC,\lambda\,)$,  
a maximal dimensional integral submanifold of $\lambda$  
is equal to $n$.  
\end{Thm}

Combining Theorem \ref{thm:max-integral-submanifold-is-n}
 and Definition \ref{def-Legendre-submanifold}, 
one concludes the following. 
\begin{Thm} 
( Dimension of Legendre submanifolds  ) : 
The dimension of any Legendre submanifold of a 
$(2n+1)$-dimensional contact manifold is $n$.
\end{Thm}

The following theorem shows
the explicit local expressions of Legendre submanifolds in terms of 
canonical coordinates. 
\begin{Thm}
\label{theorem-Legendre-submanifold-theorem-Arnold}
( Local expressions of Legendre submanifolds, \cite{Arnold1976} ) :  
Let $(\,\cC,\lambda\,)$ be a $(2n+1)$-dimensional contact manifold,  
and $(x,p,z)$ the canonical coordinates such that 
$\lambda=\dr z-p_{\,a}\,\dr x^{\,a}$  
with $x=\{\,x^{\,1},\ldots,x^{\,n}\,\}$ and $p=\{\,p_{\,1},\ldots,p_{\,n}\,\}$. 
For any partition $I\cup J$ of the set of indices $\{\,1,\ldots,n\,\}$ into 
two disjoint subsets $I$ and $J$, and for a function $\phi(x^J,p_I)$ of 
$n$ variables $p_{\,i},i\in I$, and $x^{\,j},j\in J$ the $(n+1)$ equations
\beq
x^i=-\,\frac{\partial\phi}{\partial p_i},\qquad
p_j=\frac{\partial\phi}{\partial x^j},\qquad 
z=\phi-p_i\frac{\partial\phi}{\partial p_i},
\label{Legendre-submanifold-theorem-Arnold}
\eeq
define a Legendre submanifold. Conversely, every 
Legendre submanifold of $(\,\cC,\lambda\,)$ 
in a neighborhood of any point is    
defined by these equations for at least one of the $2^n$ possible choices 
of the subset $I$.
\end{Thm}
\begin{Def}
( Legendre submanifold generated by a function ) : 
The function $\phi$ used in 
Theorem\,\ref{theorem-Legendre-submanifold-theorem-Arnold}  
is referred to as a generating function of the Legendre submanifold. 
If a  Legendre submanifold $\cA$ is expressed 
as \fr{Legendre-submanifold-theorem-Arnold}, then $\cA$ is referred to as 
a Legendre submanifold generated by $\phi$. 
\end{Def}

The following are examples of local expressions for 
 Legendre submanifolds.  
\begin{Example}
\label{example-Arnold-Legendre-submanifold-psi}
Let $(\,\cC,\lambda\,)$ be a $(2n+1)$-dimensional contact manifold,
$(x,p,z)$ the canonical coordinates such that 
$\lambda=\dr z-p_{\,a}\,\dr x^{\,a}$  
with $x=\{\,x^{\,1},\ldots,x^{\,n}\,\}$ and $p=\{\,p_{\,1},\ldots,p_{\,n}\,\}$, 
and $\psi$ a function of $x$ only.   
The Legendre submanifold $\cA_{\,\psi}$ generated by $\psi$ with 
$\Phi_{\,\cC\cA\psi}:\cA_{\,\psi}\to\cC$ being an embedding 
is such that 
\beq
\Phi_{\,\cC\cA\psi}\cA_{\,\psi}
=\left\{\ (x,p,z)\in\cC \ \bigg|\ 
p_j=\frac{\partial\psi}{\partial x^{\,j}},\ \mbox{and}\ 
z=\psi(x),\quad j\in \{\,1,\ldots,n\,\}
\ \right\}. 
\label{example-psi-Legendre-submanifold}
\eeq
The relation between this $\psi$ and $\phi$ of   
\fr{Legendre-submanifold-theorem-Arnold} is $\psi(x)=\phi(x)$ with 
$J=\{\,1,\ldots,n\,\}$.
One can verify that $\Phi_{\,\cC\cA\psi}^{\ \ \ \  *}\lambda=0$. 
\end{Example}
\begin{Example}
\label{example-Arnold-Legendre-submanifold-varphi}
Let $(\,\cC,\lambda\,)$ be a $(2n+1)$-dimensional contact manifold,
$(x,p,z)$ the canonical coordinates such that 
$\lambda=\dr z-p_{\,a}\,\dr x^{\,a}$ 
with $x=\{\,x^{\,1},\ldots,x^{\,n}\,\}$ and $p=\{\,p_{\,1},\ldots,p_{\,n}\,\}$, 
and $\varphi$ a function of $p$ only. 
The Legendre submanifold $\cA_{\,\varphi}$ generated by $-\,\varphi$ 
with $\Phi_{\,\cC\cA\varphi}:\cA_{\,\varphi}\to\cC$ being 
an embedding is such that 
\beq
\Phi_{\,\cC\cA\varphi}\cA_{\,\varphi}
=\left\{\ (x,p,z)\in\cC \ \bigg|\ 
x^{\,i}=\frac{\partial\varphi}{\partial p_{\,i}},\ \mbox{and}\ 
  z=p_{\,i}\frac{\partial\varphi}{\partial p_{\,i}}-\varphi(p),\quad i\in 
\{\,1,\ldots,n\,\}
\ \right\}. 
\label{example-varphi-Legendre-submanifold}
\eeq
The relation between this $\varphi$ and $\phi$ of   
\fr{Legendre-submanifold-theorem-Arnold} is $\varphi(p)=-\,\phi(p)$ with
$I=\{\,1,\ldots,n\,\}$.
One can verify that $\Phi_{\,\cC\cA\varphi}^{\ \ \ \ *}\lambda=0$.
\end{Example}

One can choose a function $\psi$ in 
Example\,\ref{example-Arnold-Legendre-submanifold-psi} to generate 
$\cA_{\,\psi}$,  and 
$\varphi$ in Example \ref{example-Arnold-Legendre-submanifold-varphi} to 
generate $\cA_{\,\varphi}$  independently, 
and in this case there is no relation between 
$\cA_{\,\psi}$ and $\cA_{\,\varphi}$ in general. 
On the other hand, when $\psi$ is strictly convex, and $\varphi$ 
is carefully chosen, it can be shown that 
there is a relation between $\cA_{\,\psi}$ and $\cA_{\,\varphi}$. 
To discuss such a case, 
the following transform should be introduced. The convention is 
adopted to that in information geometry. 
Note that several conventions exist in the literature. 
\begin{Def}
( Total Legendre transform )  : 
Let $\cM$ be an $n$-dimensional manifold,
$x=\{\,x^{\,1},\ldots,x^{\,n}\,\}$ coordinates, 
and $\psi$ a function of $x$.
Then the total Legendre transform of $\psi$ with respect to $x$ 
is defined to be  
\beq
\Leg[\psi](p)
:=\sup_{x}\left[\,x^{\,a}p_{\,a}-\psi(x)\,\right],
\label{def-total-Legendre-transform}
\eeq
where $p=\{\,p_{\,1},\ldots,p_{\,n}\,\}$.
\end{Def}

From this definition, one has several formulae that will be used in 
Secs.\ref{sec-contact-geometric-description-dually-flat-space}--\ref{sec-generating-function-preserving-lifts}.
\begin{Thm}
\label{theorem-Legendre-tranform-formula}
( Formulae involving the total Legendre transform ) : 
Let $\cM$ be an $n$-dimensional manifold, 
$x=\{\,x^{\,1},\ldots,x^{\,n}\,\}$ coordinates,
$\psi\in\GamLamM{0}$ a strictly convex function of $x$ only, and 
$\varphi$ the function of $p$ obtained by the total Legendre 
transform of $\psi$ with respect to $x$ where $p=\{\,p_{\,1},\ldots,p_{\,n}\,\}$ :
$\varphi(p)=\Leg[\psi](p)$. Then, for each $a$ and fixed $p$, the equation 
$$
p_{\,a}
=\left.\frac{\partial\psi (x)}{\partial x^{\,a}}\right|_{x=x_*}
=\frac{\partial\psi (x_{\,*})}{\partial x_{\,*}^{\,a}},
$$
has the unique solution
$x_{\,*}^{\,a}=x_{\,*}^{\,a}(p), (a\in\{\,1,\ldots,n\,\})$. 
In addition it follows that 
$$
\varphi(p)
=x_{\,*}^{\,a}p_{\,a}-\psi(x_{\,*}),\qquad
\frac{\partial\varphi}{\partial p_{\,a}}
=x_{\,*}^{\,a},\qquad
\delta_{\,b}^{\,a}
=\frac{\partial^2\psi}{\partial x_{\,*}^{\,b}\partial x_{\,*}^{\,l}}
\frac{\partial^2\varphi}{\partial p_{\,a}\partial p_{\,l}},
$$
and 
$$
\det\left(\frac{\partial^2\,\psi}{\partial x^{\,a}\partial x^{\,b}}\right)>0,
\qquad
\det\left(\frac{\partial^2\,\varphi}{\partial p_{\,a}\partial p_{\,b}}\right)>0.
$$
\end{Thm}

A way to describe dynamics on a contact manifold is to introduce a 
continuous diffeomorphism with a parameter. 
First, one defines a diffeomorphism on a contact manifold.
\begin{Def}
\label{def-contact-diffeomorphism}
( Contact diffeomorphism ) : 
Let $(\,\cC,\lambda\,)$ be a $(2n+1)$-dimensional 
contact manifold, 
and $\Phi:\cC\to\cC$  a diffeomorphism. If it follows that 
$$
\Phi^*\lambda
=f\,\lambda,
$$  
where $f\in\GamLamC{0}$ is a function that does not vanish at  
any point of $\cC$, then the map $\Phi$ is referred to as 
a contact diffeomorphism. 
\end{Def}

\begin{Remark}
It follows
that $\Phi$ preserves the contact structure, 
$\ker\lambda:=\{\,X\in\GTC\, |\,\ii_X\lambda=0\,\}$, 
but does not preserve the original contact form.
\end{Remark}

In addition to this diffeomorphism, one 
can introduce  one-parameter groups as follows.
\begin{Def}
( One-parameter group of continuous contact transformations ) : 
Let $(\,\cC,\lambda\,)$ be a $(2n+1)$-dimensional contact manifold, 
and $\Phi_{t}:\cC\to\cC$ a diffeomorphism with $t\in\mathbb{R}$ that satisfies
$\Phi_0=\Id_{\,\cC}$ and $\Phi_{t+s}=\Phi_{t}\circ\Phi_s, (t,s\in\mathbb{R})$,  
where $\Id_{\,\cC}$ is such that $\Id_{\,\cC}\xi=\xi$ for all $\xi\in\cC$. 
If it follows that  
$$
\Phi_{t}^*\lambda
=f_t\,\lambda,
$$  
where $f_{\,t}\in\GamLamC{0}$ is a function that does not vanish at  
any point of $\cC$, then the $\Phi_{t}$ is referred to as 
a one-parameter group of continuous contact transformations. 
If $t,s\in \mbbT$ with some $\mbbT\subset \mathbb{R}$, 
then it is referred to as 
a one-parameter local transformation group of continuous 
contact transformations.
\end{Def}
A contact vector field is defined as follows. 
\begin{Def}
\label{def-contact-vector}
( Contact vector field ) : 
Let $(\,\cC,\lambda\,)$ be a contact manifold, and $X$ a vector field on $\cC$.
If $X$ satisfies 
$$
\cL_X\lambda
=f\,\lambda,
$$
where $f$ is 
a function
on $\cC$, then 
$X$ is referred to as a contact vector field. 
\end{Def}

A  one-parameter (local) transformation groups
is realized by integrating the following vector field.
\begin{Def}
( Contact vector field associated to a contact Hamiltonian ) : 
Let $(\,\cC,\lambda\,)$ be a contact manifold, 
$\Reeb$ the Reeb vector field, 
$h$  a function on $\cC$, and 
$X_{\,h}$  a vector field.   
If $X_{\,h}\in\GTC$ satisfies 
\beq
\ii_{\,X_{\,h}}\lambda
=h,\qquad\mbox{and}\qquad 
\ii_{\,X_{\,h}}\dr \lambda
=-\,(\,\dr h-(\,\Reeb h\,)\,\lambda\,),
\label{def-contact-vector-Hamiltonian}
\eeq
then $X_{\,h}$ 
is referred to as a contact vector field \underline{associated} to 
a function $h$ or a contact Hamiltonian vector field. 
In addition $h$ is referred to as a contact Hamiltonian. 
\end{Def}

\begin{Remark}
The definition \fr{def-contact-vector-Hamiltonian} 
and the Cartan formula give 
\beq
\cL_{\,X_{\,h}}\lambda
=(\,\Reeb h\,)\,\lambda.
\label{cartan-contact-Hamiltonian-vector}
\eeq
\end{Remark}
\begin{Remark}
With \fr{def-contact-vector-Hamiltonian}, 
\fr{cartan-contact-Hamiltonian-vector}, 
and the formula 
$\cL_{\,X}\ii_{\,X}\alpha=\ii_{\,X}\cL_{\,X}\alpha$, 
where $X$ is an arbitrary vector field and $\alpha$ an arbitrary $q$-form 
field ($q\in\{0,1,\ldots\}$),  
one has that 
\beq
X_{\,h}h
=\cL_{\,X_{\,h}}h
=\cL_{\,X_{\,h}}(\,\ii_{\,X_{\,h}}\lambda\,)
=\ii_{\,X_{\,h}}(\,\cL_{\,X_{\,h}}\lambda\,)
=(\,\Reeb\,h\,)(\,\ii_{\,X_{\,h}}\lambda\,)
=(\,\Reeb\,h\,)\,h.
\label{Xh=hRh}
\eeq
\end{Remark}

From \fr{cartan-contact-Hamiltonian-vector} and 
Definition\,\ref{def-contact-vector}, one has the following.  
\begin{Thm}
( Relation between a contact Hamiltonian vector field and a 
contact vector field ) :   
Let $(\,\cC,\lambda\,)$ be a contact manifold, $h$ a contact Hamiltonian,  
and $X_h$ a contact Hamiltonian vector field. 
Then $X_h$ is a contact vector field.
\end{Thm}

Local expressions of a contact Hamiltonian vector field
\fr{def-contact-vector-Hamiltonian} are  calculated 
as follows.
\begin{Proposition}
( Local expression of contact Hamiltonian vector field ) : 
Let $(\,\cC,\lambda\,)$ be a $(2n+1)$-dimensional contact manifold, 
$h$ a contact Hamiltonian,  
$X_h$ a contact Hamiltonian vector field, and $(x,p,z)$ 
the canonical coordinates 
such that $\lambda=\dr z-p_{\,a}\,\dr x^{\,a}$ 
with $x=\{\,x^{\,1},\ldots,x^{\,n}\,\}$ and $p=\{\,p_{\,1},\ldots,p_{\,n}\,\}$. 
Then  
$$
X_h
=\dot{x}^{\,a}\frac{\partial}{\partial x^{\,a}}
+\dot{p}_{\,a}\frac{\partial}{\partial p_{\,a}}
+\dot{z}\frac{\partial}{\partial z},
$$
where $\dot{}$ denotes the differential 
with respect to a parameter $t\in\mathbb{R}$, or $t\in \mbbT$ 
with some $\mbbT\subset\mathbb{R}$, 
and 
\beq
\dot{x}^{\,a}
=-\,\frac{\partial h}{\partial p_{\,a}},\qquad
\dot{p}_{\,a}
=\frac{\partial h}{\partial x^{\,a}}
+p_{\,a}\frac{\partial h}{\partial z},\qquad
\dot{z}
=h-p_{\,a}\frac{\partial h}{\partial p_{\,a}},\qquad a\in\{\,1,\ldots,n\,\}.
\label{contact-Hamiltonian-vector-components}
\eeq
\end{Proposition}
\begin{Remark}
When $\lambda=\dr z+p_{\,a}\,\dr x^{\,a}$,  signs in   
\fr{contact-Hamiltonian-vector-components} are changed. 
\end{Remark}

The following theorem is well-known, and has been 
used in the literature of geometric thermodynamics.  
\begin{Thm}
\label{theorem-Mrugala-contact-Hamiltonian}
( Tangent vector field of a Legendre submanifold realized 
by a contact Hamiltonian vector field, \cite{Mrugala1991} ) :  
Let $(\,\cC,\lambda\,)$ be a contact manifold, $\cA$ a Legendre submanifold, and 
$h$ a contact Hamiltonian. Then 
the contact Hamiltonian vector field is tangent to $\cA$ if and only if 
$h$ vanishes on $\cA$. 
\end{Thm}

The following definitions have been used mainly in information geometry, and 
are used in this paper as well.   
\begin{Def}
\label{definition-affine-coordinate-flat-connection}
( Affine-coordinate and flat connection, \cite{AN} ) : 
Let $\cM$ be an $n$-dimensional manifold, 
$x=\{\,x^{\,1},\ldots,x^{\,n}\,\}$
coordinates, $\nabla$ a connection, $\{\Gamma_{ab}^{\ \ c}\}$ 
connection coefficients such that 
$\nabla_{\partial_a}\partial_b=\Gamma_{ab}^{\ \ c}\partial_c,$ 
$(\partial_{\,a}:=\partial/\partial x^{\,a})$. If 
$\{\,\Gamma_{ab}^{\ \ c }\,\}\equiv 0$ hold for all $\xi\in\cM$, then 
$x$
is referred to as a $\nabla$-affine coordinate system, 
or affine coordinates. 
If it is the case, then $\nabla$ 
is referred to as a flat connection. 
\end{Def}
\begin{Def}
\label{definition-dual-connection}
( Dual connection, \cite{AN} ) :  
Let $(\cM,g)$ be an $n$-dimensional Riemannian or pseudo-Riemannian manifold,
and $\nabla$ a connection. 
If a connection $\nabla^{\, *}$ satisfies 
\beq
Z\,\left[\,g(\,X,Y\,)\,\right]
=g(\,\nabla_ZX,Y\,)+g(\,X,\nabla_Z^{\,*}Y\,),\qquad \forall\, X,Y,Z\in\GTM
\label{dual-connection-general}
\eeq
then $\nabla$ and $\nabla^{\,*}$ are referred to as dual connections, also 
$\nabla^{\,*}$ 
is referred to as a dual connection of $\nabla$ with respect to $g$. 
\end{Def}

\begin{Lemma}
\label{lemma-uniquely-detemined-dual-connection}
( Existence of a unique dual connection, \cite{AN} ) : 
Given a metric tensor field and a connection, 
there exists a unique dual connection .  
\end{Lemma}
\begin{Proof}
Let $g$ and $\nabla$ be the given metric tensor field and connection, 
respectively. 
Then in what follows the explicit form of the dual connection $\nabla^{\,\prime}$ 
is shown.  
Let $x=\{\,x^{\,1},\ldots,x^{\,n}\,\}$ be local coordinates, $\{\,g_{\,ab}\,\}$ 
the components of the metric tensor field such that 
$g=g_{\,ab}\,\dr x^{\,a}\otimes\dr x^{\,b}$, 
$\{\,\Gamma_{ab}^{\ \ c}\,\}$ connection coefficients for $\nabla$ 
such that $\nabla_{\partial_{\,a}}\partial_{\,b}=\Gamma_{ab}^{\ \ c}\partial_{\,c}$ with 
$\partial_{\,a}:=\partial/\partial x^{\,a}$, and $\{\,\Gamma_{ab}^{\,\prime \ c}\,\}$ 
connection coefficients for $\nabla^{\,\prime}$ such that 
$\nabla_{\partial_{\,a}}^{\,\prime}\partial_{\,b}=\Gamma_{ab}^{\,\prime \ c}\partial_{\,c}$.
Then, defining $\Gamma_{abc}:=g_{\,ck}\Gamma_{ab}^{\ \ k}$ and 
$\Gamma_{abc}^{\,\prime}:=g_{\,ck}\Gamma_{ab}^{\,\prime \ k}$, one can determine  
$\{\,\Gamma_{abc}^{\,\prime}\,\}$ uniquely with \fr{dual-connection-general} 
as 
$\Gamma_{abc}^{\,\prime}=\partial_{\,c}\,g_{\,ab}-\Gamma_{abc}$.    
\qed
\end{Proof}
\begin{Def}
\label{definition-dual-coordinate}
( Dual coordinate, \cite{AN} ) : 
Let $(\cM,g)$ be an $n$-dimensional Riemannian or pseudo-Riemannian manifold, 
$x=\{\,x^{\,1},\ldots,x^{\,n}\,\}$ a set of local coordinates, and 
$p=\{\,p_{\,1},\ldots,p_{\,n}\,\}$ 
another set of local coordinates.  
If 
\beq
g\left(\,\frac{\partial}{\partial x^{\,a}},\frac{\partial}{\partial p_{\,b}}\,\right)
=\delta_{\,a}^{\,b},
\label{dual-coordinate-general} 
\eeq
then $p$ is referred to as the dual coordinate system. If it is the case, 
then $x$ and $p$ are referred to as being mutually dual with respect to $g$.  
\end{Def}

Combining Definitions\,\ref{definition-affine-coordinate-flat-connection}, 
\ref{definition-dual-connection} and 
\ref{definition-dual-coordinate}, one has the following.
\begin{Lemma}
\label{lemma-dual-affine-coordinates}
( Dual coordinate and affine coordinate ) : 
Let $(\cM,g)$ be an $n$-dimensional Riemannian or pseudo-Riemannian manifold,
$\nabla$ a connection, $x=\{\,x^{\,1},\ldots,x^{\,n}\,\}$ 
a set of $\nabla$-affine coordinates, and 
$p=\{\,p_{\,1},\ldots,p_{\,n}\,\}$ another set of coordinates. 
If $x$ and $p$ are mutually dual with respect to $g$, then  
$p$ is a $\nabla^{\,*}$-affine coordinate system. 
\end{Lemma}
\begin{Proof}
There exists the connection $\nabla^{\,*}$ being dual of $\nabla$ due to 
Lemma\,\ref{lemma-uniquely-detemined-dual-connection}.   
It follows from \fr{dual-coordinate-general} that 
$$
Z\left[\,g\left(\,\frac{\partial}{\partial x^{\,a}},\frac{\partial }{\partial p_{\,b}}\,\right)\,\right]
=0,
$$
for $\forall\,Z\in\GTM$.
Then, with this and \fr{dual-connection-general}, one has 
$$
Z\left[\,
g\left(\,\frac{\partial}{\partial x^{\,a}},\frac{\partial}{\partial p_{\,b}}\right)\,
\right]
=g\left(\,\nabla_{\,Z}\frac{\partial}{\partial x^{\,a}},
\frac{\partial}{\partial p_{\,b}}\right)
+g\left(\,\frac{\partial}{\partial x^{\,a}},
\nabla_{\,Z}^{\,*}\frac{\partial}{\partial p_{\,b}}\right)
=0.
$$
Since $x$ is a $\nabla$-affine coordinate system, 
$\nabla_{\,Z}(\,\partial/\partial x^{\,a}\,)=0,(a\in\{\,1,\ldots,n\,\})$, 
one can conclude that 
$\nabla_{\,Z}^{\,*}(\,\partial/\partial p_{\,b}\,)=0, (b\in\{\,1,\ldots,n\,\})$, 
from which $p$ is a $\nabla^{\,*}$-affine coordinate system.   
\qed
\end{Proof}

\begin{Def}
\label{def-dually-flat-space}
( Dually flat space, \cite{AN} ) : 
Let $(\cM,g)$ be an $n$-dimensional Riemannian or pseudo-Riemannian manifold, 
$\nabla$ and $\nabla^{\, *}$ dual connections.
If there exist $\nabla$-affine coordinates 
and $\nabla^{\,*}$-affine ones,
then $(\cM,g,\nabla,\nabla^*)$ is 
referred to as a dually flat space. 
\end{Def}

From these definitions, one can show the following relation between 
pairings and inner products.
\begin{Proposition}
( Inner products and pairings on a dually flat space ) : 
Let $(\cM,g,\nabla,\nabla^{\,*})$ be an $n$-dimensional 
dually flat space, $x=\{\,x^{\,1},\ldots,x^{\,n}\,\}$ a set of 
$\nabla$-affine coordinates, $p=\{\,p_{\,1},\ldots,p_{\,n}\,\}$ a set of 
$\nabla^{\,*}$-affine coordinates.
If the inner products 
$T_{\xi}\cM\times T_{\xi}\cM\to\mbbR, (\xi\in\cM)$ between the bases
$\{\,\partial/\partial x^{\,1},\ldots,\partial/\partial x^{\,n}\,\}$ and 
$\{\,\partial/\partial p_{\,1},\ldots,\partial/\partial p_{\,n}\,\}$ are given as 
\beqa
g\left(\,\frac{\partial}{\partial x^{\,a}},\frac{\partial}{\partial x^{\,b}}\,\right)
&=&g_{\,ab},\qquad
g\left(\,\frac{\partial}{\partial x^{\,a}},\frac{\partial}{\partial p_{\,b}}\,\right)
=\delta_{\,a}^{\,b},
\non\\
g\left(\,\frac{\partial}{\partial p_{\,a}},\frac{\partial}{\partial x^{\,b}}\,\right)
&=&\delta_{\,b}^{\,a},\qquad
g\left(\,\frac{\partial}{\partial p_{\,a}},\frac{\partial}{\partial p_{\,b}}\,\right)
=g^{\,ab},
\non
\eeqa
( i.e., $x$ and $p$ are mutually dual with respect to $g$ ), then 
one has 
the following pairings  
$T_{\xi}^{\,*}\cM\times T_{\xi}\cM\to\mbbR, (\xi\in\cM)$
\beqa
\dr x^{\,a}\left(\,\frac{\partial}{\partial x^{\,b}}\,\right)
&=&\delta^{\,ab},\qquad
\dr x^{\,a}\left(\,\frac{\partial}{\partial p_{\,b}}\,\right)
=g^{\,ab},
\non\\
\dr p_{\,a}\left(\,\frac{\partial}{\partial x^{\,b}}\,\right)
&=&g_{\,ab},\qquad
\dr p_{\,a}\left(\,\frac{\partial}{\partial p_{\,b}}\,\right)
=\delta_{\,ab}.
\non
\eeqa
\end{Proposition}

As shown in Ref.\cite{Goto2015}, 
a contact manifold and a strictly convex function 
induce a dually flat space. The proof below is slightly different to 
that in Ref.\cite{Goto2015}.   
\begin{Thm}
\label{theorem-goto-2015}
( Contact manifold and a function induce a dually flat space,\,\cite{Goto2015} ) : 
Let $(\,\cC,\lambda\,)$ be a $(2n+1)$-dimensional contact manifold, 
$(x,p,z)$ the canonical coordinates such that $\lambda=\dr z-p_{\,a}\,\dr x^{\,a}$
with $x=\{\,x^{\,1},\ldots,x^{\,n}\,\}$ and $p=\{\,p_{\,1},\ldots,p_{\,n}\,\}$, 
and $\psi$ a strictly convex function of $x$ only. 
If the Legendre submanifold generated by $\psi$ is simply connected, then 
$(\,(\,\cC,\lambda\,),\,\psi\,)$ induces an $n$-dimensional dually flat space
$(\,\Phi_{\,\cC\cA\psi}\cA_{\,\psi},g,\nabla,\nabla^{\,*}\,)$ 
with $\Phi_{\,\cC\cA\psi}:\cA_{\,\psi}\to\cC$ being an embedding.
\end{Thm}
\begin{Proof}
In this proof, from $(\,(\,\cC,\lambda\,),\,\psi\,)$  
we explicitly specify $g$, $\nabla$, $\nabla^{\,*}$, 
$\nabla$-affine coordinates, and 
$\nabla^{\,*}$-affine coordinates on the embedded 
Legendre submanifold generated by $\psi$.

Let $\varphi$ be the total Legendre transform of $\psi$ with respect to $x$. 
With the general theory of Legendre transform and the given 
strictly convex function $\psi$,  
one can define  $x_{\,a}^{\,*}$ and have the relation such that
$$
x_{\,a}^{\,*}
:=\frac{\partial \psi}{\partial x^{\,a}},\qquad
x^{\,a}
=\frac{\partial \varphi}{\partial\,x_{\,a}^{\,*}},\qquad 
\varphi\,(\,x^{\,*}\,)
=\Leg[\,\psi\,](\,x^{\,*}\,).
$$
On the embedded Legendre submanifold 
$\cA_{\,\psi}^{\,\cC}:=\Phi_{\,\cC\cA\psi}\cA_{\,\psi}^{\,\cC}$, one has that 
$x_{\,a}^{*}=p_{\,a}$, due to 
\fr{example-psi-Legendre-submanifold}. 
In what follows $g$, $\nabla$, $\nabla^{\,*}$, $\nabla$-affine coordinates, 
and $\nabla^{\,*}$-affine coordinates  are explicitly specified 
on $\cA_{\,\psi}^{\,\cC}$.  
First, one defines 
\beq
g_{\,ab}(x)
:=\frac{\partial^2\,\psi}{\partial x^{\,a}\partial x^{\,b}},
\qquad\mbox{and}\qquad 
g^{\,ab}(p)
:=\frac{\partial^2\,\varphi}{\partial p_{\,a}\partial p_{\,b}}.
\label{def-metric-components-psi-varphi-submanifolds}
\eeq
It can be shown that $g_{\,aj}g^{\,jb}=\delta_{\,a}^{\,b}$. 
It follows from $\det(\,g_{ab}(x)\,)>0,$ that 
$$
g(x):=g_{\,ab}(x)\,\dr x^{\,a}\otimes\,\dr x^{\,b},
$$
becomes a Riemannian metric tensor field. 
Let $\nabla$ be a connection such that $g=\nabla\dr\psi$. 
Then, $x$ is a set of $\nabla$-affine coordinates, 
$\{\,\Gamma_{ab}^{\ \ c}\,\}\equiv 0$, 
due to 
$$
\nabla\dr\psi
=\left(\,\frac{\partial^2\psi}{\partial x^{\,a}\partial x^{\,b}}
-\Gamma_{ab}^{\ \ c}\frac{\partial\,\psi}{\partial x^{\,c}}\,\right)\,\dr x^{\,a}\otimes \dr x^{\,b}, 
\qquad \mbox{where}\qquad
\nabla_{\frac{\partial}{\partial x^{\,a}}}\frac{\partial}{\partial x^{\,b}}
=\Gamma_{ab}^{\ \ c}\frac{\partial}{\partial x^{\,c}}.
$$
In addition, 
it follows from 
Lemma\,\ref{lemma-uniquely-detemined-dual-connection} that 
there exists a unique dual connection $\nabla^{\,*}$ for 
given $g$ and $\nabla$. 
It is straightforward to show that $x$ and $p$ are dual 
coordinates with respect to $g$ :  
$$
g\left(\,\frac{\partial }{\partial x^{\,a}},\frac{\partial}{\partial p_{\,b}}
\,\right)
=\delta_{\,a}^{\,b}.
$$
Applying Lemma\,\ref{lemma-dual-affine-coordinates}, one concludes  
that $p$ is a set of $\nabla^{\,*}$-affine coordinates. 
The connection coefficients for $\nabla^{\,*}$ such that 
$\nabla_{\partial/\partial x^{\,a}}^{\ *}(\,\partial/\partial x^{\,b}\,)=\Gamma_{ab}^{\,*\ c}(\,\partial/\partial x^{\,c}\,)$  
are 
$$
\Gamma_{ab}^{\,*\ c}
=g^{\,cj}\Gamma_{abj}^{\,*},
\qquad\mbox{with}\qquad
\Gamma_{abc}^{\,*}
=g\left(\,\nabla_{\frac{\partial}{\partial x^{\,a}}}\frac{\partial}{\partial x^{\,b}},
\frac{\partial}{\partial x^{\,c}}\,\right)
=\frac{\partial^{\,3}\,\psi}{\partial x^{\,a}\partial x^{\,b}\partial x^{\,c}}.
$$
Thus  $g$, $\nabla$, $\nabla^{\,*}$, 
$\nabla$-affine coordinates, and $\nabla^{\,*}$-affine coordinates
have been obtained.
\qed
\end{Proof}

The following is often used in information geometry.
\begin{Def}
( Canonical divergence, \cite{AN} ) : 
Let $(\cM,g,\nabla,\nabla^{\,*})$ be an $n$-dimensional dually flat space,
$\{\,x^{\,1},\ldots,x^{\,n}\,\}$ $\nabla$-affine coordinates, 
$\{\,p_{\,1},\ldots,p_{\,n}\,\}$ $\nabla^{\,*}$-affine coordinates, 
$\xi$ and $\xi^{\,\prime}$ two points of $\cM$. 
Then, the function 
$\mbbD:\cM\times\cM\to\mbbR $, 
\beq
\mbbD\,(\,\xi\,\|\,\xi^{\,\prime}\,)
:=\psi(\,\xi\,)+\varphi(\,\xi^{\,\prime}\,)
-\left.x^{\,a}\right|_{\,\xi}\left.p_{\,a}\right|_{\,\xi^{\,\prime}},
\label{def-canonical-divergence}
\eeq
is referred to as the canonical divergence. 
\end{Def}
\begin{Remark}
There is another convention for the canonical divergence
( see Ref.\,\cite{Fujiwara1995} )
. 
\end{Remark}
\begin{Remark}
The canonical divergence 
\fr{def-canonical-divergence} 
should be  distinguished from the volume-form divergence 
that has been in Definition\,\ref{def-volume-form-compressibility}.  
\end{Remark}

In information geometry, the following theorem is well-known.
\begin{Thm}
\label{theorem-Pythagorean}
( Generalized Pythagorean theorem,\,\cite{AN} ) : 
Let $(\,\cM,g,\nabla,\nabla^{\,*}\,)$ be a dually flat space,
$\mbbD:\cM\times\cM\to\mbbR $ the canonical divergence, 
$\xi^{\,\prime},\xi^{\,\prime\prime},\xi^{\,\prime\prime\prime}$ be three points of 
$\cM$ such that 
1. $\xi^{\,\prime}$ and $\xi^{\,\prime\prime}$ are 
connected with the $\nabla^{\,*}$-geodesic curve and 
2. $\xi^{\,\prime\prime}$ and $\xi^{\,\prime\prime\prime}$ are connected with 
the $\nabla$-geodesic curve.
Then, it follows that 
$$
\mbbD\,(\,\xi^{\,\prime\prime\prime}\,\|\,\xi^{\,\prime}\,)
=\mbbD\,(\,\xi^{\,\prime\prime\prime}\,\|\,\xi^{\,\prime\prime}\,)
+\mbbD\,(\,\xi^{\,\prime\prime}\,\|\,\xi^{\,\prime}\,).
$$
\end{Thm}

\section{Contact geometric descriptions of vector fields on  dually flat spaces } 
\label{sec-contact-geometric-description-dually-flat-space}
In this section,   
contact geometric descriptions of vector fields on dually flat spaces are 
shown. 
These descriptions enable one to have more links between contact geometry and 
information geometry. 

As shown in Theorem\,\ref{theorem-goto-2015}, 
a contact manifold and a strictly convex function 
induce a dually flat space. Here points of a dually flat space are 
identified with ones of the Legendre submanifold generated by the 
convex function.  
Since contact Hamiltonian vector fields 
can describe dynamics on contact manifolds, dynamics on dually flat spaces  
can also be described by such vector fields.

After fixing mathematical symbols that will be used in 
Secs.\ref{sec-contact-geometric-description-dually-flat-space}--\ref{sec-generating-function-preserving-lifts},
such vector fields on dually flat spaces are 
rewritten in a contact differential geometric language.  
\subsection{Mathematical symbols}
In this subsection mathematical symbols are fixed as follows. 
Let $(\,\cC,\lambda\,)$ be 
a $(2n+1)$-dimensional contact manifold with some 
fixed $n\in\{\,1,2,\ldots\,\}$,
$(x,p,z)$ canonical coordinates such that 
$\lambda=\dr z-p_{\,a}\,\dr x^{\,a}$ 
with $x=\{\,x^{\,1},\ldots,x^{\,n}\,\}$ and $p=\{\,p_{\,1},\ldots,p_{\,n}\,\}$, 
$\psi$ a  function on 
$\cC$ depending on $x$ only, $\Delta_{\,0}$ and 
$\{\,\Delta_{\,1},\ldots,\Delta_{\,n}\,\}$ functions such that 
\beq
\Delta_{\,0}(x,z)
:=\psi(x)-z,\qquad
\Delta_{\,a}(x,p)
:=\frac{\partial\psi}{\partial x^{\,a}}(x)-p_{\,a},\qquad 
(a\in\{\,1,\ldots,n\,\})
\label{definition-Delta_a-psi}
\eeq
and $\dot{}$ will denote differential with respect to a parameter denoted $t$.

In addition, to discuss the Legendre duality,  
other symbols are introduced as follows.
Let $\varphi$ be a function on $\cC$ depending on $p$ only, and $\Delta^{\,0}$ 
and $\{\,\Delta^{\,1},\ldots,\Delta^{\,n}\,\}$ functions such that 
\beq
\Delta^{\,0}(x,p,z)
=x^{\,j}p_{\,j}-\varphi(p)-z,\qquad
\Delta^{\,a}(x,p)
=x^{\,a}-\frac{\partial \varphi}{\partial p_{\,a}}(p),\qquad 
(a\in\{\,1,\ldots,n\,\}).
\label{definition-Delta_a-varphi}
\eeq 

The following lemma will be used.
\begin{Lemma}
  ( Equivalent local expressions of Legendre submanifolds  ) : 
The Legendre submanifold $\cA_{\,\psi}$ generated by $\psi$ 
is expressed such that  
\beq
\cA_{\,\psi}^{\,\cC}
:=\Phi_{\,\cC\cA\psi}\cA_{\,\psi}
=\{\,(x,p,z)\in\cC\,|\,\Delta_{\,0}=0,\quad\mbox{and}\quad\Delta_{\,1}=\cdots=\Delta_{\,n}=0\,\},
\label{example-psi-Legendre-submanifold-Delta}
\eeq
with $\Phi_{\,\cC\cA\psi}:\cA_{\,\psi}\to\cC$ being an embedding.   
In addition,  
the Legendre submanifold $\cA_{\,\varphi}$ generated by $-\,\varphi$ is 
expressed such that  
\beq
\cA_{\,\varphi}^{\,\cC}
:=\Phi_{\,\cC\cA\varphi}\cA_{\,\varphi}
=\left\{\,(x,p,z)\in\cC
\ |\ \Delta^{\,0}=0,\quad \mbox{and}\quad  
\Delta^{\,1}=\cdots=\Delta^{\,n}
=0
\,\right\}.
\label{example-varphi-Legendre-submanifold-Delta}
\eeq
\end{Lemma} 
\begin{Proof}
For $\cA_{\,\psi}^{\,\cC}$, substituting the definitions of 
$\Delta_{\,0},\{\,\Delta_{\,1},\ldots,\Delta_{\,n}\,\}$ 
into \fr{example-psi-Legendre-submanifold}, 
one has \fr{example-psi-Legendre-submanifold-Delta}.
For $\cA_{\,\varphi}^{\,\cC}$, one rewrites the conditions in  
\fr{example-varphi-Legendre-submanifold}. 
With 
$$
x^{\,a}=\frac{\partial\,\varphi}{\partial p_{\,a}}
\qquad\Longleftrightarrow\qquad
\Delta^{\,a}=0,
$$
for all $a\in\{\,1,\ldots,n\,\}$, 
substituting  $z=p_j\partial\varphi/\partial p_j-\varphi$ and 
$x^{\,a}=\partial\,\varphi/\partial p_{\,a}$ into $\Delta^{\,0}$, one has that 
\beqa
&&\left.\Delta^{\,0}\right|_{z=p_j\partial\varphi/\partial p_j-\varphi, x^{\,1}=\partial\varphi/\partial p_1,\ldots}
=\left(x^jp_j-\varphi-z \right)_{z=p_j\partial\varphi/\partial p_j-\varphi,\ \Delta^1=\cdots =\Delta^n=0}
\non\\
&&
=\left(\,x^jp_j-\varphi-p_j\frac{\partial\varphi}{\partial p_j}+\varphi
\,\right)_{\Delta^1=\cdots =\Delta^n=0}
=p_j\,\left(x^j-\frac{\partial\varphi}{\partial p_j}\,\right)_{\Delta^1=\cdots =\Delta^n=0}
\non\\
&&=\left(\,p_j\,\Delta^j\,\right)_{\Delta^1=\cdots =\Delta^n=0}
=0,
\non
\eeqa
from which 
$$
x^{\,1}=\frac{\partial\,\varphi}{\partial p_{\,1}},\ldots,
x^{\,n}=\frac{\partial\,\varphi}{\partial p_{\,n}},
\quad\mbox{and}\quad
z=p_{\,i}\frac{\partial\,\varphi}{\partial p_{\,i}}-\varphi,
\quad\Longrightarrow\quad
\Delta^{\,0}
=0,\quad\mbox{and}\quad
\Delta^{\,1}
=\cdots
=\Delta^{\,n}
=0. 
$$
In addition, one can easily show that 
$$
x^{\,1}=\frac{\partial\,\varphi}{\partial p_{\,1}},\ldots,
x^{\,n}=\frac{\partial\,\varphi}{\partial p_{\,n}},
\quad\mbox{and}\quad
z=p_{\,i}\frac{\partial\,\varphi}{\partial p_{\,i}}-\varphi,
\quad\Longleftarrow\quad
\Delta^{\,0}
=0,\quad\mbox{and}\quad
\Delta^{\,1}
=\cdots
=\Delta^{\,n}
=0. 
$$
Thus one has \fr{example-varphi-Legendre-submanifold-Delta}.
\qed
\end{Proof}

\subsection{Vector fields on dually flat spaces in contact geometric language}
In this subsection, it is shown that 
vector fields on dually flat spaces can be realized as a class of 
contact Hamiltonian vector fields with some restriction. 
For this purpose, functions $\psi$ and $-\,\varphi$ are strictly convex and 
the Legendre submanifolds generated by $\psi$ and $-\,\varphi$ are assumed 
simply connected in this subsection.

The following proposition can be seen as a variant of the theorem given by 
Mrugala {\it et al.} in Ref.\,\cite{Mrugala1991}.  
\begin{Proposition}
\label{Mrugala-variant-psi}
( Restricted contact Hamiltonian vector field 
as the push-forward of a vector field on 
the Legendre submanifold generated by $\psi$ ) : 
Let $\{\,F_{\,\psi}^{\,1},\ldots,F_{\,\psi}^{\,n}\,\}$ 
be a set of functions of $x$ on $\cA_{\,\psi}$ 
such that they do not identically vanish,   
and $\chX_{\,\psi}^{\,0}\in T_{\,x}\,\cA_{\,\psi}, ( x\in \cA_{\,\psi})$ 
the vector field given as   
$$
\chX_{\,\psi}^{\,0}
=\dot{x}^{\,a}\frac{\partial}{\partial x^{\,a}},\quad\mbox{where}\quad 
\dot{x}^{\,a}
=F_{\,\psi}^{\,a}(x),\qquad 
(a\in\{\,1,\ldots,n\,\}).
$$
In addition, let $X_{\,\psi}^{\,0}:=(\,\Phi_{\,\cC\cA\psi}\,)_{*}\chX_{\,\psi}^{\,0} 
\in T_{\,\xi}\cA_{\,\psi}^{\,\cC}, (\,\xi\in\cA_{\,\psi}^{\,\cC}\,)$ 
be the push-forward of
$\chX_{\,\psi}^{\,0}$,   
where $\cA_{\,\psi}^{\,\cC}:=\Phi_{\,\cC\cA\psi}\cA_{\,\psi}$ with   
$\Phi_{\,\cC\cA\psi}:\cA_{\,\psi}\to \cC$ being an embedding :  
\beqa
\Phi_{\,\cC\cA\psi} &:& \cA_{\,\psi}\to 
\cA_{\,\psi}^{\,\cC},\qquad\qquad x\mapsto (\,x,p(x),z(x)\,)
\non\\
(\,\Phi_{\,\cC\cA\psi}\,)_{\,*} &:& T_{\,x}\,\cA_{\,\psi}\to  
T_{\,\xi}\,\cA_{\,\psi}^{\,\cC},\quad \chX_{\,\psi}^{\,0}\mapsto
X_{\,\psi}^{\,0}\,.
\non
\eeqa
%
Then it follows that   
\beq
X_{\,\psi}^{\,0}
=\dot{x}^{\,a}\frac{\partial}{\partial x^{\,a}}
+\dot{p}_{\,a}\frac{\partial}{\partial p_{\,a}}
+\dot{z}\frac{\partial}{\partial z},\quad\mbox{where}\quad 
\dot{x}^{\,a}
=F_{\,\psi}^{\,a}(x),\quad
\dot{p}_{\,a}
=\frac{\dr}{\dr t}\left(\,\frac{\partial\psi}{\partial x^{\,a}}\,\right),
\quad 
\dot{z}
=\frac{\dr \psi}{\dr t}.
\label{tangent-vector-Legendre-submanifold-psi-component}
\eeq
In addition, 
one has that $X_{\,\psi}^{\,0}=X_{\,h_{\,\psi}}|_{\,h_{\,\psi}=0}$. Here 
$X_{\,h_{\,\psi}}$ is the contact Hamiltonian flow associated to 
\beq
h_{\,\psi}(x,p,z)
=\Delta_{\,a}(x,p) F_{\,\psi}^{\,a}(x) 
+\Gamma_{\,\psi}(\,\Delta_{\,0}(x,z)\,),
\label{tangent-vector-Legendre-submanifold-psi-Hamiltonian}
\eeq
where $\Gamma_{\,\psi}$ is a function of $\Delta_{\,0}$ such that 
$\Gamma_{\,\psi}(0)=0$  and $\Gamma_{\,\psi}(\Delta_{\,0})\neq 0$ 
for $\Delta_{\,0}\neq 0$.
\end{Proposition}
\begin{Proof}
To show \fr{tangent-vector-Legendre-submanifold-psi-component}, 
one has 
\beqa
X_{\,\psi}^{\,0}
&=&(\,\Phi_{\,\cC\cA\psi}\,)_{*}\chX_{\,\psi}^{\,0}
=\dot{x}^{\,a}\frac{\partial}{\partial x^{\,a}}
+\dot{x}^{\,a}\frac{\partial p_{\,b}}{\partial x^{\,a}}\frac{\partial}{\partial p_{\,b}}+\dot{x}^{\,a}\frac{\partial z}{\partial x^{\,a}}\frac{\partial }{\partial z}
=\dot{x}^{\,a}\frac{\partial}{\partial x^{\,a}}
+\left(\,\frac{\dr}{\dr t}\frac{\partial\,\psi}{\partial x^{\,a}}\,\right)
\frac{\partial}{\partial p_{\,a}}
+\frac{\dr\,\psi}{\dr t}\frac{\partial }{\partial z}
\non\\
&=&
\dot{x}^{\,a}\frac{\partial}{\partial x^{\,a}}
+\dot{p}_{\,a}\frac{\partial}{\partial p_{\,a}}
+\dot{z}\frac{\partial }{\partial z},
\non
\eeqa
from which one has 
\fr{tangent-vector-Legendre-submanifold-psi-component}. 

In the following,  
it is shown that  $X_{\,\psi}^{\,0}=X_{\,h_{\,\psi}}|_{\,h_{\,\psi}=0}$ 
with $h_{\,\psi}$ given by 
\fr{tangent-vector-Legendre-submanifold-psi-Hamiltonian}.   
First, 
one has the component expression of the contact Hamiltonian vector field 
associated to the contact Hamiltonian 
\fr{tangent-vector-Legendre-submanifold-psi-Hamiltonian} 
without any restriction as 
$$
X_{\,h_{\,\psi}}
=\dot{x}^{\,a}\frac{\partial}{\partial x^{\,a}}
+\dot{p}_{\,a}\frac{\partial}{\partial p_{\,a}}
+\dot{z}\frac{\partial}{\partial z},
$$
where
$$
\frac{\dr x^{\,a}}{\dr t}
=F_{\,\psi}^{\,a},\quad
\frac{\dr p_{\,a}}{\dr t}
=\frac{\partial^2\psi}{\partial x^{\,a}\partial x^{\,b}}\,F_{\,\psi}^{\,b}
+\left(\,\frac{\partial\psi}{\partial x^{\,b}}-p_{\,b}\,\right)
\frac{\partial F_{\,\psi}^{\,b}}{\partial x^{\,a}} 
+\left(\,\frac{\partial\psi}{\partial x^{\,a}}- p_{\,a}\,\right)\frac{\dr\Gamma_{\,\psi}}{\dr \Delta_{\,0}},\quad
\frac{\dr z}{\dr t}
=h_{\,\psi}^{\,0}+p_{\,a}F_{\,\psi}^{\,a}.
$$
Second, observe that 
the condition 
$h_{\,\psi}=\Delta_{\,a}F_{\,\psi}^{\,a}+\Gamma_{\,\psi}(\Delta_{\,0})= 0$    
is equivalent to that $\Delta_{\,0}=\cdots=\Delta_{\,n}=0$, since 
$\{\,F_{\,\psi}^{\,1},\ldots,F_{\,\psi}^{\,n}\,\}$ do not identically vanish and the 
property of $\Gamma_{\,\psi}$.  
From this and \fr{example-psi-Legendre-submanifold-Delta}, 
one has that $\cA_{\,\psi}^{\,\cC}=\{\,h_{\,\psi}=0\}$.
Finally, the contact Hamiltonian vector field restricted to $h_{\,\psi}^{\,0}=0$ 
is obtained by substituting $\Delta_{\,0}=\cdots=\Delta_{\,n}=0$ into 
$X_{\,h_{\,\psi}}$ as 
$$
\frac{\dr x^{\,a}}{\dr t}
=F_{\,\psi}^{\,a},\quad
\frac{\dr p_{\,a}}{\dr t}
=\frac{\partial^2\psi}{\partial x^{\,a}\partial x^{\,b}}\,F_{\,\psi}^{\,b}
=\frac{\partial^2\psi}{\partial x^{\,a}\partial x^{\,b}}\,
\frac{\dr x^{\,b}}{\dr t}
,\quad
\frac{\dr z}{\dr t}
=p_{\,a}F_{\,\psi}^{\,a}
=\frac{\partial\psi}{\partial x^{\,a}}\frac{\dr x^{\,a}}{\dr t},
$$
which are equivalent to 
\fr{tangent-vector-Legendre-submanifold-psi-component}. 
Thus,  one has that 
$X_{\,\psi}^{\,0}=X_{\,h_{\,\psi}}|_{h_{\,\psi}=0}$.  
\qed
\end{Proof}
\begin{Remark}
The functions $\{\,F_{\,\psi}^{\,1},\ldots,F_{\,\psi}^{\,n}\,\}$ 
need not depend on $\psi$.
\end{Remark}
\begin{Remark}
\label{remark-dually-flat-space-psi-not-preserved}
The value of the generating function 
$\psi$ is not conserved along this restricted 
contact Hamiltonian vector field, since 
$$
\cL_{X_{\,\psi}^{\,0}}\psi
=\frac{\partial\psi}{\partial x^{\,a}}\frac{\dr x^{\,a}}{\dr t}
=\frac{\partial\psi}{\partial x^{\,a}}\,F_{\,\psi}^{\,a},
$$
does not identically vanish in general.
In some cases, this vanishes. Consider the system with $n=2$, and 
$F_{\,\psi}^{\,1}=\partial\psi/\partial x^{\,2},F_{\,\psi}^{\,2}=-\,\partial\psi/\partial x^{\,1} $. Then $\cL_{X_{\,\psi}^{\,0}}\psi=0$.
\end{Remark}
\begin{Remark}
In the case of $\Gamma_{\,\psi}\equiv 0$, this contact Hamiltonian has been used 
in Ref.\,\cite{Estay2011}, and it follows 
from \fr{cartan-contact-Hamiltonian-vector} 
that 
$\cL_{X_{\,\psi}^{\,0}}\lambda=0$.  
\end{Remark}

The following is a counterpart of Proposition\,\ref{Mrugala-variant-psi}.
\begin{Proposition}
\label{Mrugala-variant-varphi}
( Restricted contact Hamiltonian vector field as the push-forward of 
vector fields on the Legendre submanifold generated by $-\,\varphi$ ) : 
Let $\{\,F_{\,1}^{\,\varphi},\ldots,F_{\,n}^{\,\varphi}\,\}$ 
be a set of functions of $p$ on $\cA_{\,\varphi}$  
such that they do not identically vanish,   
and $\chX_{\,\varphi}^{\,0}\in T_{\,p}\cA_{\,\varphi}, ( p\in\cA_{\,\varphi} )$ 
given as   
$$
\chX_{\,\varphi}^{\,0}
=\dot{p}_{\,a}\frac{\partial}{\partial p_{\,a}},\quad\mbox{where}\quad 
\dot{p}_{\,a}
=F_{\,a}^{\,\varphi}(p).
$$
In addition, let $X_{\,\varphi}^{\,0}:=(\,\Phi_{\,\cC\cA\varphi}\,)_{*}\chX_{\,\varphi}^{\,0} 
\in T_{\,\xi}\cA_{\,\varphi}^{\,\cC}, (\,\xi\in\cA_{\,\varphi}^{\,\cC}\,)$ 
be the push-forward of
$\chX_{\,\varphi}^{\,0}$,   
where $\cA_{\,\varphi}^{\,\cC}:=\Phi_{\,\cC\cA\varphi}\cA_{\,\varphi}$ with   
$\Phi_{\,\cC\cA\varphi}:\cA_{\,\varphi}\to \cC$ being an embedding : 
\beqa
\Phi_{\,\cC\cA\varphi} &:& \cA_{\,\varphi}\to 
\cA_{\,\varphi}^{\,\cC},\qquad\qquad x\mapsto (\,x(p),p,z(p)\,)
\non\\
(\,\Phi_{\,\cC\cA\varphi}\,)_{\,*} &:& T_{\,p}\,\cA_{\,\varphi}\to  
T_{\,\xi}\,\cA_{\,\varphi}^{\,\cC},\quad \chX_{\,\varphi}^{\,0}\mapsto
X_{\,\varphi}^{\,0}\,.
\non
\eeqa
Then it follows that   
\beq
X_{\,\varphi}^{\,0}
=\dot{x}^{\,a}\frac{\partial}{\partial x^{\,a}}
+\dot{p}_{\,a}\frac{\partial}{\partial p_{\,a}}
+\dot{z}\frac{\partial}{\partial z},\quad\mbox{where}\quad 
\dot{x}_{\,a}
=\frac{\dr}{\dr t}\left(\,\frac{\partial\varphi}{\partial p_{\,a}}\,\right),
\quad 
\dot{p}^{\,a}
=F_{\,a}^{\,\varphi}(p),\quad
\dot{z}
=p_{\,j}F_{\,k}^{\,\varphi}\frac{\partial^2\,\varphi}{\partial p_k\partial p_j}.
\label{tangent-vector-Legendre-submanifold-varphi-component}
\eeq
In addition, one has that $X_{\,\varphi}^{\,0}=X_{\,h_{\,\varphi}}|_{\,h_{\,\varphi}=0}$. 
Here $X_{\,h_{\,\varphi}}$ is the contact Hamiltonian associated to 
\beq
h_{\,\varphi}(x,p)
=\Delta^{\,a}(x,p) F_{\,a}^{\,\varphi}(p)
+\Gamma^{\,\varphi}(\,\Delta^{\,0}\,),
\label{tangent-vector-Legendre-submanifold-varphi-Hamiltonian}
\eeq
where $\Gamma^{\,\varphi}$ is a function of $\Delta^{\,0}$ such that 
$\Gamma^{\,\varphi}(0)=0$ and $\Gamma^{\,0}(\Delta^{\,0})\neq 0$ for 
$\Delta^{\,0}\neq 0$.
\end{Proposition}
\begin{Proof}
A proof is analogous to the proof of  
Proposition \ref{Mrugala-variant-psi}.
\qed
\end{Proof}
\begin{Remark}
The functions $\{\,F_{\,1}^{\,\varphi},\ldots,F_{\,n}^{\,\varphi}\,\}$   
need not depend on $\varphi$.
\end{Remark}
\begin{Remark}
\label{remark-dually-flat-space-varphi-not-preserved}
The value of the generating function $\varphi$
is not conserved along this restricted 
contact Hamiltonian vector field, since 
$$
\cL_{X_{\,\varphi}^{\,0}}\varphi
=\frac{\partial\varphi}{\partial p_{\,a}}\frac{\dr p_{\,a}}{\dr t}
=\frac{\partial\varphi}{\partial p_{\,a}}\,F_{\,a}^{\,\varphi},
$$
does not identically vanish in general.
In some cases, this vanishes. Consider the system with $n=2$, and 
$F_{\,1}^{\,\varphi}=\partial\varphi/\partial p_{\,2},F_{\,2}^{\,\varphi}=-\,\partial\varphi/\partial p_{\,1} $. Then $\cL_{X_{\,\varphi}^{\,0}}\varphi=0$.
\end{Remark}
\begin{Remark}
In the case of $\Gamma^{\,\varphi}\equiv 0$, 
it follows from \fr{cartan-contact-Hamiltonian-vector} 
that $\cL_{X_{\,\varphi}^{\,0}}\lambda=0$.
\end{Remark}

In \S\ref{sec-generating-function-preserving-lifts} 
another formulation will be introduced. In that formulation 
the value of the 
generating functions 
are conserved 
along  contact Hamiltonian 
vector fields for any $\{\,F_{\,\psi}^{\,1},\ldots,F_{\,\psi}^{\,n}\,\}$. 
In addition its counterpart will also be given.

In general, when connections are given on a manifold, 
one of the most important geometric 
objects is a set of geodesic curves. On a dually flat space,  
two connections are given and then two kinds of geodesic curves are 
of interest. 
In the following, it is shown that such geodesic curves are realized by the 
integral curves of 
the contact Hamiltonian vector fields by choosing 
$\{\,F_{\,\psi}^{\,1},\ldots,F_{\,\psi}^{\,n}\,\}$ in 
\fr{tangent-vector-Legendre-submanifold-psi-component} and 
$\{\,F_{\,1}^{\,\varphi},\ldots,F_{\,n}^{\,\varphi}\,\}$ in 
\fr{tangent-vector-Legendre-submanifold-varphi-component}.
\begin{Thm}
\label{theorem-psi-geodesic}
( $\nabla^{\,*}$-geodesic curves ) : 
Let $p^{\,\prime}=\{\,p_{\,1}^{\,\prime},\ldots,p_{\,n}^{\,\prime}\,\}$ 
and $p^{\,\prime\prime}=\{\,p_{\,1}^{\,\prime\prime},\ldots,p_{\,n}^{\,\prime\prime}\,\}$ be 
two sets of constants, $\psi$ a strictly convex function of $x$ only, 
$\{\,g^{\,ab}(x)\,\}$ the metric components  
\fr{def-metric-components-psi-varphi-submanifolds} where the  
arguments are written in terms of $x$. Then the restricted contact Hamiltonian 
vector field associated to  $h_{\,\psi}$ in 
\fr{tangent-vector-Legendre-submanifold-psi-Hamiltonian}
with 
\beq
F_{\,\psi}^{\,a}(x)
=g^{\,aj}(x)(\,p_{\,j}^{\,\prime\prime}-p_{\,j}^{\,\prime}\,), 
\qquad (\,a\in\{\,1,\ldots,n\,\}\,)
\label{psi-geodesic-F}
\eeq 
gives the geodesic curve connecting the two points whose 
$\nabla^{\,*}$-affine coordinates are 
$p^{\,\prime}$ and  $p^{\,\prime\prime}$.
\end{Thm}
\begin{Proof}
It follows from 
\fr{tangent-vector-Legendre-submanifold-psi-component} that
$$
\frac{\dr}{\dr t}p_{\,a}
=g_{\,aj}\frac{\dr x^{\,j}}{\dr t}
=g_{\,aj}F_{\,\psi}^{\,j}
=g_{\,aj}g^{\,jk}(\,p_{\,k}^{\,\prime\prime}-p_{\,k}^{\,\prime}\,)
=p_{\,a}^{\,\prime\prime}-p_{\,a}^{\,\prime}.
$$
The solution to this equation is obtained as 
$$
p_{\,a}(t)
=(\,p_{\,a}^{\,\prime\prime}-p_{\,a}^{\,\prime}\,)\,t+
p_{\,a}(0).
$$
Choosing the initial point $p(0)$ as $p^{\,\prime}$,
one has the straight line that connects $p^{\,\prime}$ and $p^{\,\prime\prime}$. 
This line is 
the geodesic curve in this case, and 
connects the two points 
whose $\nabla^{\,*}$-affine coordinates are 
 $p^{\,\prime}$  and $p^{\,\prime\prime}$, respectively.
\qed
\end{Proof}
Rather than $p(t)$ in the theorem, 
one way to express $x(t)$ analytically 
is to use the relation 
$x^{\,a}=\partial\varphi/\partial p_{\,a}$, where the analytic expression of 
$\varphi=\Leg[\psi]$ is required. Although such an explicit expression 
cannot be obtained in general, 
there are some cases where it is feasible. 
In Ref.\,\cite{Ohara1996},     
the explicit analytic forms of the total Legendre transform of 
given strictly convex functions and their applications in 
linear programming problems have been studied. 

There is a counterpart as follows.
\begin{Thm}
\label{theorem-varphi-geodesic}
( $\nabla$-geodesic curves ) : 
Let $x^{\,\prime}=\{\,x^{\,\prime\,1},\ldots,x^{\,\prime\,n}\,\}$ and 
$x^{\,\prime\prime}=\{\,x^{\,\prime\prime \,1},\ldots,x^{\,\prime\prime \,n}\,\}$ be 
constants, $\varphi$ a strictly convex function of $p$ only, 
$\{\,g_{\,ab}(p)\,\}$ the metric components  
\fr{def-metric-components-psi-varphi-submanifolds} where the  
arguments are written in terms of $p$. Then 
the restricted contact Hamiltonian vector field associated to 
$h_{\,\varphi}$ in 
\fr{tangent-vector-Legendre-submanifold-varphi-Hamiltonian}
with 
\beq
F_{\,a}^{\,\varphi}(p)
=g_{\,aj}(p)(\,x^{\,\prime\prime\,j}-x^{\,\prime\,j}\,), 
\qquad (\,a\in\{\,1,\ldots,n\,\}\,)
\label{varphi-geodesic-F}
\eeq 
gives the geodesic curve connecting the two points whose $\nabla$-affine 
coordinates are 
$x^{\,\prime}$ and  $x^{\,\prime\prime}$.
\end{Thm}
\begin{Proof}
A way to prove this is analogous to that of  
Theorem\,\ref{theorem-psi-geodesic}.
\qed
\end{Proof}

Similar to 
Theorems\,\ref{theorem-psi-geodesic},
one can write 
Theorem 1 of Ref.\,\cite{Fujiwara1995}
( see also Ref.\,\cite{Nakamura1994} ) 
in terms of a contact Hamiltonian vector field.
Note that the definition of the canonical divergence in 
Ref.\,\cite{Fujiwara1995} is not same as that of this paper. 
\begin{Thm}
\label{theorem-Fujiwara-Amari-psi}
( Gradient flow  of Ref.\,\cite{Fujiwara1995} 
in terms of a restricted contact Hamiltonian vector field  ) : 
Let $\psi$ be a strictly convex function of $x$ only, 
$\xi\in\cA_{\,\psi}^{\,\cC}$ a point whose coordinates are 
$(\,x,p\,(x),z\,(x)\,)$, $\xi^{\,\prime}\in\cA_{\,\psi}^{\,\cC}$ 
another point, and $\{\,g^{\,ab}(x)\,\}$  given by  
\fr{def-metric-components-psi-varphi-submanifolds} with the  
arguments rewritten in terms of $x$.
Choose $\{\,F_{\,\psi}^{\,1},\ldots,F_{\,\psi}^{\,n}\,\}$ in 
\fr{tangent-vector-Legendre-submanifold-psi-component} as  
$$
F_{\,\psi}^{\,a}(x)
=-\,g^{\,aj}(x)\frac{\partial}{\partial x^{\,j}}\mbbD\,(\,\xi\,\|\,\xi^{\,\prime}\,).
$$
Then, the flow converges to the point $\xi^{\,\prime}$ along 
the $\nabla^{\,*}$-geodesic curve. This flow  
is the restricted contact Hamiltonian vector field 
$X_{h_{\,\psi}}|_{h_{\,\psi}=0}$  
associated to $h_{\,\psi}$ given 
in \fr{tangent-vector-Legendre-submanifold-psi-Hamiltonian} 
with this set of functions $\{\,F_{\,\psi}^{\,1},\ldots,F_{\,\psi}^{\,n}\,\}$.
\end{Thm}
\begin{Proof}
A proof is similar to that for Theorems\,\ref{theorem-psi-geodesic} 
( see also Ref.\,\cite{Fujiwara1995} ). 
Let $(\,x^{\,\prime},p^{\,\prime},z^{\,\prime}\,)$ be coordinates for $\xi^{\,\prime}$. 
It follows from 
\fr{tangent-vector-Legendre-submanifold-psi-component} 
and \fr{def-canonical-divergence} 
that
\beqa
\frac{\dr}{\dr t}p_{\,a}
&=&g_{\,aj}\frac{\dr x^{\,j}}{\dr t}
=-\,g_{\,aj}g^{\,jk}(x)\frac{\partial}{\partial x^{\,k}}
\mbbD\,(\,\xi\,\|\,\xi^{\,\prime}\,).
=-\,\delta_{\,a}^{\,k}  
\frac{\partial}{\partial x^{\,k}}
\left[\,
\psi(x)+\varphi(\,p^{\,\prime}\,)-x^{\,j}p_{\,j}^{\,\prime}
\,\right]
\non\\
&=&-\,\frac{\partial}{\partial x^{\,a}}
\left[\,\psi(x)-x^{\,j}p_{\,j}^{\,\prime}\,\right]
=-\,\left(\,p_{\,a}-p_{\,a}^{\,\prime}\,\right),
\non
\eeqa
which is integrated to obtain 
$$
p_{\,a}(t)
=p_{\,a}^{\,\prime}+\left(\,p_{\,a}(0)-p_{\,a}^{\,\prime}\,\right)\,
\e^{\,-t}.
$$
This and Proposition \ref{Mrugala-variant-psi} 
prove the theorem.
\qed
\end{Proof}

Then, one has a counterpart of Theorem\,\ref{theorem-Fujiwara-Amari-psi} as 
follows. 
\begin{Thm}
\label{theorem-Fujiwara-Amari-varphi}
( Counterpart of the gradient flow of Ref.\,\cite{Fujiwara1995} in terms of a 
restricted contact Hamiltonian vector field ) : 
Let $\varphi$ be a strictly convex function of $p$ only, 
$\xi\in\cA_{\,\varphi}^{\,\cC}$  a point whose coordinates are 
$(\,x(p),p,z(p)\,)$, $\xi^{\,\prime}\in\cA_{\,\varphi}^{\,\cC}$ 
another point, and $\{\,g_{\,ab}(p)\,\}$  given by  
\fr{def-metric-components-psi-varphi-submanifolds} with the  
arguments being rewritten in terms of $p$.
Choose $\{\,F_{\,1}^{\,\varphi},\ldots,F_{\,n}^{\,\varphi}\,\}$ in 
\fr{tangent-vector-Legendre-submanifold-varphi-component} to be 
$$
F_{\,a}^{\,\varphi}(p)
=-\,g_{\,aj}(p)\frac{\partial}{\partial p_{\,j}}\mbbD\,(\,\xi^{\,\prime}\,\|\,\xi\,).
$$
Then, the flow converges to the point $\xi^{\,\prime}$ along 
the $\nabla$-geodesic curve. This flow is the restricted 
contact Hamiltonian vector field 
$X_{h_{\,\varphi}}|_{h_{\,\varphi}=0}$  
associated to $h_{\,\varphi}$ given 
in \fr{tangent-vector-Legendre-submanifold-varphi-Hamiltonian} 
with this set of functions $\{\,F_{\,1}^{\,\varphi},\ldots,F_{\,n}^{\,\varphi}\,\}$.
\end{Thm}
\begin{Proof}
A way to prove this is similar to that of  
Theorems\,\ref{theorem-Fujiwara-Amari-psi}
( see also Ref.\,\cite{Fujiwara1995} ).
\qed
\end{Proof}

With 
Theorems\,\ref{theorem-psi-geodesic} and \ref{theorem-varphi-geodesic}, 
the generalized Pythagorean theorem ( Theorem\,\ref{theorem-Pythagorean} )  
is written in terms of two contact Hamiltonian vector fields as follows.
\begin{Thm} 
( Generalized Pythagorean theorem in terms of integral curves ) : 
Let $\psi$ be a strictly convex function depending on $x$ only, $\varphi$ 
the function of $p$ that is obtained by the total Legendre transform of 
$\psi$ with respect to $x$, and   
$\xi^{\,\prime},\xi^{\,\prime\prime},\xi^{\,\prime\prime\prime}$ three points of 
$\cA_{\,\psi}^{\,\cC}\cap \cA_{\,\varphi}^{\,\cC}$ such that 
\begin{enumerate}
\item 
$\xi^{\,\prime}$ and $\xi^{\,\prime\prime}$ are 
connected with the $\nabla^{\,*}$-geodesic curve, and  any point of the 
$\nabla^{\,*}$-geodesic curve is on $\cA_{\,\psi}^{\,\cC}\cap\cA_{\,\varphi}^{\,\cC}$,
\item
$\xi^{\,\prime\prime}$ and $\xi^{\,\prime\prime\prime}$ are connected with 
the $\nabla$-geodesic curve, and  any point of the 
$\nabla$-geodesic curve is on $\cA_{\,\psi}^{\,\cC}\cap\cA_{\,\varphi}^{\,\cC}$.  
\end{enumerate}
Then 
the generalized Pythagorean theorem ( Theorem\,\ref{theorem-Pythagorean} ) 
can be written as 
$$
\mbbD\,(\,\Exp\,(\,X_{\,\varphi}^{\,0}\,)\,\Exp(\,X_{\,\psi}^{\,0}\,)\,\xi^{\,\prime}\,
\|\,\xi^{\,\prime})
=\mbbD\,(\,\Exp\,(\,X_{\,\varphi}^{\,0}\,)\,\xi^{\,\prime\prime}\,\|\,\xi^{\,\prime\prime}\,)
+\mbbD\,(\,\Exp\,(\,X_{\,\psi}^{\,0}\,)\,\xi^{\,\prime}\,\|\,\xi^{\,\prime}\,),
$$
and 
$$
\mbbD\,(\,\Exp\,(\,X_{\,\varphi}^{\,0}\,)\,\Exp(\,X_{\,\psi}^{\,0}\,)\,\xi^{\,\prime}\,
\|\,\xi^{\,\prime})
=\mbbD\,(\,\Exp\,(\,X_{\,\varphi}^{\,0}\,)\,\Exp\,(\,X_{\,\psi}^{\,0}\,)\,\xi^{\,\prime}\,\|\,\Exp\,(\,X_{\,\psi}^{\,0}\,)\,\xi^{\,\prime}\,)
+\mbbD\,(\,\Exp\,(\,X_{\,\psi}^{\,0}\,)\,\xi^{\,\prime}\,\|\,\xi^{\,\prime}\,),
$$
where $\mbbD$ is the canonical divergence defined in 
\fr{def-canonical-divergence},  
$\Exp\,(\,X_{\,\psi}^{\,0}\,) :\cA_{\,\psi}^{\,\cC}\to\cA_{\,\psi}^{\,\cC}$ 
is the exponential map of unit time 
generated by the flow 
\fr{tangent-vector-Legendre-submanifold-psi-component}
with \fr{psi-geodesic-F} 
and $\Exp\,(\,X_{\,\varphi}^{\,0}\,) :\cA_{\,\varphi}^{\,\cC}\to\cA_{\,\varphi}^{\,\cC}$ 
is the one generated by the flow 
\fr{tangent-vector-Legendre-submanifold-varphi-component}
with \fr{varphi-geodesic-F}.
\end{Thm}
\begin{Proof}
Applying Theorem\,\ref{theorem-Pythagorean}, 
Theorem\,\ref{theorem-psi-geodesic}, and 
Theorem\,\ref{theorem-varphi-geodesic}, one completes the proof.
\end{Proof}

\subsection{Applications to electric circuit models}
In this subsection, it is shown how    
the developed general 
theory is applied to engineering problems. As examples, three   
simple electric circuit models are analyzed. The first one is a series  
RC ( resistor-capacitor ) circuit model. The second one 
is a series RL ( resistor-inductor ) circuit model. These ones are 
formulated on $1$-dimensional dually flat spaces. 
The third one is a series RLC ( resistor-inductor-capacitor ) circuit model.   
This one is formulated on a $2$-dimensional dually flat space. 
For each model, after defining the  
circuit model, it is shown how the model can be seen as a dynamical 
system on a dually flat space embedded in a contact manifold.

First, a circuit model consisting of a resistor and a capacitor is 
focused. 
\begin{Def}
\label{definition-RC-circuit-model}
( RC circuit model ) : 
Let $C$ be a constant capacitor with $C>0$, $R$ a constant resistor 
with $R>0$, $Q_{\,\C}\in\mbbR$ the amount of charge, $V_{\,\C}\in\mbbR$ 
the capacitor voltage, $\cH_{\,\C}(Q_{\,\C})\in\mbbR$ 
the energy stored in the capacitor,  
and $t\in\mbbR$ time. The RC circuit model is defined as the following 
system 
\beq
R\,\frac{\dr Q_{\,\C}}{\dr t}
=-\,\frac{1}{C}Q_{\,\C},
\label{RC-circuit}
\eeq
together with 
\beq
\cH_{\,\C}(Q_{\,\C})
:=\frac{1}{2\,C}Q_{\,\C}^{\,2},\quad\mbox{and}\quad 
Q_{\,\C}
=CV_{\,\C}.
\label{RC-circuit-energy}
\eeq
\end{Def}
\begin{Remark}
The system \fr{RC-circuit} 
with \fr{RC-circuit-energy} 
can be seen as an equation for describing  
electric energy dissipation due to the existence of the resistor. 
\end{Remark}
\begin{Thm}
( RC circuit model on a dually flat space ) :  
The system \fr{RC-circuit}
can be seen as a dynamical system on a dually flat space.
In particular, 
the coordinate systems $Q_{\,\C}$ and $V_{\,\C}$ 
can be seen as being mutually dual with respect to 
a metric tensor field.
\end{Thm}
\begin{Proof}
Since $\cH_{\C}$ is a strictly convex function 
on $\mbbR$ whose coordinate is  $Q_{\,\C}$, 
one can identify $\dr^2\cH_{\,\C}/\dr Q_{\,\C}^2$  
as the component of a Riemannian metric tensor field $g_{\,\C}$. 
These are summarized as 
$$
g_{\,\C}
:=g_{\,\C 11}\,\dr Q_{\,\C}\otimes \dr Q_{\,\C},\qquad 
g_{\,\C 11}
:=\frac{\dr^2\cH_{\,\C}}{\dr Q_{\,\C}^{\,2}}
=\frac{1}{C}>0.
$$
Defining the total Legendre transform of $\cH_{\,\C}$ 
with respect to $Q_{\,\C}$ as 
$$
\cH_{\,\C}^{\,*}(V_{\,\C})
:=\sup_{Q_{\,\C}}\left(\,  Q_{\,\C}V_{\,\C}-\cH_{\,\C}(Q_{\,\C}) \,\right),
$$
one has that 
$$
\cH_{\,\C}^{\,*}(V_{\,\C})
=\frac{C}{2}V_{\,\C}^{\,2}.
$$
One can write $g_{\,\C}$ as 
$$
g_{\,\C}
=g_{\,\C 11}^{\,-1}\,\dr V_{\,\C}\otimes \,\dr V_{\,\C},\qquad 
g_{\,\C 11}^{\,-1}
:=\frac{\dr^2 \cH_{\,\C}^{\,*}}{\dr V_{\,\C}^{\,2}}
=C>0.
$$
It follows from 
$$
g_{\,\C 11}\,g_{\,\C 11}^{\,-1}
=1,\quad 
Q_{\,\C}
=\frac{\dr\, \cH_{\,\C}^{\,*}}{\dr V_{\,\C}},\quad 
V_{\,\C}
=\frac{\dr\, \cH_{\,\C}}{\dr Q_{\,\C}},\quad
g_{\,\C 11}^{\,-1}
=\frac{\dr^2  \cH_{\,\C}^{\,*}}{\dr V_{\,\C}^{\,2}}
=\frac{\dr  Q_{\,\C}}{\dr V_{\,\C}},\quad 
g_{\,\C 11} 
=\frac{\dr^2 \cH_{\,\C}}{\dr Q_{\,\C}^{\,2}}
=\frac{\dr V_{\,\C}}{\dr Q_{\,\C}},
$$
that 
$$
g_{\,\C}\left(\,
\frac{\partial}{\partial Q_{\,\C}},\frac{\partial}{\partial V_{\,\C}}\,\right)
=1.
$$
Therefore, $Q_{\,\C}$ and $V_{\,\C}$ are mutually dual with respect to $g_{\,\C}$
( see Definition\,\ref{definition-dual-coordinate} ). 

The pair $(\,\mbbR,g_{\,\C}\,)$ is a Riemannian manifold. 
Let $\nabla_{\,\C}$ be the connection such that 
$g_{\,\C}=\nabla_{\,\C}\,\dr \cH_{\,\C}$. Then,  
$Q_{\,\C}$ is the $\nabla_{\,\C}$-affine coordinate, and   
there exists 
the unique dual connection $\nabla_{\,\C}^{\,*}$ due to 
Lemma\,\ref{lemma-uniquely-detemined-dual-connection}.   
Applying 
Lemma\,\ref{lemma-dual-affine-coordinates} with 
$Q_{\,\C}$ and $V_{\,\C}$ being mutually dual with respect to $g_{\,\C}$, 
one has that \\
$(\,\mbbR,g_{\,\C},\nabla_{\,\C},\nabla_{\,\C}^{\,*}\,)$ is a $1$-dimensional 
dually flat space 
( see Definition\,\ref{def-dually-flat-space} ). 
In addition, \fr{RC-circuit} can be seen as a dynamical system on a    
$1$-dimensional dually flat space.
\qed
\end{Proof}

In what follows the dynamical system \fr{RC-circuit} 
with the condition $Q_{\,\C}=CV_{\,\C}$ in \fr{RC-circuit-energy} 
is identified with a 
vector field on the embedded Legendre submanifold generated by 
$\cH_{\,\C}$ defined in \fr{RC-circuit-energy}.
\begin{Proposition}
\label{Proposition-RC-dynamical-system}
( RC circuit model as a dynamical system on 
a Legendre submanifold 
) :  
Let $(\cC_{\,\C},\lambda_{\,\C})$ be a contact manifold
with $\cC_{\,\C}=\mbbR^{\,3}$, $(Q_{\,\C},V_{\,\C},z_{\,\C})$ 
coordinates of $\cC_{\,\C}$, and 
$\lambda_{\,\C}=\dr z_{\,\C}-V_{\,\C}\,\dr Q_{\,\C}.$
Then, the 
vector field on 
$\cC_{\,\C}$
associated with \fr{RC-circuit} 
together with \fr{RC-circuit-energy} 
is 
on the Legendre submanifold generated by $\cH_{\,\C}$, 
and its component expressions are 
$$
\frac{\dr}{\dr t}Q_{\,\C}
=-\,\frac{1}{R\,C} Q_{\,\C},\qquad 
\frac{\dr}{\dr t}V_{\,\C}
=
-\,\frac{1}{R\,C^{\,2}}\,Q_{\,\C}
,\qquad 
\frac{\dr}{\dr t}\,z_{\,\C}
=-\,\frac{1}{R\,C^2} Q_{\,\C}^{\,2},
$$
together with the conditions 
$$
Q_{\,C}
=C V_{\,\C},\qquad \mbox{and}\qquad 
z_{\,\C}
=\frac{1}{2C} Q_{\,\C}^{\,2}.
$$
\end{Proposition}
\begin{Proof}
Let $F_{\,\C}$ be the following function of $Q_{\,\C}$  
$$
F_{\,\C}(Q_{\,\C})
=-\frac{1}{R\,C}Q_{\,\C},
$$
so that \fr{RC-circuit} can be written as  
$$
\frac{\dr }{\dr t}Q_{\,\C}
=F_{\,\C}(Q_{\,\C}). 
$$
Applying Proposition\,\ref{Mrugala-variant-psi}, one has that 
\beq
\frac{\dr}{\dr t}Q_{\,\C}
=F_{\,\C}(Q_{\,\C})
=-\,\frac{Q_{\,\C}}{R\,C},
\quad 
\frac{\dr}{\dr t}V_{\,\C}
=\frac{\dr}{\dr t}\frac{\dr \cH_{\,\C}}{\dr Q_{\,\C}}
=-\,\frac{Q_{\,\C}}{R\,C^2},\quad 
\frac{\dr}{\dr t}\,z_{\,\C}
=\frac{\dr\,\cH_{\,\C}}{\dr t}
=-\,\frac{Q_{\,\C}^2}{R\,C^2}.
\label{RC-circuit-dually-flat-space-1}
\eeq
The vector field associated with this dynamical system is 
on the Legendre submanifold, where the following relations 
\beqa
\Delta_{\,0}^{\,\C}(\,Q_{\,\C},z_{\,\C}\,)
&:=&\cH_{\,\C}-z_{\,\C}
=\frac{Q_{\,\C}^{\,2}}{2C}-z_{\,\C}
=0,
\non\\
\Delta_{\,1}^{\,\C}(\,Q_{\,\C},V_{\,\C}\,)
&:=&\frac{\dr\,\cH_{\,\C}}{\dr Q_{\,\C}}-V_{\,\C}
=\frac{Q_{\,\C}}{C}-V_{\,\C}
=0,
\non
\eeqa
hold. 
From these relations, one has that 
$Q_{\,\C}=C\,V_{\,\C}$ and $z_{\,\C}=Q_{\,\C}^{\,2}/(2C)$.
In addition, this vector field is written as $X_{\,h_{\,\C}}|_{\,h_{\,\C}=0}$, where 
$X_{\,h_{\,\C}}$ is 
the contact Hamiltonian vector field associated to 
$h_{\,\C}(Q_{\,\C},V_{\,\C},z_{\,\C})=\Delta_{\,1}^{\,\C}(Q_{\,\C},z_{\,\C})F_{\,\C}(Q_{\,\C})+\Gamma_{\,\C}(\Delta_{\,0}^{\,\C})$, with $\Gamma_{\,\C}$ being a function such that 
$\Gamma_{\,\C}(0)=0$  and  
$\Gamma_{\,\C}(\Delta_{\,0}^{\,\C})\neq 0$ for $\Delta_{\,0}^{\,\C}\neq 0$.     
The condition $h_{\,\C}=0$ is equivalent to the conditions,  
$\Delta_{\,0}^{\,\C}=0$ and $\Delta_{\,1}^{\,\C}=0$.
\qed
\end{Proof}
\begin{Remark}
The discussions above on the RC circuit model can be extended to those on 
a class where 
the energy stored in the capacitor  
$$
\cH_{\,\C}(Q_{\,\C})
=\int Q_{\,\C}(V_{\,\C})\,\dr V_{\,\C},
$$
is a strictly convex function of $Q_{\,\C}$ with $Q_{\,\C}=Q_{\,\C}(V_{\,\C})$ 
being not necessary to be $Q_{\,\C}=CV_{\,\C}$. 
\end{Remark}

Second, a circuit model consisting of a resistor and an inductor is 
focused. 
\begin{Def}
\label{definition-RL-circuit-model}
( RL circuit model ) : 
Let $L$ be a constant inductor with $L>0$, 
$R$ a constant resistor with $R>0$,   
$I_{\,\L}\in\mbbR$ the current, $N_{\,\L}\in\mbbR$  
magnetic flux due to $I_{\,\L}$,  
$\cH_{\,\L}^{\,*}(I_{\,\L})\in\mbbR$
the  magnetic energy,
and $t\in\mbbR$ time.    
The RL circuit model is defined as the following system 
\beq
L\frac{\dr}{\dr t}I_{\,\L}
=-\,R\,I_{\,\L},
\label{RL-circuit}
\eeq
together with 
\beq
\cH_{\,\L}^{\,*}(I_{\,\L})
:=\frac{L}{2}I_{\,\L}^{\,2},\quad\mbox{and}\quad 
N_{\,\L}
=LI_{\,\L}.
\label{RL-circuit-energy}
\eeq
\end{Def}
\begin{Remark}
The system \fr{RL-circuit} 
with \fr{RL-circuit-energy} 
can be seen as an equation for describing  
magnetic energy dissipation due to the existence of the resistor. 
\end{Remark}

\begin{Thm}
( RL circuit model on a dually flat space ) :  
The system \fr{RL-circuit}
can be seen as a dynamical system on a dually flat space.
In particular, 
the coordinate systems $N_{\,\L}$ and $I_{\,\L}$   
can be seen as being mutually dual with respect to a metric tensor field. 
\end{Thm}
\begin{Proof}
Since $\cH_{\,\L}^{\,*}$ is a strictly convex function 
on $\mbbR$ whose coordinate is $I_{\,\L}$, 
one can identify $\dr^2\cH_{\,\L}^{\,*}/\dr I_{\,\L}^2$ 
as the component of a Riemannian metric tensor field $g_{\,\L}$. 
These are summarized 
as   
$$
g_{\,\L}
:=g_{\,\L 11}^{\,-1}\,\dr I_{\,\L}\otimes \dr I_{\,\L},
\qquad 
g_{\,\L 11}^{\,-1}
:=\frac{\dr^2 \cH_{\,\L}^{\,*}}{\dr I_{\,\L}^{\,2}}
=L>0.
$$ 
Defining the total Legendre transform of $\cH_{\,\L}^{\,*}$ 
with respect to $I_{\,\L}$ as 
$$
\cH_{\,\L}(N_{\,\L})
:=\sup_{I_{\,\L}}\left(\,
I_{\,\L}N_{\,\L}-\cH_{\,\L}^{\,*} 
\,\right),
$$
one has that 
$$
\cH_{\,\L}(N_{\,\L})
=\frac{1}{2L}N_{\,\L}^{\,2}.
$$
One can write $g_{\,\L}$ as 
$$
g_{\,\L}
=g_{\,\L 11}\,\dr N_{\,\L}\otimes \,\dr N_{\,\L},\qquad 
g_{\,\L 11}
:=\frac{\dr^2 \cH_{\,\L}}{\dr N_{\,\L}^{\,2}}
=\frac{1}{L}
>0.
$$
It follows from 
$$
g_{\,\L 11}\,g_{\,\L 11}^{\,-1}
=1,\quad 
N_{\,\L}
=\frac{\dr\, \cH_{\,\L}^{\,*}}{\dr I_{\,\L}},\quad
I_{\,\L}
=\frac{\dr \cH_{\,\L}}{\dr N_{\,\L}},\quad 
g_{\,\L 11}^{\,-1}
=\frac{\dr^2\cH_{\,\L}^{\,*}}{\dr I_{\,\L}^{\,2}}
=\frac{\dr N_{\,\L}}{\dr I_{\,\L}},\quad 
g_{\,\L11}
=\frac{\dr^2\cH_{\,\L}}{\dr N_{\,\L}^{\,2}}
=\frac{\dr I_{\,\L}}{\dr N_{\,\L}},
$$
that 
$$
g_{\,\L}\left(\,
\frac{\partial}{\partial N_{\,\L}},\frac{\partial}{\partial I_{\,\L}}\,\right)
=1.
$$
Therefore, $N_{\,\L}$ and $I_{\,\L}$ are mutually dual with respect to $g_{\,\L}$ 
( see Definition\,\ref{definition-dual-coordinate} ). 

The pair $(\,\mbbR,g_{\,\L}\,)$ is a Riemannian manifold. Let 
$\nabla_{\,\L}^{\,*}$ be the connection such that 
$g_{\,\L}=\nabla_{\,\L}^{\,*} \dr \cH_{\,\L}^{\,*}$. Then, 
$I_{\,\L}$ is the $\nabla_{\,\L}^{\,*}$-affine coordinate, and 
there exists the unique dual connection $\nabla_{\,\L}$ due to an analogue of  
Lemma\,\ref{lemma-uniquely-detemined-dual-connection}.   
Applying 
Lemma\,\ref{lemma-dual-affine-coordinates} with 
$N_{\,\L}$ and $I_{\,\L}$ being mutually dual with respect to $g_{\,\L}$, 
one has that 
$(\,\mbbR,g_{\,\L},\nabla_{\,\L},\nabla_{\,\L}^{\,*}\,)$ is a $1$-dimensional 
dually flat space 
( see Definition\,\ref{def-dually-flat-space} ). 
In addition, \fr{RL-circuit} can be seen as a dynamical system on a  
$1$-dimensional dually flat space.
\qed
\end{Proof}

In what follows the dynamical system \fr{RL-circuit} 
with the condition $N_{\L}=LI_{\L}$ in \fr{RL-circuit-energy} 
is identified with a 
vector field on the embedded Legendre submanifold 
generated by $\cH_{\,\L}^{\,*}$ defined in \fr{RL-circuit-energy}.
\begin{Proposition}
( RL circuit model as a dynamical system on a 
Legendre submanifold
) :  
Let $(\cC_{\,\L},\lambda_{\,\L})$ be a contact manifold with 
$\cC_{\,\L}=\mbbR^{\,3}$, 
$(N_{\L},I_{\L},z_{\L})$ coordinates of $\cC_{\L}$, and 
$\lambda_{\,\L}=\dr z_{\L}-I_{\L}\dr N_{\L}.$
Then, the vector field on $\cC_{\,\L}$ 
associated with \fr{RL-circuit} is 
on the Legendre submanifold generated by $\cH_{\L}^*$, and its 
component expressions are 
$$
\frac{\dr}{\dr t}N_{\,\L}
=-\,R I_{\,\L},\qquad 
\frac{\dr}{\dr t}I_{\,\L}
=-\,\frac{R}{L}\,I_{\,\L},\qquad 
\frac{\dr}{\dr t}\,z_{\,\L}
=-\,R I_{\,\L}^{\,2},
$$
together with the conditions 
$$
N_{\,L}
=L I_{\,\L},\qquad \mbox{and}\qquad 
z_{\,\L}
=\frac{L}{2} I_{\,\L}^{\,2}.
$$
\end{Proposition}
\begin{Proof}
Let $F_{\,\L}$ be the following function of $I_{\,\L}$  
$$
F_{\,\L}(I_{\,\L})
=-\frac{R}{L}I_{\,\L},
$$
so that  \fr{RL-circuit} can be written as  
$$
\frac{\dr }{\dr t}I_{\,\L}
=F_{\,\L}(I_{\,\L}). 
$$
Applying Proposition\,\ref{Mrugala-variant-varphi}, one has that 
\beq
\frac{\dr}{\dr t}N_{\,\L}
=\frac{\dr}{\dr t}\frac{\dr \cH_{\,\L}^{\,*}}{\dr I_{\,\L}}
=-\,R\, I_{\,\L},
\quad 
\frac{\dr}{\dr t}I_{\,\L}
=F_{\,\L}(I_{\,\L})
=-\,\frac{R}{L}I_{\,L},\quad 
\frac{\dr}{\dr t}\,z_{\,\L}
=I_{\,\L}F_{\,\L}(I_{\,\L})\frac{\dr^2\cH_{\,\L}^{\,*}}{\dr I_{\,\L}^2}
=-R\,I_{\,\L}^2.
\label{RL-circuit-dually-flat-space-1}
\eeq
The vector field associated with this dynamical system is 
on the Legendre submanifold, where the following relations 
\beqa
\Delta_{\,\L}^{\,0}(\,N_{\,\L},I_{\,\L},z_{\,\L}\,)
&:=&N_{\,L}I_{\,\L}-\cH_{\,\L}^{\,*}(I_{\,\L})-z_{\,\L}
=N_{\,L}I_{\,\L}-\frac{L}{2}I_{\,\L}^2-z_{\,\L}
=0,
\non\\
\Delta_{\,\L}^{\,1}(\,N_{\,\L},I_{\,\L}\,)
&:=&N_{\,\L}-\frac{\dr\,\cH_{\,\L}^{\,*}}{\dr I_{\,\L}}
=N_{\,\L}-L\,I_{\,\L}
=0,
\non
\eeqa
hold. 
From these relations, one has that 
$N_{\,\L}=L\,I_{\,\L}$ and $z_{\,\L}=L\,I_{\,\L}^{\,2}/2$. 
In addition, this vector field is written as $X_{\,h_{\,\L}}|_{\,h_{\,\L}=0}$, where 
$X_{\,h_{\,\L}}$ is 
the contact Hamiltonian vector field associated to 
$h_{\,\L}(N_{\,\L},I_{\,\L},z_{\,\L})=\Delta_{\,\L}^{\,1}(N_{\,\L},I_{\,\L})F_{\,\L}(I_{\,\L})+\Gamma_{\,\L}(\Delta_{\,\L}^{\,0})$, with $\Gamma_{\,\L}$ being a function such that 
$\Gamma_{\,\L}(0)=0$  and  
$\Gamma_{\,\L}(\Delta_{\,\L}^{\,0})\neq 0$ for $\Delta_{\,\L}^{\,0}\neq 0$.     
The condition $h_{\,\L}=0$ is equivalent to the conditions,  
$\Delta_{\,\L}^{\,0}=0$ and $\Delta_{\,\L}^{\,1}=0$.
\qed
\end{Proof}
\begin{Remark}
The discussions above on the RL circuit model can be extended to those on 
a class where 
the magnetic energy 
$$
\cH_{\,\L}(N_{\,\L})
=\int N_{\,\L}(I_{\,\L})\,\dr I_{\,\L},
$$
is a strictly convex function of $N_{\,\L}$ with $N_{\,\L}=N_{\,\L}(I_{\,\L})$ 
being not necessary to be $N_{\,\L}=LI_{\,\L}$. 
\end{Remark}

Third, a circuit model consisting of a resistor, inductor, and capacitor 
is focused. 
\begin{Def}
\label{definition-RLC-circuit-model}
( RLC circuit model ) : 
Let $C$ be a constant capacitor with $C>0$, $L$ a constant inductor with $L>0$, 
$R$ a constant resistor with $R>0$, 
$V\in\mbbR$ the capacitor voltage, 
$I\in\mbbR$ the current, 
$Q\in\mbbR$ the amount of charge in the capacitor,     
$N\in\mbbR$ magnetic flux due to $I$, 
$\cH(Q,N)\in\mbbR$ 
the total energy stored in the capacitor and the inductor, 
and $t\in\mbbR$ time.    
The RLC circuit model is defined as the following system 
\beq
C\frac{\dr}{\dr t}V
=I,\qquad
L\frac{\dr}{\dr t}I
=-V-\,R\,I, 
\label{RLC-circuit}
\eeq
together with 
\beq
\cH(Q,N)
:=\frac{Q^{\,2}}{2C}+\frac{N^{\,2}}{2L},\quad\mbox{and}\quad 
Q=CV,\quad 
N=LI.
\label{RLC-circuit-energy}
\eeq
\end{Def}
\begin{Remark}
The dynamical system \fr{RLC-circuit} with  
\fr{RLC-circuit-energy} 
can be derived from the energy balance 
equation. 
\end{Remark}
\begin{Thm}
( RLC circuit model on a dually flat space ) :  
The system \fr{RLC-circuit}
can be seen as a dynamical system on a dually flat space.
In particular, 
the coordinate systems $\{\,Q,N\,\}$ and $\{\,V,I\,\}$   
can be seen as being mutually dual with respect to a metric tensor field. 
\end{Thm}
\begin{Proof}
Since $\cH$ is a strictly convex function 
on $\mbbR^2$ whose coordinates are 
$Q$ and $N$, 
one can identify $\{\,\partial^2\cH/\partial x^{\,a}\partial x^{\,b}\,\}$  
as the components of a Riemannian metric tensor field $g$, where 
$x^{\,1}=Q$ and $x^{\,2}=N$. 
These are summarized as   
$$
g
:=g_{\,11}\,\dr Q\otimes \dr Q+g_{\,22}\,\dr N\otimes \dr N,
\qquad 
g_{\,11}
:=\frac{\partial^2 \cH}{\partial Q^{\,2}}
=\frac{1}{C}>0,\quad 
g_{\,22}
:=\frac{\partial^2 \cH}{\partial N^{\,2}}
=\frac{1}{L}>0.
$$ 
Defining the total Legendre transform of $\cH$ 
with respect to $\{\,Q,N\,\}$ as 
$$
\cH^{\,*}(V,I)
:=\sup_{Q,N}\left(\,
QV+NI-\cH 
\,\right),
$$
one has that 
$$
\cH^{\,*}(V,I)
=\frac{C}{2}V^{\,2}
+\frac{L}{2}I^{\,2}.
$$
One can write $g$ as 
$$
g
=g_{\,11}^{\,-1}\,\dr V\otimes \,\dr V+
g_{\,22}^{\,-1}\,\dr I\otimes \,\dr I,\qquad 
g_{\,11}^{-1}
:=\frac{\partial^2 \cH^{\,*}}{\partial V^{\,2}}
=C
>0,\quad 
g_{\,22}^{-1}
:=\frac{\partial^2 \cH^{\,*}}{\partial I^{\,2}}
=L
>0.
$$
It follows from 
$$
g_{\,11}\,g_{\,11}^{\,-1}
=1,\quad 
g_{\,22}\,g_{\,22}^{\,-1}
=1,\quad
V
=\frac{\partial\, \cH}{\partial Q},\quad
I
=\frac{\partial \cH}{\partial N},
$$
and 
$$
g_{\,11 }
=\frac{\partial^2\cH}{\partial Q^{\,2}}
=\frac{\partial V}{\partial Q},\quad 
g_{\,11}^{\,-1}
=\frac{\partial^2\cH^{\,*}}{\partial V^{\,2}}
=\frac{\partial Q}{\partial V},\quad 
g_{\,22 }
=\frac{\partial^2\cH}{\partial N^{\,2}}
=\frac{\partial V}{\partial N},\quad 
g_{\,22}^{\,-1}
=\frac{\partial^2\cH^{\,*}}{\partial I^{\,2}}
=\frac{\partial N}{\partial I},
$$
that 
$$
g\left(\,
\frac{\partial}{\partial Q},\frac{\partial}{\partial V}\,\right)
=
g\left(\,
\frac{\partial}{\partial N},\frac{\partial}{\partial I}\,\right)
=1,\qquad 
g\left(\,
\frac{\partial}{\partial Q},\frac{\partial}{\partial I}\,\right)
=g\left(\,
\frac{\partial}{\partial N},\frac{\partial}{\partial V}\,\right)
=0.
$$
Therefore, $\{\,Q,N\,\}$ and $\{\,V,I\,\}$ are mutually dual with respect 
to $g$ ( see Definition\,\ref{definition-dual-coordinate} ). 

The pair $(\,\mbbR^{2},g\,)$ is a Riemannian manifold. Let 
$\nabla$ be the connection such that 
$g=\nabla \dr \cH$. Then, 
$\{\,Q,N\,\}$ is a set of $\nabla$-affine coordinates, and  
there 
exists the unique dual connection $\nabla^{\,*}$ due to an analogue of  
Lemma\,\ref{lemma-uniquely-detemined-dual-connection}.   
Applying 
Lemma\,\ref{lemma-dual-affine-coordinates} with 
$\{\,Q,N\,\}$ and $\{\,V,I\,\}$ being mutually dual with respect to $g$, 
one has that 
$(\,\mbbR^{2},g,\nabla,\nabla^{\,*}\,)$ is a $2$-dimensional 
dually flat space 
( see Definition\,\ref{def-dually-flat-space} ). 
In addition, \fr{RLC-circuit} can be seen as a dynamical system on a  
$2$-dimensional dually flat space.
\qed
\end{Proof}

In what follows the dynamical system \fr{RLC-circuit} 
with the conditions $Q=CV$ and $N=LI$ in \fr{RLC-circuit-energy} 
is identified with a 
vector field on the embedded Legendre submanifold
generated by $\cH$ in \fr{RLC-circuit-energy}. 
\begin{Proposition}
( RLC circuit model as a dynamical system on a Legendre submanifold ) :  
Let $(\cC_0,\lambda_{\,0})$ be a contact manifold with 
$\cC_0=\mbbR^{\,5}$, 
$(Q,N,V,I,z)$ 
coordinates of $\cC_0$, and   
$\lambda_0=\dr z-V\dr Q-I\,\dr N$.
The vector field on $\cC_0$ associated with \fr{RLC-circuit} is expressed as 
$$
\frac{\dr\,Q}{\dr t}
=I,\quad 
\frac{\dr\, N}{\dr t}
=-\,V-R\,I,\quad 
\frac{\dr\,V}{\dr t}
=\frac{I}{C},\quad
\frac{\dr\,I}{\dr t}
=-\,\frac{V}{L}-\frac{R}{L}I,\quad 
\frac{\dr\,z}{\dr t}
=-R\,I^{\,2}
$$
together with the conditions 
$$
Q=CV,\quad
N=LI,\quad \mbox{and}\qquad 
z
=\frac{CV^{\,2}}{2}+\frac{L\,I^{\,2}}{2}.
$$
\end{Proposition}
\begin{Proof}
Let $\{\,F_{\,1},F_{\,2}\,\}$ be the following functions of $V$ and $I$   
$$
F_{\,1}(V,I)
=\frac{1}{C}I,\qquad 
F_{\,2}(V,I)
=-\frac{V}{L}-\frac{R}{L}I,
$$
so that  \fr{RLC-circuit} can be written as  
$$
\frac{\dr }{\dr t}V
=F_{\,1}(V,I),\qquad \mbox{and}\quad 
\frac{\dr }{\dr t}I
=F_{\,2}(V,I).
$$
Applying Proposition\,\ref{Mrugala-variant-varphi}, one has that 
\beq
\frac{\dr\,Q}{\dr t}
=\frac{\dr}{\dr t}\frac{\partial \cH^{\,*}}{\partial V}
=I,\quad 
\frac{\dr\,N}{\dr t}
=\frac{\dr}{\dr t}\frac{\partial \cH^{\,*}}{\partial I}
=-\, V-RI,
\label{RLC-circuit-dually-flat-space-Q-N}
\eeq
and 
\beq
\frac{\dr V}{\dr t}
=F_{\,1}
=\frac{I}{C},
\quad 
\frac{\dr I}{\dr t}
=F_{\,2}
=-\,\frac{V}{L}-\frac{R}{L}I,\quad 
\frac{\dr\,z}{\dr t}
=VF_{\,1}\frac{\partial^2\cH^{\,*}}{\partial V^2}
+IF_{\,2}\frac{\partial^2\cH^{\,*}}{\partial I^2}
=-RI^2.
\label{RLC-circuit-dually-flat-space-V-I-z}
\eeq
The vector field associated with this dynamical system is 
on the Legendre submanifold, where the following relations 
\beqa
\Delta^{\,0}
&:=&QV+NI-\cH^{\,*}-z
=QV+NI-\left(\,\frac{C}{2}V^{2}+\frac{L}{2}I^2\,\right)-z
=0,
\non\\
\Delta^{\,1}
&:=&Q-\frac{\partial\,\cH^{\,*}}{\partial V}
=Q-CV
=0,\qquad 
\Delta^{\,2}
:=N-\frac{\partial\,\cH^{\,*}}{\partial I}
=N-LI
=0,\non
\eeqa
hold. 
From these relations, one has that 
$Q=CV$, $N=L\,I$ and $z=(CV^{\,2}+L\,I^{\,2})/2$. 
In addition, this vector field is written as $X_{\,h}|_{\,h=0}$, where 
$X_{\,h}$ is 
the contact Hamiltonian vector field associated to 
$h=\Delta^{\,1}F_{\,1}+\Delta^{\,2}F_{\,2}+\Gamma(\Delta^{\,0})$, with $\Gamma$ 
being a function such that 
$\Gamma(0)=0$  and  
$\Gamma(\Delta^{\,0})\neq 0$ for $\Delta^{\,0}\neq 0$.     
The condition $h=0$ is equivalent to the conditions,  
$\Delta^{\,0}=0$ and $\Delta^{\,1}=\Delta^{\,2}=0$.
\qed
\end{Proof}
\begin{Remark}
The discussions above on the RLC circuit model can be extended to those on 
a class where 
the energy 
$$
\cH(Q,N)
=\int Q(V)\,\dr V +\int N(I)\,\dr I,
$$
is a strictly convex function of $\{Q,N\}$ with $Q=Q(V)$ and $N=N(I)$ 
being not necessary to be $Q=CV$ and $N=LI$, respectively.  
\end{Remark}

In Ref.\,\cite{Eberard2006}, 
$\cH_{\,\C}^{\,*}$, $\cH_{\,\L}^{\,*}$ and $\cH^{\,*}$ 
have been referred to as co-energy functions.

\section{Vector fields on contact manifolds lifted from Legendre submanifolds }
\label{sec-lift}
In this section, a way to give lifts of vector fields on Legendre submanifolds 
to contact manifolds, basic properties of such lifted vectors fields, 
and stability of some classes of lifted vector fields 
are discussed.  

In this paper we define the lift of  flow and that of dynamical 
system as follows.
\begin{Def}
 ( Lift of a flow, Lift of a dynamical system ) :  
Let $\cM$ and $\cN$ be manifolds with $\dim\cM<\dim\cN$, 
$\pi:\cN\to\cM$ a projection, 
$\Phi_{\,t}^{\,\cM}:\cM\to\cM$ and $\Phi_{\,t}^{\,\cN}:\cN\to\cN$ flows 
of some dynamical systems  ( $t\in \mbbT\subset \mbbR$ ). If  
$\Phi_{\,t}^{\,\cM}\circ\pi=\pi\circ\Phi_{\,t}^{\,\cN}$ is satisfied, 
then $\Phi_{\,t}^{\,\cN}$ is referred to 
as a lift of $\Phi_{\,t}^{\,\cM}$ ( see the diagram below ). 
$$
\xymatrix{
\cN\ar[r]^{\Phi_{\,t}^{\,\cN}}\ar[d]_{\pi}
& \cN\ar[d]^{\pi}
\\
\cM\ar[r]_{\Phi_{\,t}^{\cM}}
&\cM 
}
$$
\end{Def}
In this paper attention is concentrated on 
the following two cases. 
\begin{itemize}
\item
$\cN=\cC$, $\cM=\cA_{\,\psi}$, and the projection 
$\pi_{\,\psi}$ is chosen such that $\pi_{\,\psi}(x,p,z)=x$.
\item
$\cN=\cC$, $\cM=\cA_{\,\varphi}$ and the projection 
$\pi_{\,\varphi}$ is chosen such that  $\pi_{\,\varphi}(x,p,z)=p$.
\end{itemize}

\subsection{Lifted vector fields and their basic properties}
In this subsection it is shown how flows 
on the Legendre submanifold generated by $\psi$ are lifted to $\cC$. 

\begin{Proposition}
\label{lift-psi-proposition}
( Vector field on $\cC$ lifted from the Legendre submanifold generated by $\psi$ ) : 
Let $\{\,F_{\,\psi}^{\,1},\ldots,F_{\,\psi}^{\,n}\,\}$ be a set of functions 
of $x$ on $\cA_{\,\psi}$, where these functions do not identically vanish.   
The flow of the vector field associated with the dynamical system 
defined on $\cA_{\,\psi}$ as 
$$
\chX_{\,\psi}^{\,0}
=\dot{x}^{\,a}\frac{\partial}{\partial x^{\,a}},\quad\mbox{where}\quad
\frac{\dr }{\dr t}x^{\,a}
=F_{\,\psi}^{\,a}(x),\qquad a\in\{\,1,\ldots,n\,\}
$$
is lifted to $\cC$ as 
$$
X_{\,h,\psi}
=\dot{x}^{\,a}\frac{\partial}{\partial x^{\,a}}
+\dot{p}_{\,a}\frac{\partial}{\partial p_{\,a}}
+\dot{z}\frac{\partial}{\partial z},
$$ 
where $\{\,\dot{x}^{\,1},\ldots,\dot{x}^{\,n}\,\},\{\,\dot{p}_{\,1},\ldots,\dot{p}_{\,n}\,\},\dot{z}$ satisfy 
\beq
\frac{\dr }{\dr t}x^{\,a}
=F_{\,\psi}^{\,a}(x),\qquad
\frac{\dr}{\dr t}\Delta_{\,a}
=-\,\frac{\partial F_{\,\psi}^{\,b}}{\partial x^{\,a}}\Delta_{\,b}
-\,\frac{\dr\Gamma_{\,\psi}}{\dr\Delta_{\,0}}\Delta_{\,a},\qquad
\frac{\dr}{\dr t}\Delta_{\,0}
=-\,\Gamma_{\,\psi}(\,\Delta_{\,0}\,),
\label{lift-general-psi-components}
\eeq
with $\Gamma_{\,\psi}$ being a function of $\Delta_{\,0}$. 
This lifted flow $X_{\,h,\psi}$, \fr{lift-general-psi-components}, is the 
contact Hamiltonian vector field 
associated to  
the contact Hamiltonian 
\beq
h_{\,\psi}(x,p,z)
=\Delta_{\,a}(x,p)\,F_{\,\psi}^{\,a}(x)
+\Gamma_{\,\psi}\,(\,\Delta_{\,0}(x,z)\,).
\label{lift-general-psi-contact-Hamiltonian}
\eeq
In addition, if 
$\Gamma_{\,\psi}$ is such that $\Gamma_{\,\psi}(0)=0$ and 
$\Gamma_{\,\psi}(\Delta_{\,0})\neq 0$ for $\Delta_{\,0}\neq 0$,
then 
$X_{\,\psi}^{\,0}:=\left.X_{h,\psi}\right|_{\,\Delta_{\,0}=\cdots=\Delta_{\,n}=0}$ 
is tangent to $\cA_{\psi}^{\,\cC}$ 
( see the diagrams below )
$$ 
\xymatrix{
& \cC\ar[ld]_{\pi}
\\
\cA_{\,\psi}\ar[r]_{\Phi_{\,\cC\cA\psi}}
&\cA_{\,\psi}^{\,\cC}\, , 
}\qquad
\xymatrix{
&\,\, (\,x,p,z\,)\ar@{|->}[ld]_{\pi}
\\
x\ar@{|->}[r]_{\hspace*{-8mm}\Phi_{\,\cC\cA\psi}}
&\,\,(\,x,p(x),z(x)\,)\, , 
}\qquad 
\xymatrix{
&T_{\,\xi}\, \cC\ar[ld]_{\pi_{\,*}}
\\
T_{\,x}\,\cA_{\,\psi}\ar[r]_{\,\,(\Phi_{\,\cC\cA\psi})_{*}}
&\,T_{\,x^{\,\prime}}\,\cA_{\,\psi}^{\,\cC}\,, 
}\qquad 
\xymatrix{
&X_{\,h,\psi}\ar@{|->}[ld]_{\pi_{\,*}}
\\
\chX_{\,\psi}^{\,0}\ar@{|->}[r]_{\quad(\Phi_{\,\cC\cA\psi})_{*}}
&\,\,X_{\,\psi}^{\,0}\, .
}
$$
\end{Proposition}
\begin{Proof}
Substituting the contact Hamiltonian $h_{\,\psi}$ into 
\fr{contact-Hamiltonian-vector-components}, one has that 
$$
\frac{\dr x^{\,a}}{\dr t}
=F_{\,\psi}^{\,a},\quad 
\frac{\dr p_{\,a}}{\dr t}
=\frac{\dr}{\dr t}\left(\,\frac{\partial\psi}{\partial x^{\,a}}\,\right)
+\frac{\partial F_{\,\psi}^{\,b}}{\partial x^{\,a}}\Delta_{\,b}
+\frac{\dr\Gamma_{\,\psi}}{\dr \Delta_{\,0}}\Delta_{\,a},\quad
\frac{\dr z}{\dr t}
=\frac{\dr\psi}{\dr t}+\Gamma_{\,\psi}(\,\Delta_{\,0}\,),
$$
from which one has \fr{lift-general-psi-components}.
Next, consider systems where $\Gamma_{\,\psi}(0)=0$ 
and $\Gamma_{\,\psi}(\Delta_{\,0})\neq 0$ for $\Delta_{\,0}\neq 0$ hold. 
It follows that 
$$
\left.X_{h,\psi}\right|_{\,\Delta_{\,0}=\cdots=\Delta_{\,n}=0}
=\left.X_{h,\psi}\right|_{\,h_{\,\psi}=0}
=\left.X_{h,\psi}\right|_{\,\cA_{\,\psi}^{\,\cC}}.
$$
\qed
\end{Proof}
\begin{Remark}
The functions $\{\,F_{\,\psi}^{\,1},\ldots,F_{\,\psi}^{\,n}\,\}$ 
need not depend on $\psi$. 
\end{Remark}

There is a counterpart of this Proposition.
\begin{Proposition}
\label{lift-varphi-proposition}
( Vector field on $\cC$ lifted from the Legendre submanifold generated by $-\,\varphi$ ) : 
Let $\{\,F_{\,1}^{\,\varphi},\ldots,F_{\,n}^{\,\varphi}\,\}$ 
be a set of functions of $p$ on 
$\cA_{\,\varphi}^{\,\cC}$, where these functions do not identically vanish.  
The flow of the vector field associated with the 
dynamical system defined on $\cA_{\,\varphi}$ as 
$$
X_{\,\varphi}^{\,0}
=\dot{p}_{\,a}\frac{\partial}{\partial p_{\,a}},\quad\mbox{where}\quad 
\frac{\dr }{\dr t}p_{\,a}
=F_{\,a}^{\,\varphi}(p),\qquad a\in\{1,\ldots,n\,\}
$$
is lifted to $\cC$ as 
$$
X_{\,h,\varphi}
=\dot{x}^{\,a}\frac{\partial}{\partial x^{\,a}}
+\dot{p}_{\,a}\frac{\partial}{\partial p_{\,a}}
+\dot{z}\frac{\partial}{\partial z},
$$
where $\{\,\dot{x}^{\,1},\ldots,\dot{x}^{\,n}\,\},\{\,\dot{p}_{\,1},\ldots,\dot{p}_{\,n}\,\},\dot{z}$ satisfy  
\beq
\frac{\dr}{\dr t}\Delta^{\,a}
=-\,\frac{\partial F_{\,b}^{\,\varphi}}{\partial p_{\,a}}\Delta^{\,b}
-\,\frac{\dr\Gamma^{\,\varphi}}{\dr\Delta^{\,0}}\Delta^{\,a},
\qquad
\frac{\dr }{\dr t}p_{\,a}
=F_{\,a}^{\,\varphi}(p),
\qquad
\frac{\dr}{\dr t }\Delta^{\,0}
=-\,\Gamma^{\,\varphi}(\,\Delta^{\,0}\,),
\label{lift-general-varphi-components}
\eeq
with $\Gamma^{\,\varphi}$ being a function of $\Delta^{\,0}$.  
This lifted flow $X_{\,h,\varphi}$, \fr{lift-general-varphi-components}, is the  
the contact Hamiltonian vector field 
associated to the contact Hamiltonian 
$$
h^{\,\varphi}(x,p,z)
=\Delta^{\,a}(x,p)\,F_{\,a}^{\,\varphi}(p)
+\Gamma^{\,\varphi}\,(\,\Delta^{\,0}(x,p,z)\,).
$$
In addition, if 
$\Gamma^{\,\varphi}$ is such that $\Gamma^{\,\varphi}(0)=0$ and 
$\Gamma^{\,\varphi}(\Delta^{\,0})\neq 0$ for $\Delta^{\,0}\neq 0$, then 
$X_{\,\varphi}^{\,0}:=\left.X_{h,\varphi}\right|_{\,\Delta^{\,0}=\cdots=\Delta^{\,n}=0}$  
is tangent to $\cA_{\varphi}^{\,\cC}$
( see the diagrams below )
$$ 
\xymatrix{
& \cC\ar[ld]_{\pi}
\\
\cA_{\,\varphi}\ar[r]_{\Phi_{\,\cC\cA\varphi}}
&\cA_{\,\varphi}^{\,\cC}\, , 
}\qquad
\xymatrix{
&\,\, (\,x,p,z\,)\ar@{|->}[ld]_{\pi}
\\
p\ar@{|->}[r]_{\hspace*{-8mm}\Phi_{\,\cC\cA\varphi}}
&\,\,(\,x(p),p,z(p)\,)\, , 
}\qquad 
\xymatrix{
&T_{\,\xi}\, \cC\ar[ld]_{\pi_{\,*}}
\\
T_{\,p}\,\cA_{\,\varphi}\ar[r]_{\,\,(\Phi_{\,\cC\cA\varphi})_{*}}
&\,T_{\,p^{\,\prime}}\,\cA_{\,\varphi}^{\,\cC}\,, 
}\qquad 
\xymatrix{
&X_{\,h,\varphi}\ar@{|->}[ld]_{\pi_{\,*}}
\\
\chX_{\,\varphi}^{\,0}\ar@{|->}[r]_{\quad(\Phi_{\,\cC\cA\varphi})_{*}}
&\,\,X_{\,\varphi}^{\,0}\, .
}
$$
\end{Proposition}
\begin{Proof}
A way to prove this is analogous to that of  
Proposition \ref{lift-psi-proposition}.
\qed
\end{Proof}
\begin{Remark}
The functions $\{\,F_{\,1}^{\,\varphi},\ldots,F_{\,n}^{\,\varphi}\,\}$ 
need not depend on $\varphi$.  
\end{Remark}

The following is a property of the contact Hamiltonian that 
describes lifted vector fields.
\begin{Proposition}
\label{lift-contact-Hamiltonian-monotonically-decreasing-function-psi}
( Differentiation of $h_{\,\psi}$ with respect to $t$ ) : 
Consider the system stated in Proposition\,\ref{lift-psi-proposition}. 
Then, it follows that 
$$
\frac{\dr}{\dr t}h_{\,\psi}
=-\,\frac{\dr\,\Gamma_{\,\psi}}{\dr\Delta_{\,0}}h_{\,\psi}.
$$
In the case where $\Gamma_{\,\psi}(\Delta_{\,0})=\gamma_{\,0}\Delta_{\,0}$ 
with $\gamma_{\,0}$ being a non-zero constant, it follows that\\ 
$h_{\,\psi}(t)=h_{\,\psi}(0)\,\exp(-\,\gamma_{\,0}t)$.
\end{Proposition}
\begin{Proof}
It follows from \fr{lift-general-psi-components} that 
$$
\frac{\dr}{\dr t}h_{\,\psi}
=\dot{\Delta}_{\,a}F_{\,\psi}^{\,a}
+\Delta_{\,a}\dot{F}_{\,\psi}^{\,a}
+\frac{\dr\,\Gamma_{\,\psi}}{\dr\Delta_{\,0}}\dot{\Delta}_{\,0}
=-\,\frac{\dr\,\Gamma_{\,\psi}}{\dr\Delta_{\,0}}h_{\,\psi}.
$$
In the case where $\Gamma_{\,\psi}(\Delta_{\,0})=\gamma_{\,0}\Delta_{\,0}$, 
the solution to this differential equation is 
$h_{\,\psi}(t)=h_{\,\psi}(0)\,\exp(-\,\gamma_{\,0}t)$. 
\qed
\end{Proof}
\begin{Remark}
\label{invariant-subset-contact-Hamiltonian-psi}
For systems with $\Gamma_{\,\psi}(\Delta_{\,0})=\gamma_{\,0}\Delta_{\,0}$,
the subspaces $\{(x,p,z)|h_{\,\psi}>0\}$ and 
$\{(x,p,z)|h_{\,\psi}<0\}\subset\cC$ are invariant ones. 
\end{Remark}
\begin{Remark}
Choose a system where $F_{\,\psi}^{\,a}\equiv 0,(a\in\{1,\ldots,n\})$, 
$\dr\Gamma_{\,\psi}/\dr\Delta_{\,0}>0$, 
and $h_{\,\psi}(x(0),p(0),z(0))\geq 0$ with the equality holds 
when $\Delta_{\,0}=0$. Then, the contact Hamiltonian $h_{\,\psi}$ 
is a Lyapunov function ( see Ref.\,\cite{Goto2015} ).  
\end{Remark}

There is a counter part of this Proposition. 
\begin{Proposition}
  ( Differentiation of $h_{\,\varphi}$ with respect to $t$ ) : 
Consider the system stated in Proposition\,\ref{lift-varphi-proposition}. 
Then, it follows that 
$$
\frac{\dr}{\dr t}h_{\,\varphi}
=-\,\frac{\dr\,\Gamma^{\,\varphi}}{\dr\Delta^{\,0}}h_{\,\varphi}.
$$
In the case where $\Gamma^{\,\varphi}(\Delta^{\,0})=\gamma^{\,0}\Delta^{\,0}$ 
with 
$\gamma^{\,0}$ being a non-zero constant, it follows that \\
$h_{\varphi}(t)=h_{\varphi}(0)\exp(-\,\gamma^{\,0}t)$.  
\end{Proposition}
\begin{Proof}
A way to prove this is analogous to that of 
Proposition\,\ref{lift-contact-Hamiltonian-monotonically-decreasing-function-psi}.
\qed
\end{Proof}
\begin{Remark}
\label{invariant-subset-contact-Hamiltonian-varphi}
For systems with $\Gamma^{\,\varphi}(\Delta^{\,0})=\gamma^{\,0}\Delta^{\,0}$,
the subspaces $\{(x,p,z)|h_{\,\varphi}>0\}\subset\cC$ and 
$\{(x,p,z)|h_{\,\varphi}<0\}\subset\cC$ are invariant ones. 
\end{Remark}
\begin{Remark}
Choose a system where $F_{\,a}^{\,\varphi}\equiv 0,(a\in\{1,\ldots,n\})$, 
$\dr\Gamma^{\,\varphi}/\dr\Delta^{\,0}>0$, 
and $h_{\,\varphi}\geq 0$ with the equality holds 
when $\Delta^{\,0}=0$. Then, the contact Hamiltonian $h_{\,\varphi}$ 
is a Lyapunov function ( see Ref.\,\cite{Goto2015} ).  
\end{Remark}

As shown below, 
the phase compressibility for a class of the lifted vector fields 
with $\Omega_{\,\lambda}$   
does not depend on $\{\,F_{\,\psi}^{\,1},\ldots,F_{\,\psi}^{\,n}\,\}$.
\begin{Lemma}
\label{phase-compressibility-psi}
( Phase compressibility for $X_{\,h,\psi}$ with $\Omega_{\,\lambda}$ ) : 
Consider the system stated in Proposition\,\ref{lift-psi-proposition}. Choose   
$\Gamma_{\,\psi}(\Delta_{\,0})=\gamma_{\,0}\Delta_{\,0}$ with $\gamma_{\,0}$ being 
a constant. 
Then, the phase compressibility  
defined in \fr{def-phase-compressibility}
is calculated as 
$$
\kappa_{\Omega_{\,\lambda}}(X_{h,\psi})
=-\,(n+1)\gamma_{\,0},
$$
where $\Omega_{\,\lambda}$ has been defined in 
\fr{def-standard-volume-form-contact-manifold}.
\end{Lemma}
\begin{Proof}
With $\cL_{X_{h,\psi}}\lambda=(\Reeb h_{\,\psi})\lambda=-\,\gamma_{\,0}\lambda$ and 
$\cL_{X_{h,\psi}}\dr\lambda=\dr\cL_{X_{h,\psi}}\lambda=-\,\gamma_{\,0}\dr\lambda$, 
one has that 
\beq
\cL_{X_{h,\psi}}\Omega_{\,\lambda}
=-\,(n+1)\,\gamma_{\,0}\,\Omega_{\,\lambda}.
\label{phase-compressibility-Lie-derivative-psi}
\eeq
Comparing  \fr{phase-compressibility-Lie-derivative-psi} with 
\fr{def-phase-compressibility}, one completes the proof.
\qed
\end{Proof}

There is a counterpart of this Lemma.
\begin{Lemma}
( Phase compressibility for $X_{\,h,\varphi}$ with $\Omega_{\,\lambda}$ ) : 
Consider the system stated in Proposition\,\ref{lift-varphi-proposition}. 
Choose   
$\Gamma^{\,\varphi}(\Delta^{\,0})=\gamma^{\,0}\Delta^{\,0}$ with $\gamma^{\,0}$ being 
a constant. 
Then, the phase compressibility  
defined in \fr{def-phase-compressibility}
is calculated as 
$$
\kappa_{\Omega_{\,\lambda}}(X_{h,\varphi})
=-\,(n+1)\gamma^{\,0},
$$
where $\Omega_{\,\lambda}$ has been defined in 
\fr{def-standard-volume-form-contact-manifold}.
\end{Lemma}
\begin{Proof}
A way to prove this is analogous to that of 
Lemma\,\ref{phase-compressibility-psi}.
\qed
\end{Proof}

The following theorem and its counterpart are keys to deal with 
statistical behaviour of the lifted vector fields.
\begin{Thm}
\label{theorem-inivariant-measure-psi}
( Invariant measure for $X_{\,h,\psi}$ with $\Omega_{\,\lambda}$ ) : 
Consider the system stated in Proposition\,\ref{lift-psi-proposition} with 
$h_{\,\psi}(x(0),p(0),z(0))> 0$.  
Choose   
$\Gamma_{\,\psi}(\Delta_{\,0})=\gamma_{\,0}\Delta_{\,0}$ with $\gamma_{\,0}$ being 
a non-zero constant.
An invariant measure in the sense of  
Definition\,\ref{def-invariant-measure-general} 
is obtained as 
$$
\Omega_{h,\psi}^{\Inv}
=f_{h,\psi}^{\Inv}\,\Omega_{\,\lambda},\qquad\mbox{where}\quad 
f_{h,\psi}^{\Inv}
=\frac{1}{\cZ_{h,\psi}}h_{\,\psi}^{-(n+1)},
$$
for the class of  the form $f_{h,\psi}^{\Inv}=f_{h,\psi}^{\Inv}(h_{\,\psi})$ 
with  
$\cZ_{h,\psi}$ being a non-zero constant. 
\end{Thm}
\begin{Proof}
A way to prove this is to verify that $\cL_{X_{h,\psi}}\Omega_{h,\psi}^{\Inv}=0$.
First, one has that 
\beqa
\cL_{X_{h,\psi}}f_{h,\psi}^{\Inv}
&=&X_{h,\psi}f_{h,\psi}^{\Inv}
=\frac{\dr f_{h,\psi}^{\Inv}}{\dr h_{\,\psi}}(X_{h,\psi}h_{\,\psi})
=\frac{\dr f_{h,\psi}^{\Inv}}{\dr h_{\,\psi}}(\Reeb h_{\,\psi} )\,h_{\,\psi}
\non\\
&=&\frac{\dr f_{h,\psi}^{\Inv}}{\dr h_{\,\psi}}\frac{\partial h_{\,\psi}}{\partial z}h_{\,\psi}
=-\,\frac{\dr f_{h,\psi}^{\Inv}}{\dr h_{\,\psi}}\frac{\dr \Gamma_{\,\psi}}{\dr \Delta_{\,0}}h_{\,\psi}
=-\,\gamma_{\,0}\frac{\dr f_{h,\psi}^{\Inv}}{\dr h_{\,\psi}}h_{\,\psi},
\non
\eeqa
where  \fr{coordinate-expression-Reeb}, \fr{Xh=hRh}, 
$f_{h,\psi}^{\Inv}=f_{h,\psi}^{\Inv}(h_{\,\psi})$, 
and 
$\Gamma_{\,\psi}(\Delta_{\,0})=\gamma_{\,0}\Delta_{\,0}$ have been used.
Second, substituting this expression and 
\fr{phase-compressibility-Lie-derivative-psi}, one has that     
\beqa
\cL_{X_{h,\psi}}\Omega_{h,\psi}^{\Inv}
&=&\cL_{X_{h,\psi}}\left(\,f_{h,\psi}^{\Inv}\,\Omega_{\,\lambda}\,\right)
=\left(\,\cL_{X_{h,\psi}}f_{h,\psi}^{\Inv}\,\right)\,\Omega_{\,\lambda}
+f_{h,\psi}^{\Inv}\,\left(\,\cL_{X_{h,\psi}}\,\Omega_{\,\lambda}\,\right)
\non\\
&=&
-\,\gamma_{\,0}\left[\,h_{\,\psi}\frac{\dr f_{h,\psi}^{\Inv}}{\dr h_{\,\psi}}
+(n+1)f_{h,\psi}^{\Inv}
\right]\,\Omega_{\,\lambda}.
\non
\eeqa
Finally, one observes that the assumed form of $f_{h,\psi}^{\Inv}$ satisfies 
the equation 
$$
\frac{\dr f_{h,\psi}^{\Inv}}{\dr h_{\,\psi}}
+\frac{n+1}{h_{\,\psi}}f_{h,\psi}^{\Inv}
=0.
$$ 
Thus, one has that $\cL_{X_{h,\psi}}\Omega_{h,\psi}^{\Inv}=0$.
\qed
\end{Proof}
\begin{Remark}
This invariant measure is the same as that in Ref.\,\cite{Bravetti-JPhysA2014}. 
\end{Remark}
\begin{Remark}
Due to Remark\,\ref{invariant-subset-contact-Hamiltonian-psi}, 
the subset $\{(x,p,z)|h_{\,\psi}>0\}\subset \cC$ is invariant one, 
one has that $f_{\,h,\psi}^{\Inv}>0$. 
\end{Remark}

There is a counterpart of this Theorem.
\begin{Thm}
( Invariant measure for $X_{\,h,\varphi}$ with $\Omega_{\,\lambda}$ ) : 
Consider the system stated in Proposition\,\ref{lift-varphi-proposition} with 
$h_{\,\varphi}(x(0),p(0),z(0))> 0$.  
Choose   
$\Gamma^{\,\varphi}(\Delta^{\,0})=\gamma^{\,0}\Delta^{\,0}$ with $\gamma^{\,0}$ being 
a non-zero constant.
An invariant measure in the sense of  
Definition\,\ref{def-invariant-measure-general} is obtained as 
$$
\Omega_{\,h,\varphi}^{\Inv}
=f_{\,h,\varphi}^{\Inv}\,\Omega_{\,\lambda},\qquad\mbox{where}\quad 
f_{\,h,\varphi}^{\Inv}
=\frac{1}{\cZ_{\,h,\varphi}}h_{\,\varphi}^{-(n+1)},
$$
for the class of  the form $f_{\,h,\varphi}^{\Inv}=f_{\,h,\varphi}^{\Inv}(h_{\,\varphi})$ 
with $\cZ_{\,h,\varphi}$ being a non-zero constant. 
\end{Thm}
\begin{Proof}
A way to prove this is analogous to that of  
Theorem\,\ref{theorem-inivariant-measure-psi}.
\qed
\end{Proof}
\begin{Remark}
Due to Remark\,\ref{invariant-subset-contact-Hamiltonian-varphi}, 
the subset $\{(x,p,z)|h_{\,\varphi}>0\}\subset \cC$ is invariant one, 
one has $f_{\,h,\varphi}^{\Inv}>0$. 
\end{Remark}
\subsection{Stability theorems}
In this subsection, 
we show 
 several classes where lifted flows asymptotically approach to 
the Legendre submanifold.

When $\{\,F_{\,\psi}^{\,1},\ldots,F_{\,\psi}^{\,n}\,\}$ and 
$\Gamma_{\,\psi}$ in \fr{lift-general-psi-components}
have some properties, it is shown below  
that flows asymptotically approach to $\cA_{\,\psi}^{\,\cC}$.
\begin{Thm}
\label{theorem-controlled-relaxation-psi-exponential}
( Stability theorem 1 ) : 
Let $\gamma_{\,0}$  and $\Lambda_{\,0}$ be non-zero constants. 
Consider the dynamical system \fr{lift-general-psi-components} with 
$$
\frac{\partial F_{\,\psi}^{\,b}}{\partial x^{\,a}}
=\Lambda_{\,0}\,\delta_{\,a}^{\,b},\quad\mbox{and}\quad
\Gamma_{\,\psi}(\,\Delta_{\,0}\,)
=\gamma_{\,0}\,\Delta_{\,0}.
$$
If the conditions 
$$
\Lambda_{\,0}+\gamma_{\,0}>0,\quad \mbox{and}\quad 
\gamma_{\,0}>0,
$$
are satisfied, then the flow asymptotically approaches to $\cA_{\,\psi}^{\,\cC}$. 
\end{Thm} 
\begin{Proof}
The solutions $\Delta_{\,0}(t)$ and $\Delta_{\,a}(t)$  
are obtained by solving \fr{lift-general-psi-components} as 
$$
\Delta_{\,0}(t)
=\e^{-\,\gamma_{\,0}\,t}\Delta_{\,0}(0),\quad \mbox{and}\quad 
\Delta_{\,a}(t)
=\e^{-\,(\,\gamma_{\,0}+\Lambda_{\,0}\,)\,t}\Delta_{\,a}(0).
$$
From these equations one has that 
$$
\lim_{t\,\to\,\infty}\Delta_{\,0}(t)
=0,\quad \mbox{and}\quad
\lim_{t\,\to\,\infty}\Delta_{\,a}(t)
=0,
$$
or equivalently 
$$
\psi(x(\infty))
=z(\infty),\quad\mbox{and}\quad
\frac{\partial\psi}{\partial x^{\,a}}(\,x(\infty)\,) 
=p_{\,a}(\infty),
$$
which are the conditions that the flow asymptotically approaches to 
$\cA_{\,\psi}^{\,\cC}$.
\qed
\end{Proof}
\begin{Remark}
When $\Lambda_{\,0}=0$, this dynamics is identical to the relaxation process 
discussed in Ref.\,\cite{Goto2015}.
\end{Remark}

There is a counterpart of this theorem.  
\begin{Thm}
\label{theorem-controlled-relaxation-varphi-exponential}
( Stability theorem 2 ) : 
Let $\gamma^{\,0}$  and $\Lambda^{\,0}$ be non-zero constants. 
Consider the dynamical system \fr{lift-general-varphi-components} with 
$$
\frac{\partial F_{\,b}^{\,\varphi}}{\partial p_{\,a}}
=\Lambda^{\,0}\,\delta_{\,b}^{\,a},\quad\mbox{and}\quad
\Gamma^{\,\varphi}(\,\Delta^{\,0}\,)
=\gamma^{\,0}\,\Delta^{\,0}.
$$
If the conditions 
$$
\Lambda^{\,0}+\gamma^{\,0}>0,\quad \mbox{and}\quad 
\gamma^{\,0}>0,
$$
are satisfied, then the flow asymptotically approaches to 
$\cA_{\,\varphi}^{\,\cC}$. 
\end{Thm} 
\begin{Proof}
A way to prove this is analogous to that of  
Theorem \ref{theorem-controlled-relaxation-psi-exponential}.
\qed
\end{Proof}

In Theorem \ref{theorem-controlled-relaxation-psi-exponential},  
asymptotic behaviour is stated for    
the case where $(\,\partial F_{\,\psi}^{\,a}/\partial x^{\,b}\,\,)$ is diagonal and 
$\{\,x^{\,1},\ldots,x^{\,n}\,\}$ 
is exponentially increasing or decreasing in $t$.  
In the following, asymptotic behavior is stated for a class of systems where 
$\{\,x^{\,1}(t),\ldots,x^{\,n}(t)\,\}$ is oscillatory in $t$.

\begin{Thm}
\label{theorem-controlled-relaxation-psi-oscillatory}
( Stability theorem 3 ) : 
Let $n=2$, and $\gamma_{\,0}>0,\omega$ constants. 
Consider the dynamical system \fr{lift-general-psi-components} with 
$$
F_{\,\psi}^{\,1}\left(\,x^{\,1},x^{\,2}\,\right)
=\omega\,x^{\,2},\quad 
F_{\,\psi}^{\,2}\left(\,x^{\,1},x^{\,2}\,\right)
=-\,\omega\,x^{\,1},\quad \mbox{and}\quad
\Gamma_{\,\psi}(\,\Delta_{\,0}\,)
=\gamma_{\,0}\,\Delta_{\,0}.
$$
Then 
the flow asymptotically approaches to $\cA_{\,\psi}^{\,\cC}$. 
\end{Thm}
\begin{Proof}
The contact Hamiltonian flow is expressed as 
$$
\frac{\dr x^{\,a}}{\dr t}
=F_{\,\psi}^{\,a}.\quad
\frac{\dr}{\dr t}\left(
\begin{array}{cc}
\Delta_{\,1}\\
\Delta_{\,2}
\end{array}
\right)
=\left(
\begin{array}{cc}
-\,\gamma_{\,0}&\omega\\
-\,\omega&-\,\gamma_{\,0}
\end{array}
\right)
\left(
\begin{array}{cc}
\Delta_{\,1}\\
\Delta_{\,2}
\end{array}
\right),
\qquad
\frac{\dr}{\dr t }\Delta_{\,0}
=-\,\gamma_{\,0}\,\Delta_{\,0}.
$$
The solutions for $\Delta_{\,0}, \Delta_{\,1}$ and $\Delta_{\,2}$ are given by 
\beqa
\Delta_{\,0}(t)
&=&\e^{-\,\gamma_{\,0}\,t}\,\Delta_{\,0}(0),
\non\\
\Delta_{\,1}(t)
&=&\e^{-\,\gamma_{\,0}\,t}\,\left[\,\Delta_{\,1}(0)\cos\,(\,\omega\, t\,)
+\Delta_{\,2}(0)\sin\,(\,\omega\, t\,)\,\right]
,\non\\
\Delta_{\,2}(t)
&=&\e^{-\,\gamma_{\,0}\,t}\,\left[\,-\,\Delta_{\,1}(0)\sin\,(\,\omega\, t\,)
+\Delta_{\,2}(0)\cos\,(\,\omega\, t\,)\,\right].
\non
\eeqa
From these equations one has that 
$$
\lim_{t\,\to\,\infty}\Delta_{\,0}(t)
=\lim_{t\,\to\,\infty}\Delta_{\,1}(t)
=\lim_{t\,\to\,\infty}\Delta_{\,2}(t)
=0,
$$
or, equivalently 
$$
\psi(x(\infty))
=z(\infty),\quad\mbox{and}\quad
\frac{\partial\psi}{\partial x^{\,a}}(\,x(\infty)\,) 
=p_{\,a}(\infty),\quad a\in\{\,1,2\,\}
$$
which are the conditions that the flow asymptotically approaches to 
$\cA_{\,\psi}^{\,\cC}$.
\qed
\end{Proof}

There is a counterpart of this theorem.
\begin{Thm}
\label{theorem-controlled-relaxation-varphi-oscillatory}
( Stability theorem 4 ) : 
Let $n=2$, and $\gamma^{\,0}>0,\omega$ constants. 
Consider the dynamical system \fr{lift-general-varphi-components} with 
$$
F_{\,1}^{\,\varphi}\left(\,p_{\,1},p_{\,2}\,\right)
=\omega\,p_{\,2},\quad 
F_{\,2}^{\,\varphi}\left(\,p_{\,1},p_{\,2}\,\right)
=-\,\omega\,p_{\,1},\quad \mbox{and}\quad
\Gamma^{\,\varphi}(\,\Delta^{\,0}\,)
=\gamma^{\,0}\,\Delta^{\,0}.
$$
Then 
the flow asymptotically approaches to $\cA_{\,\varphi}^{\,\cC}$. 
\end{Thm}
\begin{Proof}
A way to prove this  is 
analogous to that of   
Theorem\,\ref{theorem-controlled-relaxation-psi-oscillatory}.
\qed
\end{Proof}

So far the given theorems above do not involve any fixed point for $x$.  
However, the following involves a fixed point for $x$.

\begin{Thm}
\label{theorem-controlled-relaxation-psi-gradient-flow}
( Stability theorem 5 ) : 
Let $\gamma_{\,0}>0$ be a constant, 
$(\,L^{\,ab}\,)$ 
a constant symmetric positive definite matrix, 
and $\cU_{\,\psi}$ a positive function of $x$ only. 
Consider the dynamical  system \fr{lift-general-psi-components} with 
$$
F_{\,\psi}^{\,a}(x)
=-\,L^{\,aj}\frac{\partial\,\cU_{\,\psi}}{\partial x^{\,j}},
\quad\mbox{and}\quad 
\Gamma_{\,\psi}(\,\Delta_{\,0}\,)
=\gamma_{\,0}\,\Delta_{\,0}.
$$
If the condition 
$$
\gamma_{\,0}L^{\,ab}-
L^{\,ak}\frac{\partial^2\,\cU_{\,\psi}}{\partial x^{\,k}\partial x^{\,j}}
L^{\,jb}
\quad
\mbox{are components of a positive definite matrix for all $x$}
$$
is satisfied, then the flow asymptotically approaches to 
a fixed point in $\cA_{\,\psi}^{\,\cC}$. Here the fixed point 
$(\,\ol{x},\ol{p},\ol{z}\,)\in\cA_{\,\psi}^{\,\cC}$ is determined by 
$$
L^{\,aj}\frac{\partial\, \cU_{\,\psi}}{\partial x^{\,j}}\left(\ol{x}\right)
=0,\quad 
\ol{\Delta_{\,0}}
=\ol{\Delta_{\,1}}
=\cdots
=\ol{\Delta_{\,n}}
=0.
$$
\end{Thm}
\begin{Proof}
It is shown below that there exists a Lyapunov function. 
Then  
applying the stability theorem, one can complete the proof  
 ( see Ref.\,\cite{HS1974} ). 
Define the function $V$ on $\cC$ as 
$$
V(x,p,z)
=\cU_{\,\psi}(x)+\frac{1}{2}\Delta_{\,a}(x,p)\,L^{\,ab}\,\Delta_{\,b}(x,p)
+\frac{1}{2}\left(\,\Delta_{\,0}(x,z)\,\right)^2.
$$
Then it follows that $V\geq 0$, and 
$$
\dot{V}
=-\,\frac{\partial\,\cU_{\,\psi}}{\partial x^{\,b}}\,L^{\,ba}\,
\frac{\partial\,\cU_{\,\psi}}{\partial x^{\,a}}
-\Delta_{\,b}\left(\,
\gamma_{\,0}L^{\,ba}-L^{\,bk}\frac{\partial^2\,\cU_{\,\psi}}{\partial x^{\,k}\partial x^{\,j}}L^{\,ja}
\,\right)\Delta_{\,a}
-\,\gamma_{\,0}\,\left(\,\Delta_{\,0}\,\right)^2.
$$
Thus if the condition is satisfied, then 
$\dot{V}\leq 0$. 
On $(\ol{x},\ol{\Delta_{\,0}},\ol{\Delta_{\,1}},\ldots,\ol{\Delta_{\,n}})$, 
the equality holds. Thus, $V$ is a Lyapunov function.
\qed
\end{Proof}
\begin{Remark}
The function $\cU_{\,\psi}$ 
need not depend on $\psi$. 
\end{Remark}

There is a counterpart of this Theorem. 
\begin{Thm}
\label{theorem-controlled-relaxation-varphi-gradient-flow}
( Stability theorem 6 ) : 
Let $\gamma^{\,0}>0$ be a constant, 
$(\,L_{\,ab}\,)$ components of 
a constant symmetric positive definite matrix, 
and $\cU^{\,\varphi}$ a positive function of $p$ only.
Consider the dynamical  system \fr{lift-general-varphi-components} with 
$$
F_{\,a}^{\,\varphi}(p)
=-\,L_{\,aj}\frac{\partial\,\cU^{\,\varphi}}{\partial p_{\,j}},
\quad\mbox{and}\quad 
\Gamma^{\,\varphi}(\,\Delta^{\,0}\,)
=\gamma^{\,0}\,\Delta^{\,0}.
$$
If the condition 
$$
\gamma^{\,0}L_{\,ab}-
L_{\,ak}\frac{\partial^2\,\cU^{\,\varphi}}{\partial p_{\,k}\partial p_{\,j}}
L_{\,jb}
\quad
\mbox{are components of a positive definite matrix for all $p$}
$$
is satisfied, then the flow asymptotically approaches to 
a fixed point in $\cA_{\,\varphi}^{\,\cC}$. Here the fixed point 
$(\,\ol{x},\ol{p},\ol{z}\,)\in\cA_{\,\psi}^{\,\cC}$ is determined by 
$$
L_{\,aj}\frac{\partial\, \cU^{\,\varphi}}{\partial p_{\,j}}\left(\ol{p}\right)
=0,\quad 
\ol{\Delta^{\,0}}
=\ol{\Delta^{\,1}}
=\cdots
=\ol{\Delta^{\,n}}
=0.
$$
\end{Thm}
\begin{Proof}
A way to prove this  is 
analogous to that of   
Theorem\,\ref{theorem-controlled-relaxation-psi-gradient-flow}.
\qed
\end{Proof}
\begin{Remark}
The function $\cU_{\,\varphi}$ 
need not depend on $\varphi$. 
\end{Remark}


\subsection{Application  to a spin system under time-dependent controlled magnetic field}
In this subsection it is shown how useful the general theory is in physics by 
analyzing a simple model.
Since contact geometry is used for geometrization of thermodynamics and 
statistical mechanics,  
a variety of examples are expected to be found for this purpose. 
In this paper a  nonequilibrium system is focused  
since it involves naturally time-development physical quantities. 
We choose as an example a spin system 
in contact with a time-dependent external magnetic field and  
time-independent heat bath, since   
this example is simple enough. In the case where 
the system is in contact with 
time-independent heat bath and time-independent magnetic field, 
the equilibrium state and a relaxation process 
have been studied in Ref.\,\cite{Goto2015}. Thus one can compare 
the present case and the previous one. 
Here relaxation process is defined such that every physical quantity 
approaches to the one at the equilibrium state as time develops, 
and equilibrium state is defined such that the canonical distribution of  
microscopic variables is realized. If the equilibrium state is realized 
for a given system and 
the corresponding cumulant generating function is analytically obtained, 
then an expectation value is in principle 
calculated by differentiating 
the cumulant generating function with respect to the thermodynamic
conjugate variable ( see Definition\,\ref{definition-spin-equilibrium} ).   

Symbols for physical variables 
used in this subsection are summarized in 
Table \ref{Symbols-for-physical-quantities}.  
\begin{table}[htbc]
\begin{center}
\begin{tabular}{|c|l|}
\hline
symbol&interpretation\\
\hline
$t$&time\\
$T_{\abs}$&absolute temperature\\
$k_{\,\B}$&Boltzmann constant\\
$H$&magnetic field whose dimension is an energy\\
$\sigma$&spin that takes values $\pm 1$\\
$\cF$& Helmholtz free energy\\
\hline
\end{tabular}
\end{center}
\caption{Symbols for physical quantities}
\label{Symbols-for-physical-quantities}
\end{table}

In the following the definition of a spin system at 
the equilibrium state is given. 
\begin{Def}
\label{definition-spin-equilibrium}
( Equilibrium state of a spin system with an external constant magnetic field 
in contact with a heat bath  ) :  
Let $\sigma=\pm 1$ be a spin variable, $H$ a spatially homogeneous and 
time-independent external magnetic field 
whose physical dimension is an energy, and $\theta:=H/(\,k_{\,B}T_{\abs}\,)$. 
Then the canonical distribution function for this system is defined to be 
$$
\mbbP_{\,\theta}^{\,\can}(\sigma)
:=\exp(\,\,\theta\,\sigma-\psi\,(\,\theta\,)\,),
$$
where $\psi(\theta)$ is to normalize  $\mbbP_{\,\theta}^{\,\can}(\sigma)$ such that 
$\sum_{\sigma=\pm1}\mbbP_{\,\theta}^{\,\can}(\sigma)=1$ as 
$$
\psi\,(\,\theta\,)
=\ln\,\cosh\theta+\ln 2.
$$
In addition, the equilibrium value of the magnetization is defined to be
$$
\eta(\theta)
:=\frac{\partial\,\psi}{\partial\theta}
=
\avgg{\sigma}{\can}
=\tanh\theta,
$$
where 
$$
\avgg{\cO}{\can}
:=\sum_{\sigma=\pm 1}\cO(\sigma)\,\mbbP_{\,\theta}^{\,\can}(\sigma),
$$
for a given function $\cO$ of $\sigma$. 
The macroscopic equilibrium state for this system is defined to be the 
triplet 
$(\,\theta,\eta\,(\,\theta\,),\psi(\,\theta\,)\,)$.
\end{Def} 
The physical interpretation of $\psi$ is the cumulant generating function or 
the negative dimensionless Helmholtz 
free energy : $\psi=-\cF/(\,k_{\B}\,T_{\abs}\,)$ ( see Ref.\,\cite{Goto2015} ). 
Instead of $\psi$, a partition function $Z(\,\theta\,)$ 
is used in physics to normalize a distribution function. The relation between 
$Z(\,\theta\,)$
and $\psi(\,\theta\,)$ is given by $\psi(\,\theta\,)=\ln\,Z(\,\theta\,)$.

To geometrically describe a relaxation dynamics, where the system is 
in contact with time-dependent external magnetic field 
and time-independent heat bath, the following postulates are 
made.
\begin{Postulate}
1. Coordinates $(x,p,z)$ of a $3$-dimensional contact manifold 
$(\,\cC,\lambda\,)$ can be introduced such that $\lambda=\dr z-p\,\dr x$, 
and that 
$$
(\,x,p,z\,)|_{\cA_{\,\psi}^{\,\cC}}
=(\,\theta,\eta\,(\,\theta\,),\psi\,(\,\theta\,)\,).
$$
2. The domain of $\psi$ can be extended such that one can write $\psi(x)$. 
\end{Postulate}
The physical meaning of the Legendre submanifold generated by $\psi$ 
is the subspace where the 
thermodynamic relation $\eta(\theta)=\partial\psi/\partial \theta$ holds and 
the value of the negative dimensionless Helmholtz free energy 
is the same as that calculated by 
$\psi$, $z(\theta)=\psi(\theta)$.   
Furthermore, when $T_{\abs}$ does not depend of $t$, 
physical interpretations of canonical coordinates $(x,p,z)$ 
for the contact manifold  
are given as follows. 
\begin{itemize}
\item  
$x(t)$ can express the time-dependent magnetic field. \\
At equilibrium, it follows that  
$x|_{\,\cA_{\,\psi}^{\,\cC}}=\theta=H/(k_{\B}T_{\,\abs})$. 
\item 
$p(t)$ is the time-dependent magnetization.\\
At equilibrium, it follows that   
$p|_{\,\cA_{\,\psi}^{\,\cC}}=\avgg{\sigma}{\can}=\eta=\partial\psi(\theta)/\partial\theta$.
\item 
$z(t)$ is the time-dependent negative dimensionless Helmholtz free energy.\\
At equilibrium, it follows that   
$z|_{\,\cA_{\,\psi}^{\,\cC}}=-\cF(\theta)/(k_{\B}\,T_{\abs})=\psi(\,\theta\,)$.  
\end{itemize}

In the special case where $x$ is constant in time, 
the contact Hamiltonian vector field associated 
with $h=\gamma_{\,0}(\psi(x)-z)$ has been studied in Ref.\,\cite{Goto2015}.  
The vector field is expressed in this case as 
\beq
\frac{\dr\,x}{\dr t}
=0,\qquad
\frac{\dr\,p}{\dr t}
=\gamma_{\,0}\left(\,\frac{\dr\psi}{\dr x}(x)-p\,\right),\qquad 
\frac{\dr z}{\dr t}
=\gamma_{\,0}(\psi(x) -z),
\label{contact-Hamilton-vector-kinetic-spin-model-free}
\eeq
where $\psi(x)=\ln\cosh(x)+\ln 2$.
This vector field has been shown to include a 
kinetic spin model without spin-coupling, where that model is obtained 
from a master equation under the detailed balance condition.

The system where 
$x$ is controlled such that $x$ approaches to a fixed value 
$\theta$ as time develops  
can be considered 
by means of the developed general theory of this paper. 
As an application of 
Theorem\,\ref{theorem-controlled-relaxation-psi-exponential},
one has the model which shows a relaxation process 
in contact with a controlled magnetic field as follows.  
\begin{Thm}
( Relaxation dynamics of a spin system in controlled magnetic field ) : 
Let 
$(\cC,\lambda)$ be a contact manifold with $\cC=\mbbR^3$ 
whose coordinates are $(x,p,z)$, and $\lambda=\dr z-p\,\dr x$. In addition, 
$\Lambda_{\,0}, \theta$ and $\gamma_{\,0}$ be positive constants.
Consider the system where $T_{\abs}$ is constant in time,   
the magnetic field divided by 
$k_{\,B}T_{\abs}$, denoted $x(t)$, is controlled as 
$$
x(t)=(\,x(0)-\theta\,)\,\e^{-\Lambda_{\,0}\,t}+\theta,
$$
and the condition $\gamma_{\,0}>\Lambda_{\,0}$ is satisfied.  
Then  the contact Hamiltonian vector field 
associated with \fr{lift-general-psi-contact-Hamiltonian}
\beqa
\frac{\dr}{\dr\, t}\, x
&=&-\,\Lambda_{\,0}\, (\,x-\theta\,),
\non\\
\frac{\dr}{\dr\, t}\, p
&=&-\,\Lambda_{\,0}\,(x-\theta)\frac{\dr^2\psi}{\dr x^2}(x)
-\Lambda_{\,0}\left(\frac{\dr\psi}{\dr x}(x)-p\right)+\gamma_{\,0}
\left(\frac{\dr\psi}{\dr x}(x)-z\right),
\non\\
\frac{\dr}{\dr\, t}\,z
&=&-\,\Lambda_{\,0}(x-\theta)\frac{\dr\psi}{\dr x}(x)+\gamma_{\,0}
(\,\psi(x)-z\,),
\non
\eeqa
with $\psi(x)=\ln\cosh(x)+\ln 2$, or equivalently, 
$$
\frac{\dr }{\dr t}x
=-\,\Lambda_{\,0}\, (\,x-\theta\,),\qquad
\frac{\dr}{\dr t}\Delta_{\,1}
=\Lambda_{\,0}\,\Delta_{\,1}-\gamma_{\,0}\,\Delta_{\,1},\qquad 
\frac{\dr}{\dr t}\Delta_{\,0}
=-\,\gamma_{\,0}\Delta_{\,0},
$$
with 
$$
\Delta_{\,0}(x,z)
=\psi(x)-z,\qquad
\Delta_{\,1}(x,p)
=\frac{\dr\,\psi}{\dr x}(x)-p,
$$
models a relaxation dynamics under the controlled magnetic field : 
$$
x(\infty)
=\theta,\qquad 
\psi(x(\infty))
=z(\infty),\qquad 
\frac{\partial\psi}{\partial x}(x(\infty))
=p(\infty).
$$
\end{Thm}
\begin{Proof}
Applying Theorem\,\ref{theorem-controlled-relaxation-psi-exponential}, 
one completes the proof.
\qed
\end{Proof}
In the case $\Lambda_{\,0}\equiv 0$, this system reduces to 
\fr{contact-Hamilton-vector-kinetic-spin-model-free}
that includes the kinetic spin model without control. 
Although it interesting to have the corresponding 
kinetic model in contact with the controlled magnetic field, where  
the model is constructed based on a master equation, 
such a model has not been known.

\subsection{Application of the general theory to a class of 
phenomenological equations in nonequilibrium thermodynamics}
The aim of this subsection is to show how 
Theorem\,\ref{theorem-controlled-relaxation-psi-gradient-flow} is applied 
to nonequilibrium thermodynamics. 
In particular, the so-called phenomenological equations are considered. 

In nonequilibrium thermodynamics, it has been proposed that 
time evolution of some macroscopic  
variables $x=\{\,x^{\,1 },\ldots,x^{\,n}\,\}\in\mbbR^{n}$ 
is described by the following gradient system 
\beq
\frac{\dr}{\dr t}x^{\,a}
=-\,L^{aj}\frac{\partial}{\partial x^{\,j}}\,\cU(\,x\,)
=:F^{\,a}(\,x\,),
\label{phenomenological-equation-general}
\eeq
where $\{\,L^{\,ab}\,\}$ is identified with a set of 
the Onsager coefficients, $\cU$ 
a free energy or entropy\cite{Onsager1953}.
These systems are referred to as {\it phenomenological equations} and 
often assumed to be valid near the equilibrium state. 
In this paper, we assume that this class of systems 
is described at the state   
where  the 
following thermodynamic relations 
\beq
p_{\,a}
=\frac{\partial\psi}{\partial x^{\,a}}(\,x\,),\qquad 
a\in\{\,1,\ldots,n\,\}
\label{phenomenological-equation-equilibrium-condition}
\eeq
are satisfied,  with $p=\{\,p_{\,1},\ldots,p_{\,n}\,\}$ being a 
set of thermodynamic conjugate variables of $x$, and  
$\psi$ some potential. 
Note that under a similar assumption,  
a similar system has been studied in Ref.\,\cite{Schaft2005},
where a dynamical system has been derived 
{\it on} the Legendre submanifold generated 
by a function whose physical interpretation is an energy.  
In their system it has been shown that 
the energy is conserved and the entropy increases as time 
develops.  
Note also that in general, 
if a dynamical system is on the Legendre submanifold, then its 
standard physical interpretation   
in the contact geometric thermodynamics is that the system 
is at equilibrium\cite{MrugalaX}.  

In contact geometry under our assumption, and introducing a 
new variable $z$ such that $z=\psi(x)$,  
\fr{phenomenological-equation-general} can be seen as a dynamical system on 
the $n$-dimensional Legendre submanifold generated by $\psi$ of a 
$(2n+1)$-dimensional contact manifold. 
With our framework on how to lift a dynamical system to the contact manifold, 
such a lifted dynamical system is immediately obtained as 
\fr{lift-general-psi-components} with 
$F^{\,a}(\,x\,)$ given by \fr{phenomenological-equation-general}. 
This lifted dynamical system 
can be a candidate of a dynamical system that describes 
some nonequilibrium thermodynamics. 
The stability of this lifted system is understood as 
Theorem\,\ref{theorem-controlled-relaxation-psi-gradient-flow}.  

If the system  is 
restricted to the state where 
\fr{phenomenological-equation-equilibrium-condition} is satisfied, and  
has some special properties, one has the following.
\begin{Proposition}
( Dually flat space for phenomenological equations at equilibrium ) :  
Consider the system \fr{phenomenological-equation-general}, where 
all the following conditions are satisfied
\begin{itemize} 
\item
The relation \fr{phenomenological-equation-equilibrium-condition} holds.
\item 
The matrix $(\,L_{\,ab}\,)$ 
is a constant symmetric positive definite matrix 
\item
The introduced potential is identical to $\psi$ and is of the form 
$$
\cU(x)
=\psi(x)
=\frac{1}{2}\,x^{\,a}\,M_{\,ab}\,x^{\,b},
$$
where $(\,M_{\,ab}\,)$ is the inverse matrix of $(\,L^{\,ab}\,)$. 
\end{itemize}
Then, this system can be seen as a dynamical system on 
a dually flat space 
with the metric tensor field $g=L_{\,ab}\,\dr x^{\,a}\otimes \,\dr x^{\,b}$, and 
$x$ and $p$ are mutually dual with respect to $g$. In addition, 
integral curves in $p$-coordinates trace $\nabla^{\,*}$-geodesic curves 
connecting 
$p(0)$ and $p=0$ where 
$\nabla^{\,*}$ is the dual connection of $\nabla$ that is  defined 
such that $\nabla\dr\psi=g$. 
\end{Proposition}
\begin{Proof}
It immediately follows from $\nabla\dr\psi=g$ 
that $x$ is a set of $\nabla$-affine coordinates.  
Introducing   
$\varphi(\,p\,):=\Leg[\,\psi\,](\,p\,)$, which is 
a strictly convex function, 
one has 
$$
x^{\,a}
=\frac{\partial\,\varphi}{\partial p_{\,a}},\quad 
\mbox{and}\quad 
\dr x^{\,a}\left(\,\frac{\partial}{\partial p_{\,b}}\,\right)
=M^{\,ab},
$$
from which 
one concludes that $x$ and $p$ are mutually dual with respect to $g$ :   
$$
g\left(\,\frac{\partial}{\partial x^{\,a}},\frac{\partial}{\partial p_{\,b}}\,
\right)
=\delta_{\,a}^{\,b}.
$$
In addition, the dual connection $\nabla^{\,*}$ exists due to 
Lemma\,\ref{lemma-uniquely-detemined-dual-connection}, and 
the connection coefficients $\{\,\Gamma_{\,ab}^{\,\prime\ c}\,\}$ such that 
$\nabla_{\partial_{\,a}}^{\,\prime}\partial^{\,b}=\Gamma_{\,ab}^{\,\prime \ c}\partial_{\,c}, 
(\partial_{\,a}:=\partial/\partial x^{\,a})$ are 
$$
\Gamma_{\,ab}^{\,\prime\ c}
=g^{\,cj}\,\Gamma_{\,abj},\quad\mbox{with}\quad
\Gamma_{\,abc}^{\,\prime}
=\frac{\partial^3\psi}{\partial x^{\,a}\partial x^{\,b}\partial x^{\,c}}.
$$
It follows from Lemma\,\ref{lemma-dual-affine-coordinates}  
that $p$ is a set of $\nabla^{\,*}$-affine coordinates.   
Therefore, 
$(\,\mbbR^{\,n},g,\nabla,\nabla^{\,*}\,)$ is a dually flat space, where 
$\nabla$-affine coordinates are $x$, 
and $\nabla^{\,*}$-affine coordinates $p$.  
In addition, it follows that 
$$
\frac{\dr}{\dr t}p_{\,a}
=\frac{\dr}{\dr t}\left(\frac{\partial\psi}{\partial x^{\,a}}\right)
=M_{\,ab}\frac{\dr\,x^{\,b}}{\dr t}
=-\,M_{\,ab}L^{\,bj}\,\frac{\partial \psi}{\partial x^{\,j}}
=\,-p_{\,a},
$$
whose solution for each $a$ is obtained as 
$$
p_{\,a}(t)
=p_{\,a}(0)\,\e^{-t}.
$$
\qed
\end{Proof}

 \section{Generating function 
conserving 
lifting scheme for 
contact Hamiltonian vector fields}
\label{sec-generating-function-preserving-lifts}
Consider the vector field $X$  
on the embedded Legendre submanifold 
obtained by the 
push-forward of a vector field $X_{\,0}$, where the Legendre submanifold 
is generated by a function $\psi$, and 
$X=\Phi_{*}X_{\,0}$ holds with $\Phi$ being an embedding.  
As mentioned in Remark\,\ref{remark-dually-flat-space-psi-not-preserved}, 
the value of the generating function $\psi$ is not conserved 
in the sense that 
$\cL_{\,X}\psi\nequiv 0$. If one requires that 
the value of the generating 
function is conserved, 
then another formulation is needed. 
In this section, such a formulation is shown with  
a higher dimensional contact manifold. Note that this 
formulation can be seen
as a generalization of the work in Ref.\,\cite{Schaft2005}.

After fixing mathematical symbols that will be used in 
Subsections\,\ref{subsection-Vector_fields_on_dually_flat_space_and_their_lifts_in_contact geometric_language}--\ref{subsection-Applications_to_electric_circuit_models_in_a_thermal_environment}, 
such vector fields on 
dually flat spaces and their lifts will be  investigated. 

\subsection{Mathematical symbols}
In this subsection mathematical symbols are fixed as follows. 
Let $(\,\wt{\cC},\wt{\lambda}\,)$ be 
a $(2n+3)$-dimensional contact manifold with 
some fixed $n\in\{\,1,2,\ldots\,\}$, and 
$(x,x^{\,n+1},p,p_{\,n+1},z)$ canonical coordinates such that 
$\wt{\lambda}=\dr z-p_{\,a}\,\dr x^{\,a}-p_{\,n+1}\,\dr x^{\,n+1}$  
with $x=\{\,x^{\,1},\ldots,x^{\,n}\,\}$ and $p=\{\,p_{\,1},\ldots,p_{\,n}\,\}$.

\begin{Proposition}
( Coordinate expression of the Reeb vector field on $\wt{\cC}$ ) : 
The coordinate expression of the Reeb vector field 
on $(\wt{\cC},\wt{\lambda})$ is given by 
$$
\wt{\Reeb}
=\frac{\partial}{\partial z}.
$$
\end{Proposition}
\begin{Proof}
This is proved by substituting this expression of $\wt{\Reeb}$ into \fr{def-Reeb-vector}.
\qed
\end{Proof}
To discuss phase compressibilities of vector fields and invariant measures, 
the following $(2n+3)$-form will be used. 
\begin{Def}
( Standard volume-form on a high-dimensional contact manifold ) : 
$$
\wt{\Omega}_{\wt{\lambda}}
:=\wt{\lambda}\wedge
\underbrace{\dr\wt{\lambda}\wedge\cdots\wedge\dr\wt{\lambda}}_{n+1}.
$$
\end{Def}

Define a function on $\wt{\cC}$ 
$$
\wt{\psi}(x,x^{\,n+1})
=\psi(x)+p_{\,n+1}^{(0)}x^{\,n+1},
$$
where $p_{\,n+1}^{\,(0)}$ is a non-zero constant, and $\psi$ a function.
Then, applying Theorem\,\ref{theorem-Legendre-submanifold-theorem-Arnold}, 
one has the $(n+1)$-dimensional Legendre submanifold $\wt{\cA}_{\,\wt{\psi}}$ 
generated by $\wt{\psi}$, such that 
\beq
\wt{\cA}_{\,\wt{\psi}}^{\,\wt{\cC}}
:=\Phi_{\,\wt{\cC}\wt{\cA}\wt{\psi}}\wt{\cA}_{\,\wt{\psi}}
=\left\{\,(x,x^{\,n+1},p,p_{\,n+1},z)\in\wt{\cC}\,\bigg|\,
p_{\,j}=\frac{\partial\psi}{\partial x^{\,j}},\,
p_{\,n+1}=p_{\,n+1}^{(0)},\,
z=\wt{\psi},\,j\in\{1,\ldots,n\}
\,\right\},
\label{(n+1)-dimensional-Legendre-submanifold}
\eeq  
where $\Phi_{\,\wt{\cC}\wt{\cA}\wt{\psi}}:\wt{\cA}_{\,\wt{\psi}}\to\wt{\cC}$ is 
an embedding. The following functions 
$$
\wt{\Delta}_{\,0}(x,x^{\,n+1},z)
:=\wt{\psi}(x,x^{\,n+1})-z,\qquad
\wt{\Delta}_{\,a}(x,p,p_{\,n+1})
:=\frac{p_{\,n+1}}{p_{\,n+1}^{\,(0)}}
\frac{\partial\psi}{\partial x^{\,a}}(x)-p_{\,a}
$$
will be used. It follows that  
$$
\left.\wt{\Delta}_{\,0}\right|_{\wt{\cA}_{\,\wt{\psi}}^{\,\wt{\cC}}}
=0,\qquad\mbox{and}\qquad 
\left.\wt{\Delta}_{\,a}\right|_{\wt{\cA}_{\,\wt{\psi}}^{\,\wt{\cC}}}
=0,\quad a\in\{\,1,\ldots,n\,\}.
$$
In addition, since the matrix components which are 
differentiation of $\wt{\psi}$ with respect to $x$ and $x^{\,n+1}$ are obtained 
as 
$$
\wt{g}_{\,\mu\nu}
=
\left(
\begin{array}{cc}
\frac{\partial^{\,2}\psi}{\partial x^{\,a}\partial x^{\,b}}&0\\
0&0
\end{array}
\right),\qquad 
a,b\in\{1,\ldots,n\},\quad 
\mu,\nu\in\{\,1,\ldots,n+1\,\},
$$
the matrix $(\,\wt{g}_{\,\mu\nu}\,)$ provides neither a 
Riemannian metric tensor field nor a pseudo-Riemannian metric tensor 
field on $\wt{\cA}_{\,\wt{\psi}}$.

Similarly one has the counterpart. 
Define a function on $\wt{\cC}$ 
$$
\wt{\varphi}(p,p_{\,n+1})
=\varphi(p)+p_{\,n+1}x_{\,(0)}^{\,n+1},
$$
where $x_{\,(0)}^{\,n+1}$ is a non-zero constant, and $\varphi$ a function.
Then, applying Theorem\,\ref{theorem-Legendre-submanifold-theorem-Arnold}, 
one has the $(n+1)$-dimensional Legendre submanifold $\wt{\cA}_{\,\wt{\varphi}}$ 
generated by $-\,\wt{\varphi}$, such that 
$$
\wt{\cA}_{\,\wt{\varphi}}^{\,\wt{\cC}}
:=\Phi_{\,\wt{\cC}\wt{\cA}\wt{\varphi}}\wt{\cA}_{\,\wt{\varphi}}
=\left\{\,(x,x^{\,n+1},p,p_{\,n+1},z)\in\wt{\cC}\,\bigg|\,
x^{\,i}=\frac{\partial\varphi}{\partial p_{\,i}},\,
x^{\,n+1}=x_{\,(0)}^{\,n+1},\,
z=p_{\,i}\frac{\partial\varphi}{\partial p_{\,i}}-\varphi,\,i\in\{1,\ldots,n\}
\,\right\},
$$
The following functions
$$
\wt{\Delta}^{\,0}(x,x^{\,n+1},p,p_{\,n+1},z)
:=x^{\,i}p_{\,i}+(\,x^{\,n+1}-x_{\,(0)}^{\,n+1}\,)\, p_{\,n+1}-\varphi(p)-z,
\quad
\wt{\Delta}^{\,a}(x,x^{\,n+1},p)
:=x^{\,a}-
\frac{x^{\,n+1}}{x_{\,(0)}^{\,n+1}}\frac{\partial\varphi}{\partial p_{\,a}}(p),
$$
will be used. It follows that 
$$
\left.\wt{\Delta}^{\,0}\right|_{\wt{\cA}_{\,\wt{\varphi}}^{\,\wt{\cC}}}
=0,\qquad\mbox{and}\qquad
\left.\wt{\Delta}^{\,a}\right|_{\wt{\cA}_{\,\wt{\varphi}}^{\,\wt{\cC}}}
=0.
$$
In addition, since the matrix components which are 
differentiation of $\wt{\varphi}$ with respect to $p$ and $p_{\,n+1}$ 
are obtained as 
$$
\wt{g}^{\,\mu\nu}
=
\left(
\begin{array}{cc}
\frac{\partial^{\,2}\varphi}{\partial p_{\,a}\partial p_{\,b}}&0\\
0&0
\end{array}
\right),\qquad a,b\in\{1,\ldots,n\},\quad \mu,\nu\in\{\,1,\ldots,n+1\,\},
$$
the matrix $(\,\wt{g}_{\,\mu\nu}\,)$ provides neither a 
Riemannian metric tensor field nor a pseudo-Riemannian metric tensor 
field on $\wt{\cA}_{\,\wt{\varphi}}$.
\subsection{Vector fields on dually flat space and their lifts in 
contact geometric language }
\label{subsection-Vector_fields_on_dually_flat_space_and_their_lifts_in_contact geometric_language}
Consider an $n$-dimensional dually flat space 
$(\,\cM,\nabla\dr\psi,\nabla,\nabla^{\,*}\,)$, 
where $\psi$ is a strictly convex function giving a metric tensor field 
$g=\nabla\dr\psi$. 
Given $\nabla$-affine coordinates $x=\{\,x^{\,1},\ldots,x^{\,n}\,\}$, 
$\nabla$-affine coordinates 
$p=\{\,p_{\,1},\ldots,p_{\,n}\,\}$
are obtained by $p_{\,a}=\partial\psi/\partial x^{\,a},(a\in\{\,1,\ldots,n\,\})$. 
As shown below, under the condition that 
the value of the function generating the Legendre submanifold is conserved, 
one way to write a dynamical system  on $\cM$
\beq
\dot{x}^{\,a}
=F_{\,\psi}^{\,a}(x),\quad\mbox{ and }\quad
\dot{p}_{\,a}
=\frac{\dr}{\dr t}\left(\,\frac{\partial\psi}{\partial x^{\,a}}\,\right),
\label{high-dimension-dually-flat-space-dynamical-system}
\eeq
with some $\{\,F_{\,\psi}^{\,1},\ldots,F_{\,\psi}^{\,n}\,\}$ in a 
contact geometric language is to introduce a $(2n+3)$-dimensional 
contact manifold $\wt{\cC}$ and another 
convex function $\wt{\psi}$ that generates a new Legendre submanifold.
One then needs to find an appropriate contact Hamiltonian 
such that some components of the 
contact Hamiltonian vector field being restricted to 
the new Legendre submanifold are identical to 
\fr{high-dimension-dually-flat-space-dynamical-system}.
The lift of the dynamical system to $\wt{\cC}$ is then immediately obtained as
the unrestricted contact Hamiltonian vector field. 

The following and its counterpart are fundamental in this subsection.
\begin{Proposition}
\label{lift-preserving-scheme-psi}
( Vector field on $\wt{\cC}$ lifted from the Legendre submanifold generated by $\psi$ ) :  
Let $\{\,F_{\,\psi}^{\,1},\ldots,F_{\,\psi}^{\,n}\,\}$ 
be a set of functions of $x=\{\,x^{\,1},\ldots,x^{\,n}\,\}$ on $\cA_{\,\psi}$,   
where these functions do not identically vanish.   
The flow of the vector field associated with the dynamical system 
defined on $\cA_{\,\psi}$ as 
$$
X_{\,\psi}^{\,0}
=\dot{x}^{\,a}\frac{\partial}{\partial x^{\,a}},\qquad\mbox{where}\qquad
\frac{\dr }{\dr t}x^{\,a}
=F_{\,\psi}^{\,a}(x),\qquad a\in\{\,1,\ldots,n\,\}
$$
is lifted to $\wt{\cC}$ as 
$$
\wt{X}_{\,\wt{h},\psi}
=\dot{x}^{\,a}\frac{\partial}{\partial x^{\,a}}
+\dot{x}^{\,n+1}\frac{\partial}{\partial x^{\,n+1}}
+\dot{p}_{\,a}\frac{\partial}{\partial p_{\,a}}
+\dot{p}_{\,n+1}\frac{\partial}{\partial p_{\,n+1}}
+\dot{z}\frac{\partial}{\partial z},
$$ 
where 
\beqa
\hspace*{-8mm}
&&\frac{\dr x^{\,a}}{\dr t}
=F_{\,\psi}^{\,a}(x),\quad 
\frac{\dr x^{\,n+1}}{\dr t}
=-\,\frac{1}{p_{\,n+1}^{\,(0)}}\frac{\partial\psi}{\partial x^{\,j}}
F_{\,\psi}^{\,j},\,\,
\frac{\dr p_{\,n+1}}{\dr t}
=\frac{\dr \wt{\Gamma}_{\,\psi}}{\dr \wt{\Delta}_{\,0}}
\left(p_{\,n+1}^{(0)}-p_{\,n+1}\right),\,\,
\frac{\dr z}{\dr t}
=\wt{\Gamma}_{\,\psi}(\wt{\Delta}_{\,0}),
\label{lift-general-psi-components-high-dimension-01}\\
\hspace*{-8mm}
&&\frac{\dr p_{\,a}}{\dr t}
=\frac{p_{\,n+1}}{p_{\,n+1}^{\,(0)}}\left(\,
\frac{\partial^2\psi}{\partial x^{\,a}\partial x^{\,j}}F_{\,\psi}^{\,j}
+\frac{\partial \psi}{\partial x^{\,j}}
\frac{\partial F_{\,\psi}^{\,j}}{\partial x^{\,a}}
\,\right)-p_{\,j}\frac{\partial F_{\,\psi}^{\,j}}{\partial x^{\,a}}
+\frac{\dr \wt{\Gamma}_{\,\psi}}{\dr \wt{\Delta}_{\,0}}\left(
\frac{\partial\psi}{\partial x^{\,a}}-p_{\,a}
\right),\quad a,j\in\{1,\ldots,n\},
\label{lift-general-psi-components-high-dimension-02}
\eeqa
with $\wt{\Gamma}_{\,\psi}$ being a function of $\wt{\Delta}_{\,0}$. 
Equivalently, these can be written as 
\beqa
&&\frac{\dr x^{\,a}}{\dr t}
=F_{\,\psi}^{\,a}(x),\quad
\frac{\dr x^{\,n+1}}{\dr t}
=\frac{-\,1}{p_{\,n+1}^{\,(0)}}\frac{\partial\psi}{\partial x^{\,j}}
F_{\,\psi}^{\,j},\quad
\frac{\dr p_{\,n+1}}{\dr t}
=\frac{\dr \wt{\Gamma}_{\,\psi}}{\dr \wt{\Delta}_{\,0}}
\left(\,p_{\,n+1}^{(0)}-p_{\,n+1}\,\right),
\label{lift-general-psi-components-high-dimension-1}
\\
&&\frac{\dr \wt{\Delta}_{\,a}}{\dr t}
=-\,\frac{\partial F_{\,\psi}^{\,j}}{\partial x^{\,a}}\wt{\Delta}_{\,j}
-\,\frac{\dr\wt{\Gamma}_{\,\psi}}{\dr\wt{\Delta}_{\,0}}\wt{\Delta}_{\,a},
\qquad
\frac{\dr \wt{\Delta}_{\,0}}{\dr t}
=-\,\wt{\Gamma}_{\,\psi}(\wt{\Delta}_{\,0}),
\label{lift-general-psi-components-high-dimension-2}
\eeqa
This lifted flow $\wt{X}_{\,\wt{h},\psi}$, 
\fr{lift-general-psi-components-high-dimension-1} and 
\fr{lift-general-psi-components-high-dimension-2}, 
is the contact Hamiltonian vector field associated to  
the contact Hamiltonian 
\beq
\wt{h}_{\,\psi}(x,x^{\,n+1},p,p_{\,n+1},z)
=\wt{\Delta}_{\,a}(x,p,p_{\,n+1})\,F_{\,\psi}^{\,a}(x)
+\wt{\Gamma}_{\,\psi}\,(\,\wt{\Delta}_{\,0}(x,x^{\,n+1},z)\,),
\label{lift-general-psi-contact-Hamiltonian-high-dimension}
\eeq
and has the property 
$$
\cL_{\,\wt{X}_{\,\wt{h},\psi}}\wt{\psi}
=0.
$$
\end{Proposition}
\begin{Proof}
Substituting \fr{lift-general-psi-contact-Hamiltonian-high-dimension}
into the high-dimensional analogue of 
\fr{contact-Hamiltonian-vector-components}, 
one has the component expression of the contact Hamiltonian vector field as
\fr{lift-general-psi-components-high-dimension-01} 
and
\fr{lift-general-psi-components-high-dimension-02}.
Furthermore, differentiating $\wt{\Delta}_{\,0}$ and 
$\{\,\wt{\Delta}_{\,1},\ldots,\wt{\Delta}_{\,n}\,\}$
with respect to $t$, and using 
\fr{lift-general-psi-components-high-dimension-01} 
and
\fr{lift-general-psi-components-high-dimension-02}, one has 
\fr{lift-general-psi-components-high-dimension-1} and 
\fr{lift-general-psi-components-high-dimension-2}. 
To show $\cL_{\,\wt{X}_{\,\wt{h},\psi}}\wt{\psi}=0$, one has that 
$$
\cL_{\,\wt{X}_{\,\wt{h},\psi}}\wt{\psi}
=\cL_{\,\wt{X}_{\,\wt{h},\psi}}\left(\,\psi(x)\,+p_{\,n+1}^{\,(0)}\,x^{n+1}\,\right)
=\dot{x}^{\,j}\frac{\partial\psi}{\partial x^{\,j}}
+p_{\,n+1}^{(0)}\dot{x}^{\,n+1}
=F_{\,\psi}^{\,j}\frac{\partial\psi}{\partial x^{\,j}}
+p_{\,n+1}^{(0)}\dot{x}^{\,n+1}
=0,
$$
where \fr{lift-general-psi-components-high-dimension-1} 
has been used. 
\qed
\end{Proof}
\begin{Remark}
The functions $\{\,F_{\,\psi}^{\,1},\ldots,F_{\,\psi}^{\,n}\,\}$  
need not depend on $\psi$.
\end{Remark}
\begin{Remark}
  If $F_{\,\psi}^{\,a}(x)=(J^{\,aj}(x)-R^{\,aj}(x))\,\partial\,\psi/\partial x^{\,j}$ 
where $(\,J^{\,ab}\,)$ is a skew-symmetric matrix ( $J^{\,ab}=-\,J^{\,ba}$ ),  
$(\,R^{\,ab}\,)$ is a symmetric definite matrix, and $p_{\,n+1}^{(0)}>0$, then 
$$
\frac{\dr}{\dr t}x^{\,n+1}
>0. 
$$
This inequality corresponds to the statement ``the entropy is increasing'' 
found in Ref.\,\cite{Eberard2006}. 
\end{Remark}
There is a counterpart of this Proposition.
\begin{Proposition}
\label{lift-preserving-scheme-varphi} 
( Vector field on $\wt{\cC}$ lifted 
from the Legendre submanifold generated by $-\,\varphi$  ) :  
Let $\{\,F_{\,1}^{\,\varphi},\ldots,F_{\,n}^{\,\varphi}\,\}$ 
be a set of functions of $p=\{\,p_{\,1},\ldots,p_{\,n}\,\}$ on $\cA_{\,\varphi}$,  
where these functions  do not identically vanish.   
The flow of the vector field associated with the dynamical system 
defined on $\cA_{\,\varphi}$ as  
$$
X_{\,\varphi}^{\,0}
=\dot{p}_{\,a}\frac{\partial}{\partial p_{\,a}},\qquad\mbox{where}\qquad
\frac{\dr }{\dr t}p_{\,a}
=F_{\,a}^{\,\varphi}(p),\qquad a\in\{\,1,\ldots,n\,\}
$$
is lifted to $\wt{\cC}$ as 
$$
\wt{X}_{\,\wt{h},\varphi}
=\dot{x}^{\,a}\frac{\partial}{\partial x^{\,a}}
+\dot{x}^{\,n+1}\frac{\partial}{\partial x^{\,n+1}}
+\dot{p}_{\,a}\frac{\partial}{\partial p_{\,a}}
+\dot{p}_{\,n+1}\frac{\partial}{\partial p_{\,n+1}}
+\dot{z}\frac{\partial}{\partial z},
$$
where 
\beqa
&&\frac{\dr x^{\,a}}{\dr t}
=\frac{x^{\,n+1}}{x_{\,(0)}^{\,n+1}}
\frac{\partial^2\varphi}{\partial p_{\,a}\partial p_{\,i}}
F_{\,i}^{\,\varphi}
-\wt{\Delta}^{\,i}\frac{\partial F_{\,i}^{\,\varphi}}{\partial p_{\,a}}
+\frac{\dr \wt{\Gamma}^{\,\varphi}}{\dr \wt{\Delta}^{\,0}}\left(
\frac{\partial\varphi}{\partial p_{\,a}}-x^{\,a}
\right),\quad a,i\in\{\,1,\ldots,n\,\},
\non\\
&&\frac{\dr x^{\,n+1}}{\dr t}
=\frac{\dr \wt{\Gamma}^{\,\varphi}}{\dr \wt{\Delta}^{\,0}}
\left(\,x_{(0)}^{\,n+1}-x^{\,n+1}\,\right),\,
\frac{\dr p_{a}}{\dr t}
=F_{\,a}^{\,\varphi},\,\,
\frac{\dr p_{\,n+1}}{\dr t}
=\frac{-\,1}{x_{\,(0)}^{\,n+1}}\frac{\partial\varphi}{\partial p_{\,i}}
F_{\,i}^{\,\varphi},\,\,
\non
\eeqa
$$
\frac{\dr z}{\dr t}
=\wt{\Delta}^{i}F_{\,i}^{\,\varphi}+\wt{\Gamma}^{\,\varphi}
+\frac{x^{\,n+1}}{x_{\,(0)}^{\,n+1}}p_{\,i}
\frac{\partial^2\varphi}{\partial p_{\,i}\partial p_{\,a}}F_{\,a}^{\,\varphi}
-p_{\,i}\wt{\Delta}^{\,a}\frac{\partial F_{\,a}^{\,\varphi}}{\partial p_{\,i}}
-\left[x^{\,i}p_{\,i}+(x^{\,n+1}-x_{\,(0)}^{\,n+1})p_{\,n+1}
-p_{\,i}\frac{\partial\varphi}{\partial p_{\,i}}
\right]\frac{\dr\wt{\Gamma}^{\,\varphi}}{\dr\wt{\Delta}^{\,0}},
$$
with $\wt{\Gamma}^{\,\varphi}$ being a function of $\wt{\Delta}^{\,0}$. 
Equivalently, these can be written as 
\beqa
&&\frac{\dr p_{\,a}}{\dr t}
=F_{\,a}^{\,\varphi}(p),\quad
\frac{\dr p_{\,n+1}}{\dr t}
=\frac{-\,1}{x_{\,(0)}^{\,n+1}}\frac{\partial\varphi}{\partial p_{\,i}}
F_{\,i}^{\,\varphi},\quad
\frac{\dr x^{\,n+1}}{\dr t}
=\frac{\dr \wt{\Gamma}_{\,\psi}}{\dr \wt{\Delta}^{\,0}}
\left(\,x_{(0)}^{\,n+1}-x^{\,n+1}\,\right),
\label{lift-general-varphi-components-high-dimension-1}
\\
&&\frac{\dr \wt{\Delta}^{\,a}}{\dr t}
=-\,\frac{\partial F_{\,i}^{\,\varphi}}{\partial p_{\,a}}\wt{\Delta}^{\,i}
-\,\frac{\dr\wt{\Gamma}^{\,\varphi}}{\dr\wt{\Delta}^{\,0}}\wt{\Delta}^{\,a},
\qquad
\frac{\dr \wt{\Delta}^{\,0}}{\dr t}
=-\,\wt{\Gamma}_{\,\psi}(\wt{\Delta}^{\,0}),
\label{lift-general-varphi-components-high-dimension-2}
\eeqa
This lifted flow $\wt{X}_{\,\wt{h},\varphi}$, 
\fr{lift-general-varphi-components-high-dimension-1} and 
\fr{lift-general-varphi-components-high-dimension-2}, 
is the contact Hamiltonian vector field associated to  
the contact Hamiltonian 
\beq
\wt{h}_{\,\varphi}(x,x^{\,n+1},p,p_{\,n+1},z)
=\wt{\Delta}^{\,a}(x,x^{\,n+1},p)\,F_{\,a}^{\,\varphi}(p)
+\wt{\Gamma}^{\,\varphi}\,(\,\wt{\Delta}^{\,0}(x,x^{\,n+1},p,p_{\,n+1},z)\,),
\label{lift-general-varphi-contact-Hamiltonian-high-dimension}
\eeq
and has the property 
$$
\cL_{\,\wt{X}_{\,\wt{h},\varphi}}\wt{\varphi}
=0.
$$
\end{Proposition}
\begin{Proof}
A way to prove this proposition is analogous to that of 
Preposition\,\ref{lift-preserving-scheme-psi}. 
\qed
\end{Proof}
\begin{Remark}
The functions $\{\,F_{\,1}^{\,\varphi},\ldots,F_{\,n}^{\,\varphi}\,\}$ 
need not depend on $\varphi$. 
\end{Remark}
\begin{Remark}
If 
$F_{\,a}^{\,\varphi}(p)=(J_{\,ai}(p)-R_{\,ai}(p))\,\partial\,\varphi/\partial p_{\,i}$ 
where $(\,J_{\,ab}\,)$ is a skew-symmetric matrix ( $J_{\,ab}=-J_{\,ba}$ ),  
$(\,R^{\,ab}\,)$ is a symmetric definite matrix, and $x_{(0)}^{\,n+1}>0$, then 
$$
\frac{\dr}{\dr t}p_{\,n+1}
>0. 
$$
\end{Remark}
\begin{Proposition}
\label{lift-contact-Hamiltonian-monotonically-decreasing-function-psi-high-dimension}
( Differentiation of $\wt{h}_{\,\psi}$ with respect to $t$ ) : 
Consider the system stated in Proposition\,\ref{lift-preserving-scheme-psi}.
Then, it follows that 
$$
\frac{\dr}{\dr t}\wt{h}_{\,\psi}
=-\,\frac{\dr\,\wt{\Gamma}_{\,\psi}}{\dr\wt{\Delta}_{\,0}}\wt{h}_{\,\psi}.
$$
\end{Proposition}
\begin{Proof}
It follows from 
\fr{lift-general-psi-components-high-dimension-2}  
that 
$$
\frac{\dr}{\dr t}\wt{h}_{\,\psi}
=\dot{\wt{\Delta}}_{\,a}F_{\,\psi}^{\,a}
+\wt{\Delta}_{\,a}\dot{F}_{\,\psi}^{\,a}
+\frac{\dr\,\wt{\Gamma}_{\,\psi}}{\dr\wt{\Delta}_{\,0}}\dot{\wt{\Delta}}_{\,0}
=-\,\frac{\dr\,\wt{\Gamma}_{\,\psi}}{\dr\wt{\Delta}_{\,0}}\wt{h}_{\,\psi}.
$$
\qed
\end{Proof}
\begin{Remark}
\label{invariant-subset-contact-Hamiltonian-psi-high-dimension}
For systems where    
$\wt{\Gamma}_{\,\psi}(\wt{\Delta}_{\,0})=\wt{\gamma}_{\,0}\wt{\Delta}_{\,0}$ with 
$\wt{\gamma}_{\,0}$ being a non-zero constant, the subsets 
$\{(x,x^{\,n+1},p,p_{n+1},z)|\wt{h}_{\,\psi}>0\}\subset\wt{\cC}$ and 
$\{(x,x^{\,n+1},p,p_{n+1},z)|\wt{h}_{\,\psi}<0\}\subset\wt{\cC}$ are invariant ones.
\end{Remark}
There is a counterpart of this Proposition.
\begin{Proposition}
( Differentiation of $\wt{h}_{\,\varphi}$ with respect to $t$ ) : 
Consider the system stated in Proposition\,\ref{lift-preserving-scheme-varphi}.
Then, 
it follows that 
$$
\frac{\dr}{\dr t}\wt{h}_{\,\varphi}
=-\,\frac{\dr\,\wt{\Gamma}^{\,\varphi}}{\dr\wt{\Delta}^{\,0}}\wt{h}_{\,\varphi}.
$$
\end{Proposition}
\begin{Proof}
A way to prove this is analogous to that of 
Proposition\,\ref{lift-contact-Hamiltonian-monotonically-decreasing-function-psi-high-dimension}
\qed
\end{Proof}
\begin{Remark}
\label{invariant-subset-contact-Hamiltonian-varphi-high-dimension}
For systems where 
$\wt{\Gamma}^{\,\varphi}(\wt{\Delta}^{\,0})=\wt{\gamma}^{\,0}\wt{\Delta}^{\,0}$ with 
$\wt{\gamma}^{\,0}$ being a non-zero constant, the subsets 
$\{(x,x^{n+1},p,p_{n+1},z)|\wt{h}_{\,\varphi}>0\}\subset\wt{\cC}$ and 
$\{(x,x^{n+1},p,p_{n+1},z)|\wt{h}_{\,\varphi}<0\}\subset\wt{\cC}$ are invariant ones.
\end{Remark}

As shown below, the phase compressibility for a class of the lifted vector 
fields with $\wt{\Omega}_{\,\wt{\lambda}}$ does not depend on 
$\{\,F_{\,\psi}^{\,1},\ldots,F_{\,\psi}^{\,n}\}$.
\begin{Lemma}
\label{phase-compressibility-psi-higher-dimension}
( Phase compressibility for $\wt{X}_{\wt{h},\psi}$ with $\wt{\Omega}_{\wt{\lambda}}$ )
 : 
Consider the system stated in Proposition\,\ref{lift-preserving-scheme-psi}. 
Choose $\wt{\Gamma}_{\,\psi}(\wt{\Delta}_{\,0})=\wt{\gamma}_{\,0}\wt{\Delta}_{\,0}$ 
with $\wt{\gamma}_{\,0}$ being a constant. 
Then, the phase compressibility is calculated as 
$$
\kappa_{\wt{\Omega}_{\,\wt{\lambda}}}\left(\wt{X}_{\,\wt{h},\psi}\right)
=-\,(n+2)\,\wt{\gamma}_{\,0}.
$$
\end{Lemma}
\begin{Proof}
With $\cL_{\wt{X}_{\wt{h},\psi}}\wt{\lambda}=(\wt{\Reeb} \wt{h}_{\,\psi})\wt{\lambda}
=-\,\wt{\gamma}_{\,0}\wt{\lambda}$ and 
$\cL_{\wt{X}_{\wt{h},\psi}}\dr\wt{\lambda}=\dr\cL_{\wt{X}_{\wt{h},\psi}}\wt{\lambda}
=-\,\wt{\gamma}_{\,0}\dr\wt{\lambda}$, one has that 
\beq
\cL_{\wt{X}_{\wt{h},\psi}}\wt{\Omega}_{\,\wt{\lambda}}
=-\,(n+2)\,\wt{\gamma}_{\,0}\,\wt{\Omega}_{\,\wt{\lambda}}.
\label{phase-compressibility-Lie-derivative-psi-higher-dimension}
\eeq
Comparing  \fr{phase-compressibility-Lie-derivative-psi-higher-dimension} with 
\fr{def-phase-compressibility}, one completes the proof.
\qed
\end{Proof}
There is a counterpart of this Lemma.
\begin{Lemma}
( Phase compressibility for $\wt{X}_{\wt{h},\varphi}$ with $\wt{\Omega}_{\wt{\lambda}}$ )
 : 
Consider the system stated in Proposition\,\ref{lift-preserving-scheme-varphi}. 
Choose $\wt{\Gamma}_{\,\varphi}(\wt{\Delta}^{\,0})=\wt{\gamma}^{\,0}\wt{\Delta}^{\,0}$ 
with $\wt{\gamma}^{\,0}$ being a constant. 
Then, the phase compressibility is calculated as 
$$
\kappa_{\wt{\Omega}_{\,\wt{\lambda}}}\left(\wt{X}_{\,\wt{h},\varphi}\right)
=-\,(n+2)\,\wt{\gamma}^{\,0}.
$$
\end{Lemma}
\begin{Proof}
A way to prove this is analogous to that of 
Lemma\,\ref{phase-compressibility-psi-higher-dimension}. 
\qed
\end{Proof}

The following and its counterpart are keys to deal with 
statistical behaviour of the lifted vector fields.
\begin{Thm}
( Invariant measure for $\wt{X}_{\,\wt{h},\psi}$ with $\wt{\Omega}_{\,\wt{\lambda}}$ ) 
:  
Consider the system stated in Proposition\,\ref{lift-preserving-scheme-psi} 
with 
$\wt{h}_{\,\psi}> 0$.  Choose   
$\wt{\Gamma}_{\,\psi}(\wt{\Delta}_{\,0})=\wt{\gamma}_{\,0}\wt{\Delta}_{\,0}$ 
with $\wt{\gamma}_{\,0}$ being a non-zero constant.
An invariant measure in the sense of  
Definition\,\ref{def-invariant-measure-general} is obtained as 
$$
\wt{\Omega}_{\,\wt{h},\psi}^{\Inv}
=\wt{f}_{\,\wt{h},\psi}^{\Inv}\,\wt{\Omega}_{\,\wt{\lambda}},
\qquad\mbox{where}\quad 
\wt{f}_{\,\wt{h},\psi}^{\Inv}
=\frac{1}{\wt{\cZ}_{\,\wt{h},\psi}}\wt{h}_{\,\psi}^{-(n+2)},
$$
for the class of the form 
$\wt{f}_{\,\wt{h},\psi}^{\Inv}=\wt{f}_{\,\wt{h},\psi}^{\Inv}(\wt{h}_{\,\psi})$ with   
$\wt{\cZ}_{\,\wt{h},\psi}$ being a non-zero constant. 
\end{Thm}
\begin{Proof}
A way to prove this is 
  analogous to that of Theorem\,\ref{theorem-inivariant-measure-psi}.
\end{Proof}
\begin{Remark}
Due to Remark\,\ref{invariant-subset-contact-Hamiltonian-psi-high-dimension} 
and the subset $\{(x,x^{\,n+1},p,p_{\,n+1},z)|\wt{h}_{\,\psi}>0\}\subset \wt{\cC}$ 
is invariant one, one has that $\wt{f}_{\,\wt{h},\psi}^{\Inv}>0$. 
\end{Remark}

There is a  counterpart of this Theorem.
\begin{Thm}
( Invariant measure for $\wt{X}_{\,\wt{h},\varphi}$ 
with $\wt{\Omega}_{\,\wt{\lambda}}$ ) :  
Consider the system stated in Proposition\,\ref{lift-preserving-scheme-varphi} 
with 
$\wt{h}_{\,\varphi}> 0$.  Choose   
$\wt{\Gamma}^{\,\varphi}(\wt{\Delta}^{\,0})=\wt{\gamma}^{\,0}\wt{\Delta}^{\,0}$ 
with $\wt{\gamma}^{\,0}$ being a non-zero constant.
An invariant measure in the sense of  
Definition\,\ref{def-invariant-measure-general} is obtained as 
$$
\wt{\Omega}_{\,\wt{h},\varphi}^{\Inv}
=\wt{f}_{\,\wt{h},\varphi}^{\Inv}\,\wt{\Omega}_{\,\wt{\lambda}},
\qquad\mbox{where}\quad 
\wt{f}_{\,\wt{h},\varphi}^{\Inv}
=\frac{1}{\wt{\cZ}_{\,\wt{h},\varphi}}\wt{h}_{\,\varphi}^{-(n+2)},
$$
for the class of the form 
$\wt{f}_{\,\wt{h},\varphi}^{\Inv}=\wt{f}_{\,\wt{h},\varphi}^{\Inv}(\wt{h}_{\,\varphi})$ with  
$\wt{\cZ}_{\,\wt{h},\varphi}$ being a non-zero constant. 
\end{Thm}
\begin{Proof}
A way to prove this is 
  analogous to that of Theorem\,\ref{theorem-inivariant-measure-psi}.
\end{Proof}
\begin{Remark}
Due to Remark\,\ref{invariant-subset-contact-Hamiltonian-varphi-high-dimension} 
and the subset $\{(x,x^{\,n+1},p,p_{\,n+1},z)|\wt{h}_{\,\varphi}>0\}\subset \wt{\cC}$ 
is invariant one, one has that $\wt{f}_{\,\wt{h},\varphi}^{\Inv}>0$. 
\end{Remark}

The vector field on the Legendre submanifold $\wt{\cA}_{\,\wt{\psi}}^{\,\wt{\cC}}$
is immediately obtained from the restriction of $\wt{X}_{\,\wt{h},\psi}$ 
as follows.
\begin{Proposition}
\label{restricted-lift-general-psi-contact-Hamiltonian-high-dimension}
( Restriction of $\wt{X}_{\,\wt{h},\psi}$ on 
$\wt{\cA}_{\,\wt{\psi}}^{\,\wt{\cC}}$
)  : 
Let $\wt{X}_{\,\wt{h},\psi}$ be 
the contact Hamiltonian vector field on $\wt{\cC}$ associated with 
$\wt{h}_{\,\psi}$ in \fr{lift-general-psi-contact-Hamiltonian-high-dimension}.
If $\wt{\Gamma}_{\,\psi}$ is such that $\wt{\Gamma}_{\,\psi}(0)=0$
and $\wt{\Gamma}_{\,\psi}(\wt{\Delta}^{\,0})\neq 0$ for $\wt{\Delta}^{\,0}\neq 0$, 
then the components of the 
vector field 
$\left.\wt{X}_{\,\wt{h},\psi}\right|_{\,\wt{\cA}_{\,\wt{\psi}}^{\,\wt{\cC}}}$ 
are written as 
$$
\frac{\dr x^{\,a}}{\dr t}
=F_{\,\psi}^{\,a}(x),\,\,
\frac{\dr x^{\,n+1}}{\dr t}
=\frac{-\,1}{p_{\,n+1}^{\,(0)}}\frac{\partial\psi}{\partial x^{\,j}}
F_{\,\psi}^{\,j},\,\,
\frac{\dr p_{\,n+1}}{\dr t}
=0,\,\,
\frac{\dr z}{\dr t}
=0,\,\,
\frac{\dr p_{\,a}}{\dr t} 
=\frac{\partial^2\psi}{\partial x^{\,a}\partial x^{\,j}}F_{\,\psi}^{\,j},
\,\, a,j\in\{1,\ldots,n\}.
$$
Equivalently, these can be written as 
$$
\frac{\dr x^{\,a}}{\dr t}
=F_{\,\psi}^{\,a}(x),\quad
\frac{\dr x^{\,n+1}}{\dr t}
=\frac{-\,1}{p_{\,n+1}^{\,(0)}}\frac{\partial\psi}{\partial x^{\,j}}
F_{\,\psi}^{\,j},\quad
\frac{\dr p_{\,n+1}}{\dr t}
=0,\,\,
\frac{\dr \wt{\Delta}_{\,0}}{\dr t}
=0,\quad
\frac{\dr \Delta_{\,a}}{\dr t}
=0,\quad a\in\{1,\ldots,n\}
$$
where $\Delta_{\,a}$ has been defined in \fr{definition-Delta_a-psi} as 
$$
\Delta_{\,a}
=\frac{\partial\psi}{\partial x^{\,a}}-p_{\,a}.
 $$
Furthermore, this restricted contact Hamiltonian vector field has the property 
\beq
\cL_{\,\wt{X}_{\,\wt{h},\psi}|_{\wt{\cA}_{\,\wt{\psi}}^{\,\wt{\cC}}}}\,
\wt{\psi}
=0.
\label{conserved-psi-high-dimension}
\eeq
\end{Proposition}
\begin{Proof}
Components of the restricted vector field are easily obtained 
from Proposition \ref{lift-preserving-scheme-psi}. 
The property \fr{conserved-psi-high-dimension}
 is proved by straightforward calculations as 
$$
\cL_{\,\wt{X}_{\,\wt{h},\psi}|_{\wt{\cA}_{\,\wt{\psi}}^{\,\wt{\cC}}}}\,\wt{\psi}
=\wt{X}_{\,\wt{h},\psi}|_{\wt{\cA}_{\,\wt{\psi}}^{\,\wt{\cC}}}
\wt{\psi}
=\wt{X}_{\,\wt{h},\psi}|_{\wt{\cA}_{\,\wt{\psi}}^{\,\wt{\cC}}}
(\,\psi(x)+p_{\,n+1}^{\,(0)}x^{\,n+1}\,)
=\frac{\partial\psi}{\partial x^{\,j}}F_{\,\psi}^{\,j}+p_{\,n+1}^{\,(0)}\dot{x}^{\,n+1}
=0.
$$
\qed
\end{Proof}

There is a counterpart of this proposition.  
The vector field on the Legendre submanifold $\cA_{\,\wt{\varphi}}^{\,\wt{\cC}}$
is immediately obtained from the restriction of $\wt{X}_{\,\wt{h},\varphi}$ 
as follows.
\begin{Proposition}
( Restriction of $\wt{X}_{\,\wt{h},\varphi}$ on 
$\wt{\cA}_{\,\wt{\varphi}}^{\,\wt{\cC}}$
) : 
Let $\wt{X}_{\,\wt{h},\varphi}$ be 
the contact Hamiltonian vector field on $\wt{\cC}$ associated with 
$\wt{h}_{\,\varphi}$ 
in \fr{lift-general-varphi-contact-Hamiltonian-high-dimension}.
If 
$\wt{\Gamma}^{\,\varphi}$ is such that $\wt{\Gamma}^{\,\varphi}(0)=0$ 
and $\wt{\Gamma}^{\,\varphi}(\wt{\Delta}^{\,0})\neq 0$ for $\wt{\Delta}^{\,0}\neq 0$, then 
the components of the vector field 
$\left.\wt{X}_{\,\wt{h},\varphi}\right|_{\,\wt{\cA}_{\,\wt{\varphi}}^{\,\wt{\cC}}}$ 
are written as 
$$
\frac{\dr p_{\,a}}{\dr t}
=F_{\,a}^{\,\varphi}(p),\,\,
\frac{\dr p_{\,n+1}}{\dr t}
=\frac{-\,1}{x_{\,(0)}^{\,n+1}}\frac{\partial\varphi}{\partial p_{\,i}}
F_{\,i}^{\,\varphi},\,\,
\frac{\dr x^{\,n+1}}{\dr t}
=0,\,\,
\frac{\dr z}{\dr t}
=0,\,\,
\frac{\dr x^{\,a}}{\dr t} 
=\frac{\partial^2\varphi}{\partial p_{\,a}\partial p_{\,i}}F_{\,i}^{\,\varphi},
\,\, a,i\in\{1,\ldots,n\}.
$$
Equivalently, these can be written as 
$$
\frac{\dr p_{\,a}}{\dr t}
=F_{\,a}^{\,\varphi}(p),\quad
\frac{\dr p_{\,n+1}}{\dr t}
=\frac{-\,1}{x_{\,(0)}^{\,n+1}}\frac{\partial\varphi}{\partial p_{\,a}}
F_{\,a}^{\,\varphi},\quad
\frac{\dr p_{\,n+1}}{\dr t}
=0,\,\,
\frac{\dr \wt{\Delta}^{\,0}}{\dr t}
=0,\quad
\frac{\dr \Delta^{\,a}}{\dr t}
=0,\quad a\in\{1,\ldots,n\}
$$
where $\Delta^{\,a}$ has been defined in \fr{definition-Delta_a-varphi} as  
$$
\Delta^{\,a}
=x^{\,a}-\frac{\partial \varphi}{\partial p_{\,a}}.
$$
Furthermore, this restricted contact Hamiltonian vector field has the property 
$$
\cL_{\,\wt{X}_{\,\wt{h},\varphi}|_{\wt{\cA}_{\,\wt{\varphi}}^{\,\wt{\cC}}}}\,
\wt{\varphi}
=0.
$$
\end{Proposition}
\begin{Proof}
A way to prove this is analogous to that of 
Proposition\,\ref{restricted-lift-general-psi-contact-Hamiltonian-high-dimension}.
\end{Proof}
\subsection{Applications to electric circuit models in a thermal environment}
\label{subsection-Applications_to_electric_circuit_models_in_a_thermal_environment}
In this subsection, 
Proposition\,\ref{restricted-lift-general-psi-contact-Hamiltonian-high-dimension} is applied to systems consisting of the RC, RL and RLC circuit models 
( see Definitions\,\ref{definition-RC-circuit-model},  
\ref{definition-RL-circuit-model},
\ref{definition-RLC-circuit-model}  ) and an environment with 
the constant temperature $T_{\,0}$.

First, an extension of the RC circuit model is considered.  
\begin{Def}
( RC circuit model in a thermal environment ) : 
Consider the RC circuit model stated in 
Definition\,\ref{definition-RC-circuit-model}. 
In addition, let $T_{\,0}>$ be a constant temperature, 
$S$ entropy, and 
$\cH_{\,\C,T_{\,0}}^{\,\tot}$ the total energy  
$$
\cH_{\,\C,T_{\,0}}^{\,\tot}(Q_{\,\C},S)
=\cH_{\,\C}(Q_{\,\C})+T_{\,0}S,
$$  
where $\cH_{\,\C}(Q_{\,\C})$ is given in 
\fr{RC-circuit-energy}. 
Then, 
the space $\mbbR^{\,5}$ whose coordinates are   
$(Q_{\,\C},S,V_{\,\C},T,z_{\,\C})$ 
is referred to as the extended thermodynamic phase space, 
where $T$ is a $1$-dimensional coordinate for expressing temperature, 
and $z_{\,\C}$ is a $1$-dimensional coordinate for expressing energy. 
In addition,  
the contact Hamiltonian vector field restricted to the Legendre submanifold 
generated by $\cH_{\,\C,T_{\,0}}^{\,\tot}$
, 
$\wt{X}_{\,\wt{h},\psi}|_{\wt{\cA}_{\wt{\psi}}^{\wt{\cC}}}$ in Proposition 
\ref{restricted-lift-general-psi-contact-Hamiltonian-high-dimension},  
is referred to as the RC circuit model 
in the thermal environment with temperature $T_{\,0}$.
Here $\psi$ and $\wt{\psi}$ are 
identified with $\cH_{\,\C}$ and $\cH_{\,\C, T_{\,0}}^{\,\tot}$, respectively, 
and $F_{\,\psi}^{\,1}$ is identified with $F_{\,\C}$ 
in Proposition \ref{Proposition-RC-dynamical-system}. 
\end{Def}
\begin{Remark}
If $T_{\,0}=0$, 
then the RC circuit model in the thermal environment with temperature $T_{\,0}$ 
reduces to the system in 
Definition\,\ref{definition-RC-circuit-model}. 
\end{Remark}
\begin{Remark}
This system proposed in this Definition is similar to that of 
Ref.\,\cite{Eberard2006}. 
\end{Remark}
\begin{Remark}
On the Legendre submanifold generated by 
$\cH_{\,\C,T_{\,0}}^{\,\tot}$,  the relations 
$$
V_{\,\C}
=\frac{\dr\cH_{\,\C}}{\dr Q_{\,\C}}
=\frac{Q_{\,\C}}{C},\qquad
T=T_{\,0},\qquad \mbox{ and }\qquad
z_{\,\C}
=\cH_{\,\C,T_{\,0}}^{\,\tot},
$$
hold in the extended thermodynamic phase space, due to 
\fr{(n+1)-dimensional-Legendre-submanifold}. 
\end{Remark}

Then, one has the following proposition.
\begin{Proposition}
( Total energy and entropy production in the RC circuit model 
in a thermal environment  ) : 
For the RC circuit model in the thermal environment with $T_{\,0}$, 
one has the equations for $Q_{\,\C}, V_{\,\C}$ and $S$ as  
$$
\frac{\dr\,Q_{\,\C}}{\dr t}
=-\,\frac{Q_{\,\C}}{R\,C},\quad 
\frac{\dr\,V_{\,\C}}{\dr t}
=-\,\frac{Q_{\,\C}}{R\,C^{\,2}},\quad 
\frac{\dr\,S}{\dr t}
=\frac{1}{T_{\,0}}\frac{Q_{\,\C}^{\,2}}{R\,C^{\,2}}
>0,\quad 
V_{\,\C}=Q_{\,\C}V_{\,\C}. 
$$ 
In addition, it follows that 
$$
\frac{\dr}{\dr t}\cH_{\,\C,T_{\,0}}^{\,\tot}
=0.
$$
\end{Proposition}
\begin{Proof}
The equations for $Q_{\,\C},V_{\,\C}$, and $S$ 
are obtained by substituting the identifications, 
$n=1$, and 
$$
x^{\,1}
=Q_{\,\C},\quad 
p_{\,1}
=V_{\,\C},\quad
x^{\,2}
=S,\quad
p_{\,2}^{\,(0)}
=T_{\,0},
$$
$$
\psi(x^{\,1})
=\cH_{\,\C}(Q_{\,\C}),\quad
\wt{\psi}(x^{\,1},x^{\,2})
=\cH_{\,\C,T_{\,0}}^{\,\tot}(Q_{\,\C},S\,),\quad 
F_{\,\psi}^{\,1}(x^{\,1})
=-\,\frac{Q_{\,\C}}{RC},
$$
into  the equations in 
Proposition\,\ref{restricted-lift-general-psi-contact-Hamiltonian-high-dimension}.
\end{Proof}

\begin{Remark}
A physical interpretation of the equation involving $\dr\,S/\dr t$ 
is the entropy production. 
\end{Remark}

Second, an extension of the RL circuit model is considered.  
\begin{Def}
( RL circuit model in a thermal environment ) : 
Consider the RL circuit model stated in 
Definition\,\ref{definition-RL-circuit-model}. 
In addition, let $T_{\,0}>$ be a constant temperature, 
$S$ entropy, and 
$\cH_{\,\L,T_{\,0}}^{\,\tot}$ the total energy  
$$
\cH_{\,\L,T_{\,0}}^{\,\tot}(N_{\,\L},S)
=\cH_{\,\L}(N_{\,\L})+T_{\,0}S,
$$  
where $\cH_{\,\L}(N_{\,\L})=\sup_{I_{\,\L}}(I_{\,\L}N_{\,\L}-\cH_{\,\L}^{\,*}),$
and $\cH_{\,\L}^{\,*}(I_{\,\L})$ is given in \fr{RL-circuit-energy}. 
Then, 
the space $\mbbR^5$ whose coordinates are  
$(N_{\,\L},S,I_{\,\L},T,z_{\,\L})$ 
is referred to as the extended thermodynamic phase space, 
where $T$ is a $1$-dimensional coordinate for expressing temperature, and 
$z_{\,\L}$ is a $1$-dimensional coordinate for expressing energy. 
In addition,  
the contact Hamiltonian vector field restricted to the Legendre submanifold 
generated by $\cH_{\,\L,T_{\,0}}^{\,\tot}$, 
$\wt{X}_{\,\wt{h},\psi}|_{\wt{\cA}_{\wt{\psi}}^{\wt{\cC}}}$ in Proposition 
\ref{restricted-lift-general-psi-contact-Hamiltonian-high-dimension},  
is referred to as the RL circuit model 
in the thermal environment with temperature $T_{\,0}$. 
Here $\psi$ and $\wt{\psi}$ are 
identified with $\cH_{\,\L}$ and $\cH_{\,\L,T_{\,0}}^{\,\tot}$, respectively, 
and $F_{\,\psi}^{\,1}(x^{\,1})=-\,RN_{\,\L}/L$. 
\end{Def}
\begin{Remark}
If $T_{\,0}=0$, then the RL circuit model 
in the thermal environment with temperature $T_{\,0}$ 
reduces to the system in 
Definition\,\ref{definition-RL-circuit-model}. 
\end{Remark}
\begin{Remark}
This system proposed in this Definition is similar to that of 
Ref.\,\cite{Eberard2006}. 
\end{Remark}
\begin{Remark}
On the Legendre submanifold generated by 
$\cH_{\,\L,T_{\,0}}^{\,\tot}$,  the relations 
$$
I_{\,\L}
=\frac{\dr\cH_{\,\L}}{\dr N_{\,\L}}
=\frac{N_{\,\L}}{L},\qquad
T=T_{\,0},\qquad \mbox{ and }\qquad
z_{\,\L}
=\cH_{\,\L,T_{\,0}}^{\,\tot},
$$
hold in the extended thermodynamic phase space, due to 
\fr{(n+1)-dimensional-Legendre-submanifold}. 
\end{Remark}

Then, one has the following proposition.
\begin{Proposition}
( Total energy and entropy production in the RL circuit model in a thermal environment  ) : 
For the RL circuit model in the thermal environment with $T_{\,0}$, 
one has the equations for $N_{\,\L}, I_{\,\L}$ and $S$ as  
$$
\frac{\dr\,N_{\,\L}}{\dr t}
=-\,R I_{\,\L},\quad 
\frac{\dr\,I_{\,\L}}{\dr t}
=-\,\frac{R}{L}I_{\,\L},\quad 
\frac{\dr\,S}{\dr t}
=\frac{RI_{\,\L}}{T_{\,0}}\frac{N_{\,\L}}{L}
=\frac{RI_{\,\L}^{\,2}}{T_{\,0}}
>0,\quad
I_{\,\L}
=\frac{N_{\,\L}}{L}. 
$$ 
In addition, it follows that 
$$
\frac{\dr}{\dr t}\cH_{\,\L,T_{\,0}}^{\,\tot}
=0.
$$
\end{Proposition}
\begin{Proof}
The equations for $N_{\,\L},I_{\,\L}$, and $S$ 
are obtained by substituting the identifications,  
$n=1$, and 
$$
x^{\,1}
=N_{\,\L},\quad 
p_{\,1}
=I_{\,\L},\quad
x^{\,2}
=S,\quad
p_{\,2}^{\,(0)}
=T_{\,0},
$$
$$
\psi(x^{\,1})
=\cH_{\,\L}(N_{\,\L}),\quad
\wt{\psi}(x^{\,1},x^{\,2})
=\cH_{\,\L,T_{\,0}}^{\,\tot}(N_{\,\L},S),\quad
F_{\,\psi}^{\,1}(x^{\,1})
=-\frac{R}{L}N_{\,\L}, 
$$
into  the equations in 
Proposition\,\ref{restricted-lift-general-psi-contact-Hamiltonian-high-dimension}.
\end{Proof}

Third, an extension of the RLC circuit model is considered. 
\begin{Def}
( RLC circuit model in a thermal environment ) : 
Consider the RLC circuit model stated in 
Definition\,\ref{definition-RLC-circuit-model}. 
In addition, let $T_{\,0}>$ be a constant temperature, 
$S$ entropy, and 
$\cH^{\,\tot}$ the total energy  
$$
\cH_{\,T_{\,0}}^{\,\tot}(Q,N,S)
=\cH(Q,N)+T_{\,0}S,
$$  
where $\cH(Q,N)$ is given in \fr{RLC-circuit-energy}. 
Then, 
the space $\mbbR^{\,7}$ whose coordinates are 
$(Q,N,S,V,I,T,z)$ 
is referred to as the extended thermodynamic phase space, 
where $T$ is a $1$-dimensional coordinate for expressing temperature, 
and $z$ is a $1$-dimensional coordinate for expressing energy. 
In addition,  
the contact Hamiltonian vector field restricted to the Legendre submanifold 
generated by $\cH_{\,T_{\,0}}^{\,\tot}$, 
  $\wt{X}_{\,\wt{h},\psi}|_{\wt{\cA}_{\wt{\psi}}^{\wt{\cC}}}$ in Proposition 
\ref{restricted-lift-general-psi-contact-Hamiltonian-high-dimension},  
is referred to as the RLC circuit model 
in the thermal environment with temperature $T_{\,0}$.
Here $\psi$ and $\wt{\psi}$ are identified with $\cH$ and 
$\cH_{\,T_{\,0}}^{\,\tot}$, respectively, and 
$F_{\,\psi}^{\,1}(x)=N/L, F_{\,\psi}^{\,2}(x)=-Q/C-RN/L$. 

\end{Def}
\begin{Remark}
If $T_{\,0}=0$, 
then the RLC circuit model in the thermal environment with temperature $T_{\,0}$ 
reduces to the system in 
Definition\,\ref{definition-RC-circuit-model}. 
\end{Remark}
\begin{Remark}
This system proposed in this Definition is similar to that of 
Ref.\,\cite{Eberard2006}. 
\end{Remark}
\begin{Remark}
On the Legendre submanifold generated by 
$\cH_{\,T_{\,0}}^{\,\tot}$,  the relations 
$$
V=\frac{\partial\cH}{\partial Q}
=\frac{Q}{C},\quad
I=\frac{\partial\cH}{\partial N}
=\frac{N}{L},\quad 
T=T_{\,0},\qquad \mbox{ and }\qquad
z
=\cH_{\,T_{\,0}}^{\,\tot},
$$
hold in the extended thermodynamic phase space, due to 
\fr{(n+1)-dimensional-Legendre-submanifold}. 
\end{Remark}

Then, one has the following proposition.
\begin{Proposition}
( Total energy and entropy production in the RLC 
circuit model in a thermal environment  ) : 
For the RLC circuit model in the thermal environment with $T_{\,0}$, 
one has the equations for $Q,N,V,I$ and $S$ as  
$$
\frac{\dr\,Q}{\dr t}
=I,\quad 
\frac{\dr\,N}{\dr t}
=-V-RI,\quad
\frac{\dr\,V}{\dr t}
=\frac{I}{C},\quad 
\frac{\dr\,I}{\dr t}
=-\,\frac{V}{L}-\frac{R}{L}I,\quad
\frac{\dr\,S}{\dr t}
=\frac{R}{T_{\,0}}I^{\,2}
>0,\quad
V=\frac{Q}{C},\quad
I=\frac{N}{L}.
$$ 
In addition, it follows that 
$$
\frac{\dr}{\dr t}\cH_{\,T_{\,0}}^{\,\tot}
=0.
$$
\end{Proposition}
\begin{Proof}
The equations for $Q,N,V,I$, and $S$ 
are obtained by substituting the identifications,  
$n=2$, and 
$$
x^{\,1}
=Q,\quad 
x^{\,2}
=N,\quad 
x^{\,3}
=S,\qquad
p_{\,1}
=V,\quad
p_{\,2}
=I,\quad 
p_{\,3}^{\,(0)}
=T_{\,0},
$$
$$
\psi(x^{\,1},x^{\,2})
=\cH(Q,N),\ 
\wt{\psi}(x^{\,1},x^{\,2},x^{\,3})
=\cH_{\,T_{\,0}}^{\,\tot}(Q,N,S\,),\quad
F_{\,\psi}^{\,1}(x^{\,1},x^{\,2})
=\frac{N}{L},\ 
F_{\,\psi}^{\,2}(x^{\,1},x^{\,2})
=-\frac{Q}{C}-\frac{R}{L}N,
$$
into  the equations in 
Proposition\,\ref{restricted-lift-general-psi-contact-Hamiltonian-high-dimension}.
\end{Proof}
\begin{Remark}
A physical interpretation of the equation involving $\dr\,S/\dr t$ 
is the entropy production. 
This equation is the same as that of 
Ref.\,\cite{Doty2011}.
When $R=0$, the entropy production vanishes. 
\end{Remark}

 \section{Concluding remarks}
\label{sec-summary}
This paper offers 
contact geometric descriptions of vector fields on dually flat spaces. 
This offer provides a view point that some ideas in information geometry 
can be written by means of contact geometric tools.
Based on these descriptions, basic properties of 
vector fields on contact manifolds lifted from 
Legendre submanifolds have been investigated.  
Throughout this paper, Legendre duality has explicitly been stated.  
In addition, some applications of the developed general theories to   
science and foundation of engineering have also been given.     

There are some potential future works that follow from this paper. 
One is to compare this work with other existing works. Such other works 
include the work on symplectic geometry adopted 
to a dually flat space discussed in Ref.\,\cite{Noda2011}, 
and relaxation dynamics on a statistical manifold in Ref.\cite{OW09}. 
In addition, it is interesting to construct 
the vector field associated with the Brayton-Moser equations 
on a dually flat space, where the Brayton-Moser equations describe 
a class of electric circuit models\cite{Eberard2006}. 
Another important future work is to import mathematical 
findings in contact geometry and topology to 
science and engineering. 
We believe 
that the elucidation of these remaining questions 
will develop the geometric theories in mathematical sciences and 
foundation of engineering.

\section*{Acknowledgments }
The author would like to thank 
Y. Shikano 
( Institute for Molecular Science ) and K. Umeno ( Kyoto University ) 
for supporting my work, 
and thank 
T. Wada, M. Koga, Y. Nakata, D. Tarama, 
and an anonymous referee 
for giving various comments on this work. 

\end{document}